\documentclass[reprint,twocolumn,superscriptaddress,aip]{revtex4-2}
\usepackage{lineno}\usepackage[normalem]{ulem}
\usepackage{xcolor}\usepackage{tikz}
\usepackage{amsmath,amssymb}
\usepackage{MnSymbol}
\usepackage{graphicx}
\usepackage{dcolumn}
\usepackage{bm}
\usepackage[sort&compress]{natbib}
\usepackage{subfigure}
\newsavebox{\measurebox}
\usepackage{cancel,pifont,wrapfig}
\usepackage{sidecap}
\newcommand{\bra}[1]{\left(#1\right)}

\newcommand{\rT}{\rho_{\text{T}}}
\newcommand{\rf}{\rho_{\text{f}}}
\newcommand{\rH}{\rho_{\text{H}}}
\newcommand{\rSN}{\rho_{\text{SN}}}

\newcommand{\vP}{{\bf P}}
\newcommand{\vp}{{\bf p}}
 \newcommand{\dx}{\text{d}{x}}

\newcommand{\dparf}[2]{\frac{\partial^2 #1}{\partial #2^2}}

\usepackage{pstricks}

\begin{document}

\title{Front propagation and global bifurcations in a multivariable reaction-diffusion model}

\author{Edgar Knobloch}
\affiliation{Department of Physics, University of California, Berkeley, California 94720, USA}

\author{Arik Yochelis}\email{yochelis@bgu.ac.il}
\affiliation{Department of Solar Energy and Environmental Physics, Swiss Institute for Dryland Environmental and Energy Research, Blaustein Institutes for Desert Research, Ben-Gurion University of the Negev, Sede Boqer Campus, Midreshet Ben-Gurion 8499000, Israel}%
\affiliation{Department of Physics, Ben-Gurion University of the Negev, Be'er Sheva 8410501, Israel}%

\date{\today}

\begin{abstract}
We study the existence and stability of propagating fronts in Meinhardt's {multivariable} reaction-diffusion model of branching in one spatial dimension. We identify a saddle-node-infinite-period (SNIPER) bifurcation of fronts that leads to episodic front propagation in the parameter region below propagation failure and show that this state is stable. Stable constant speed fronts exist only above this parameter value. We use numerical continuation to show that propagation failure is a consequence of the presence of a T-point corresponding to the formation of a {heteroclinic} cycle in a spatial dynamics description. Additional T-points are identified that are responsible for a large multiplicity of different {unstable} traveling front-peak states. The results indicate that multivariable models may support new types of behavior that are absent from typical two-variable models but may nevertheless be important in developmental processes such as branching and somitogenesis.
\end{abstract}

\maketitle
\noindent \textbf{Fronts are spatially localized interfaces separating two spatially homogeneous states such as oil and water, or different symmetry states, such as distinct crystallographic phases in a solid. Fronts can be stationary or they may propagate with one phase displacing or converting the other. Persistence of such propagating fronts or propagation failure as parameters vary are both of significance in applications. Motivated by branching phenomenology described by a reaction-diffusion model, we identify a novel mechanism responsible for front propagation failure whose details are described. Despite involving global bifurcations of different types, the mechanism is robust and hence is expected to represent a generic feature of this class of multivariable reaction-diffusion models, thereby offering a fresh view of front dynamics in biological systems exhibiting bistability.}

\section{Introduction}

Self-organization in natural systems often manifests itself in a rich variety of spatiotemporal patterns at all scales.\cite{whitesides2002self,cross2009pattern,meron2015nonlinear} In some cases, patterns are an inspiring source of mathematical complexity while in many other cases they underlie fundamental functional processes ranging from adverse physiological conditions, such as cardiac arrhythmia,\cite{alonso2016nonlinear} to essential technological properties, such as magnetic domains in computer memory devices.\cite{parkin2008magnetic} Either way, it is known that many of the observed behaviors exhibit model-independent properties and so can be understood on the basis of universal principles, see e.g. the seminal review by Cross and Hohenberg.\cite{ch93}

Our interest here is in systems that are driven far from equilibrium, specifically reaction-diffusion systems, exhibiting two distinct states corresponding, for example, to different concentrations, separated by an interface between them. Such structures are referred to as \textit{fronts} or domain walls. Typically, such fronts arise in models with two coexisting linearly stable spatially homogeneous states. In these circumstances the speed of the front is necessarily selected via nonlinear processes (such fronts are {\it pushed}) in contrast to front propagation into a linearly unstable state (a {\it pulled} front). In gradient systems the speed of pushed fronts is selected via energy considerations~\cite{burke2006localized,knobloch2016localized,champneys2021bistability} but in generic systems the front speed solves a nonlinear eigenvalue problem that must be solved numerically. Beyond their significance in phase-separating processes, fronts are instrumental in macropinocytosis mediated by the dynamics of circular dorsal ruffles in cells~\cite{bernitt2017fronts} and in the uptake of extracellular matter, for example, while front propagation failure has been associated with the transition to a cancerous phenotype.\cite{hoon2012functions,itoh2013mechanistic} Consequently, understanding the presence and/or absence of front {propagation}, even in one spatial dimension, often provides significant information even if the medium is fundamentally three-dimensional.

Geometrically, stationary or propagating fronts can be viewed as spatial heteroclinic connections in the comoving frame. Such global connections arise already in scalar models,\cite{pismen2006patterns} e.g.~Fisher or Allen-Cahn equations, and may be robust. Models of this type, including two-variable models with mass conservation,\cite{brauns2020phase,champneys2021bistability} are amenable to phase plane analysis and so are frequently employed in the study of fronts. In two-variable models without conservation fronts may become linearly unstable to an oscillatory instability, a possibility studied in detail by Ponedel and Knobloch.\cite{ponedel2016forced} Recently, however, a new type of front behavior was observed in a multivariable activator-inhibitor model: a nonlinear transition from a steady propagating front to a strongly modulated front with spatially localized oscillations along the front.\cite{yochelis2021nonlinear,knobloch2021stationary,knobloch2022instability} We show that this transition occurs via a global bifurcation in time called a saddle-node-infinite-period (SNIPER) bifurcation and is nonhysteretic. Moreover, this bifurcation occurs already in one spatial dimension (1D).

This SNIPER bifurcation is of particular interest because it represents \textit{propagation failure}: at the bifurcation a stable steadily propagating front annihilates with an unstable front, and no steadily propagating fronts are present beyond the transition. This is precisely the regime in which one finds the above-mentioned, strongly modulated fronts generated by this bifurcation. In the present work we are interested in uncovering the origin of this novel behavior. For this purpose we study Meinhardt's model of branching, a four-variable activator-inhibitor model.\cite{meinhardt1976morphogenesis} In Section~\ref{sec:model} we provide a basic overview of the model and computational methodology, followed by numerical evidence for a global SNIPER bifurcation of fronts. The organization of steady propagating fronts is key and is studied in Section~\ref{sec:fronts} using numerical continuation, revealing the presence of several distinct T-points, i.e. codimension-two points at which two distinct fronts are simultaneously present. These points are ultimately responsible for the large multiplicity of different {unstable} fronts (heteroclinic connections) and {unstable} peak states (homoclinic connections) present in this model as well as the above-mentioned propagation failure. The implications of our results are summarized in Section~\ref{sec:discussion}.

\begin{figure*}[tp]
    {\includegraphics[width=0.95\textwidth]{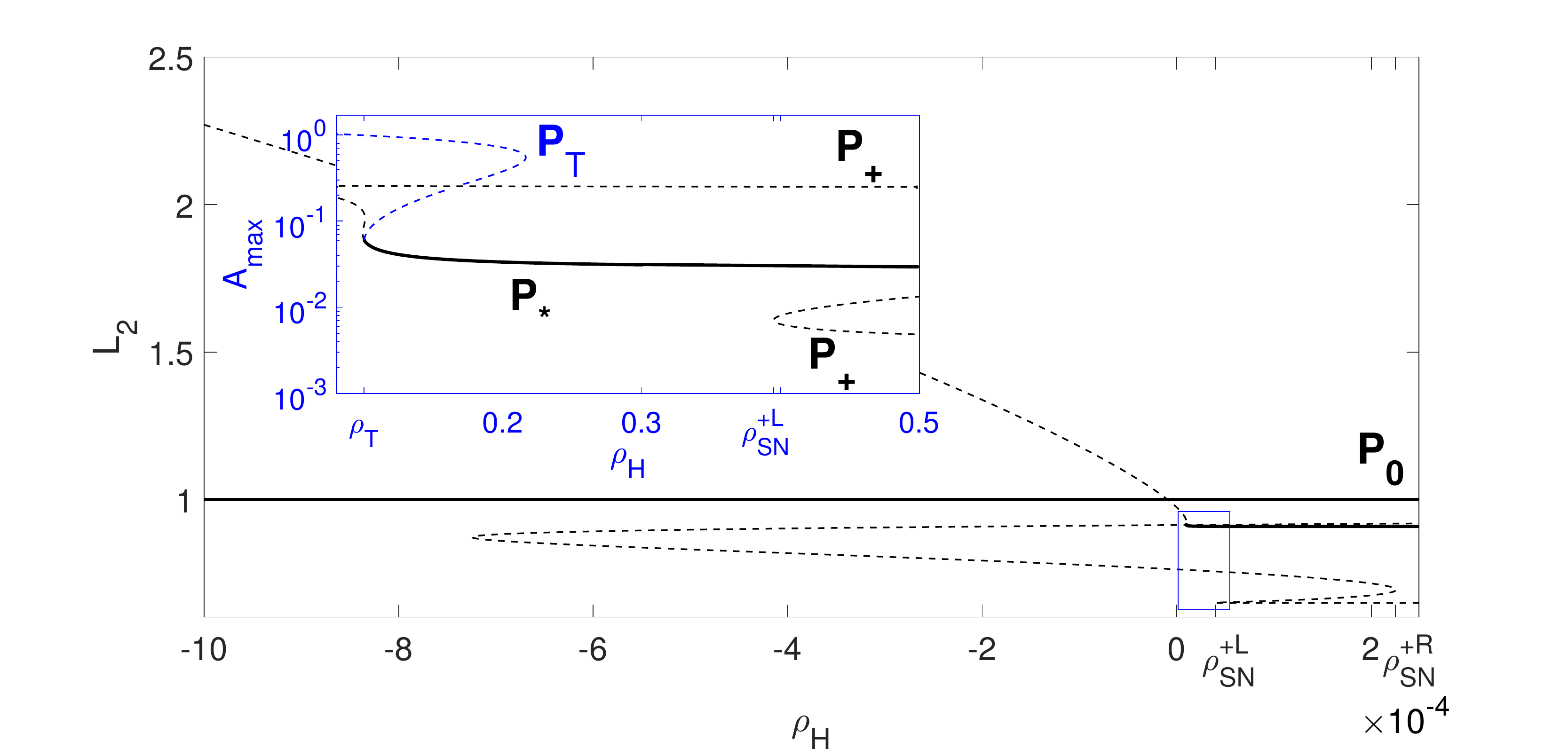}}
    \caption{Bifurcation diagram showing the coexistence of stable ($\vP_0$ and $\vP_*$, solid black lines) and unstable ($\vP_+$, dashed black lines) spatially uniform states together with the branch of unstable periodic Turing states ($\vP_{\rm T}$, dashed blue line in inset) in terms of the ${\text L}_2$ norm defined in Eq.~\eqref{eq:L2}, as computed from Eqs.~\eqref{eq:AI} with the parameters given in the text. The inset zooms into the rectangle in the lower right, showing the uniform state $\vP_*$ and Turing state $\vP_{\rm T}$, as well as a small portion of the intermediate homogeneous state $\vP_+$. {The uniform state $\vP_*$ is linearly stable with respect to uniform perturbations down to the fold at $\rH=\rho_{\rm SN}^*\approx 0.99005\times 10^{-5}$ but unstable to spatially periodic perturbations (Turing unstable) in the interval $\rho_{\rm SN}^*<\rH<\rho_{\rm T}\approx 1.0011\times 10^{-5}$.} All three homogeneous states, and in particular the saddle nodes of $\vP_+$ located at $\rH=\rho_{SN}^{+L}\approx 0.395\times 10^{-4}$ and $\rH=\rho_{SN}^{+R}\approx 2.25\times 10^{-4}$, play an important role. Reproduced with modifications to the layout from Knobloch and Yochelis, Instability mechanisms of repelling peak solutions in a multivariable activator-inhibitor system, Chaos \textbf{32}, 123129 (2022), with the permission of AIP Publishing.}
	\label{fig:bif_uni}
\end{figure*}

\section{Problem setup}\label{sec:model}

\subsection{Model equations, coexistence and bistability of uniform solutions}

Following Yochelis,\cite{yochelis2021nonlinear} we employ Meinhardt's four-variable activator-inhibitor-substrate model,\cite{meinhardt1976morphogenesis} where the four fields $A$, $H$, $S$ and $Y$ represent, {respectively}, the concentrations of an activator, an inhibitor, the substrate, and a marker for differentiation:\cite{yao2007matrix}
\begin{subequations}\label{eq:AI}
	\begin{eqnarray}
	\frac{\partial A}{\partial t}&=&c\dfrac{SA^2}{H}-\mu A+\rho_{\text{A}} Y+D_{\text{A}} \dparf{A}{x}, \\
	\frac{\partial H}{\partial t}&=&cSA^2-\nu H+\rho_{\text{H}} Y+D_{\text{H}} \dparf{H}{x}, \\
	\frac{\partial S}{\partial t}&=&c_0-\gamma S-\varepsilon Y S+D_{\text{S}} \dparf{S}{x}, \\
	\frac{\partial Y}{\partial t}&=&d A-eY+\dfrac{Y^2}{1+fY^2}+D_{\text{Y}} \dparf{Y}{x}.
    \end{eqnarray}
\end{subequations}
Although this model was suggested by Meinhardt~(\citeyear{meinhardt1976morphogenesis}) as a framework for studying branching phenomena, a fundamentally higher-dimensional process (see also Ref.\cite{yao2007matrix}), we focus here solely on front dynamics in 1D. As in Refs.\cite{yochelis2021nonlinear,knobloch2021stationary,knobloch2022instability}, we employ the inhibitor control level, $\rho_{\text{H}}$, as a control parameter while keeping all other parameters fixed and within the range of previous studies:\\ \\
\noindent
$c=0.002$, $\mu=0.16$, $\rho_{\text{A}}=0.005$, $\nu=0.04$, $c_0=0.02$, $\gamma=0.02$, $\varepsilon=0.1$, $d=0.008$, $e=0.1$, $f=10$, $D_{\text{A}}=0.001$, $D_{\text{H}}=0.02$, $D_{\text{S}}=0.01$, $D_{\text{Y}}=10^{-7}$.\\ \\
\noindent {The choice $D_{\text{Y}}=10^{-7}$ {reflects} negligible diffusion (in the spirit of the original model) but we also show results for $D_{\text{Y}}=10^{-4}$ to model moderate diffusion.} In either case $D_{\text{Y}}$ is at least an order of magnitude lower than {the diffusion coefficient} of any of the other chemical components, here $D_{\text{A}}$. 

 {The system (\ref{eq:AI}) admits two competing spatially uniform solutions, $\vP_0\equiv (A_0,H_0,S_0,Y_0)=(0,0,c_0/\gamma,0)$ and $\vP_*\equiv (A_*,H_*,S_*,Y_*)$. The former is stable but with $D_{\text{Y}}=10^{-7}$ the latter undergoes a Turing bifurcation at $\rT\approx 1.0 \times 10^{-5}$, with $\vP_*$ stable for $\rH>\rT$ and unstable for $\rH<\rT$. This bifurcation is subcritical, creating a branch of unstable spatially periodic states $\vP_{\rm T}$ in $\rH>\rT$.\cite{yochelis2021nonlinear} Figure~\ref{fig:bif_uni} shows both $\vP_0$ and $\vP_*$ (black lines, solid lines indicate stability) together the branch $\vP_{\rm T}$ of unstable Turing states (inset, dashed blue line). The figure also shows additional branches of unstable uniform solutions, collectively labeled as $\vP_+$, that are also present (dashed black lines). These states also play a role in the results that follow, despite their instability. Figure~\ref{fig:bif_uni} and subsequent figures employ the norm}
\begin{equation}\label{eq:L2}
    {\text L}_2\equiv \sqrt{L^{-1}\int_{0}^{L} \dx {\sum_{j={A,H,S,Y}} P_j^2+(\partial_x P_j)^2}}
\end{equation}
that takes into account both the solution variables and their spatial derivatives. Here $L$ is the domain length. {The maximum amplitude $A_{\rm max}$ of the field $A$ is also used to present some of our results.}

In recent work,\cite{knobloch2021stationary,knobloch2022instability} we showed that the Turing instability in~\eqref{eq:AI} is subcritical and gives rise to multiple repelling spatially localized states in both 1D and 2D. Those in 1D are organized within a bifurcation structure called foliated homoclinic snaking but are unstable. The detailed organization of the solution branches in 2D is unknown but both isolated peaks or spots and groups of spots can be stable. Next, we consider the behavior of fronts connecting the homogeneous state $\vP_*$ to the trivial state $\vP_0$ and their stability properties.

\subsection{Stable fronts and propagation failure}\label{sec:front_inst}

Since $\vP_*$ and $\vP_0$ are both stable for $\rH>\rT$, fronts connecting them in this regime are pushed fronts, and their speed is therefore determined by nonlinear processes. Moreover, any instability mode will necessarily be a consequence of spatial variation and hence be localized at the front.

A simple front $\vP(x,t)$ connecting $\vP_*$ to $\vP_0$ has the property that $\vP(x,t)\to \vP_*$ as $x\to-\infty$ and $\vP(x,t)\to \vP_0$ as $x\to \infty$, or vice versa, and propagates without change of shape with constant speed $v$ that depends on the parameter $\rH$. In the context of branching phenomena $\vP_*$ represents the differentiated state and we therefore expect that $\vP_*$ always invades $\vP_0$, i.e. $v>0$; in the model this is indeed the case.
The speed $v$ solves a nonlinear eigenvalue problem obtained by writing $\vP(x,t)=\vP(\xi)$, $\xi\equiv x-vt$. The resulting eigenvalue problem can be written as an eight-dimensional dynamical system:
\begin{subequations}\label{eq:AIode}
		\begin{eqnarray}
			A'&=&-a, \\
			a'&=&D^{-1}_{\text{A}} \bra{c\dfrac{SA^2}{H}-\mu A+\rho_{\text{A}} Y+va}, \\
			H'&=&-h, \\
			h'&=&D^{-1}_{\text{H}} \bra{cSA^2-\nu H+\rho_{\text{H}} Y+vh}, \\
			S'&=&-s, \\ 
			s'&=&D^{-1}_{\text{S}} \bra{c_0-\gamma S-\varepsilon Y S+vs}, \\ 
			Y'&=& {-y}, \\
			 {y}'&=&D^{-1}_{\text{Y}} \bra{d A-eY+\dfrac{Y^2}{1+fY^2}+ {vy}},
		\end{eqnarray}
	\end{subequations}
where the prime denotes a derivative with respect to $\xi$. The eigenvalue $v$ is determined by the requirement that there exists a heteroclinic connection between $\vP_*$ and $\vP_0$. Generically, one expects such a connection to leave $\vP_*$ along the eigenvector of the unstable spatial eigenvalue of $\vP_*$ closest to $0$ and approach $\vP_0$ along the corresponding stable eigenvector of $\vP_0$. 

The above problem can be solved using the numerical continuation software AUTO~\cite{doedel2012auto} on a large but finite domain with Neumann boundary conditions (NBC) replacing the asymptotic conditions specified above. The temporal stability of the resulting solutions can be examined on writing 
\begin{equation}
\vP(x,t)=\vP(\xi)+\epsilon\widetilde{\vp}(\xi,t),
\end{equation}
and linearizing in $\epsilon$, where $|\epsilon|\ll1$. The resulting linear eigenvalue problem is solved for the instability growth rate $\sigma$ subject to the requirement that $\widetilde{\vp}\to0$ as $\xi\to\pm\infty$. Once again the calculation is performed on a large but finite domain.

 {For some calculations we use periodic boundary conditions (PBC) in place of NBC. The former require that the solution asymptote to the same state at both ends of the domain, $\vP_0$ say, while the latter allow for different states at either end, for example, $\vP_*$ at $x=0$ and $\vP_+$ at $x=L$. Thus NBC are appropriate for studying fronts connecting $\vP_*$ to $\vP_0$, while PBC are appropriate for steady states or traveling waves of constant form. The latter may resemble a front connecting $\vP_*$ to $\vP_0$ but are necessarily accompanied by a trailing reverse connection traveling with the same speed as the front. Both boundary conditions may be used to shed light on front propagation in the present system, as discussed further below.}

\begin{figure}[ht!]
    (a)\includegraphics[height=3.65cm]{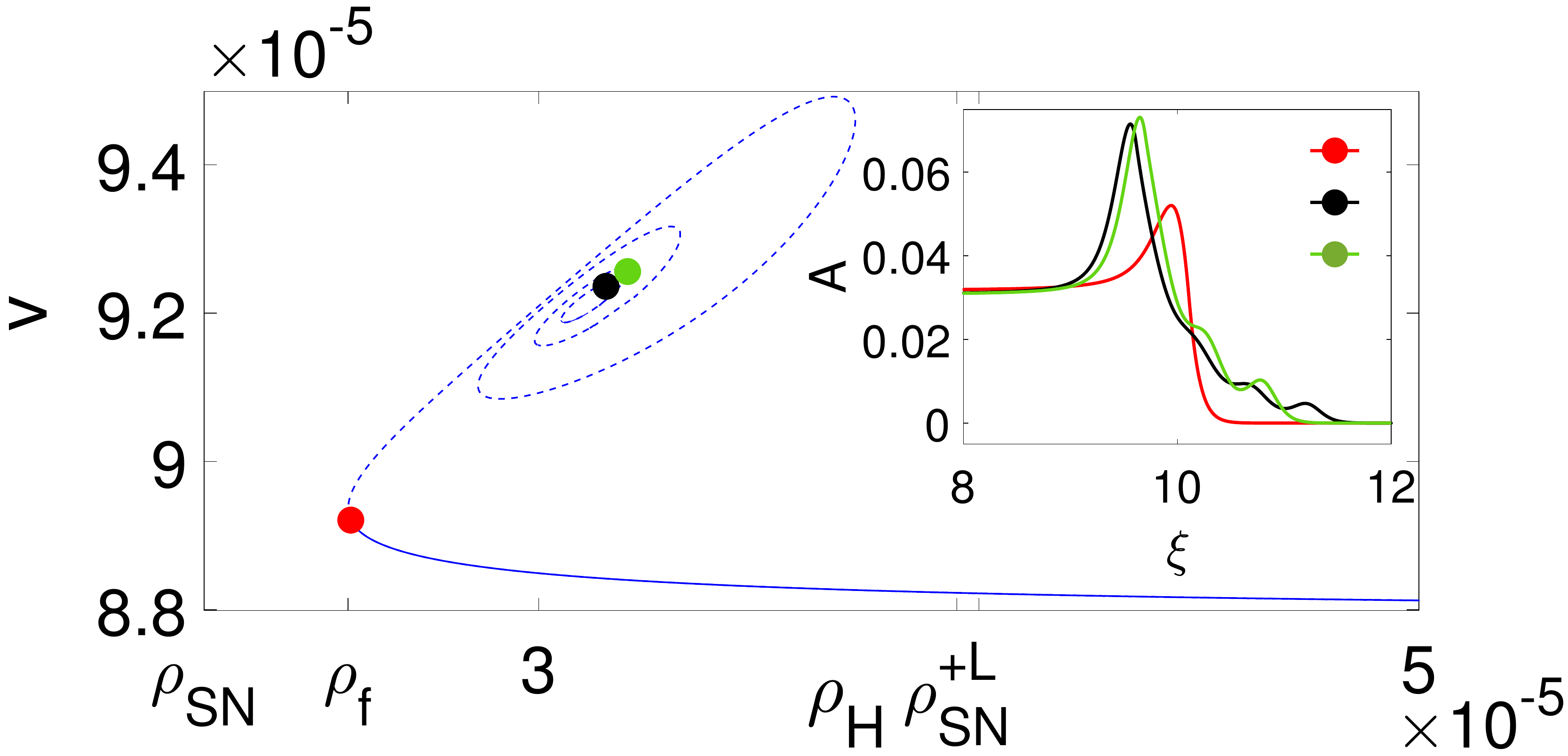}\vskip 0.15in
    (b)\includegraphics[height=3.7cm]{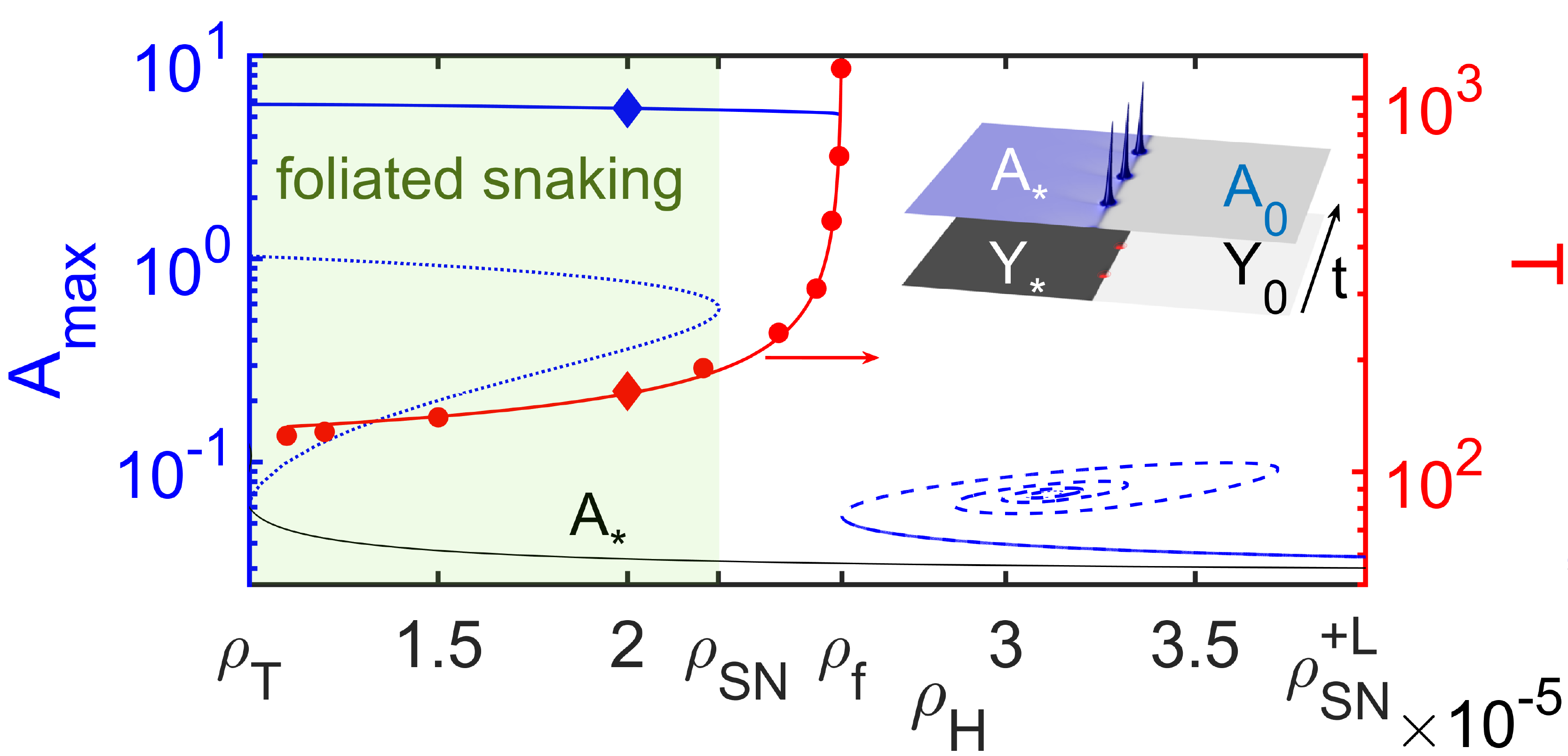}\vskip 0.1in
    (c)\includegraphics[height=3.7cm]{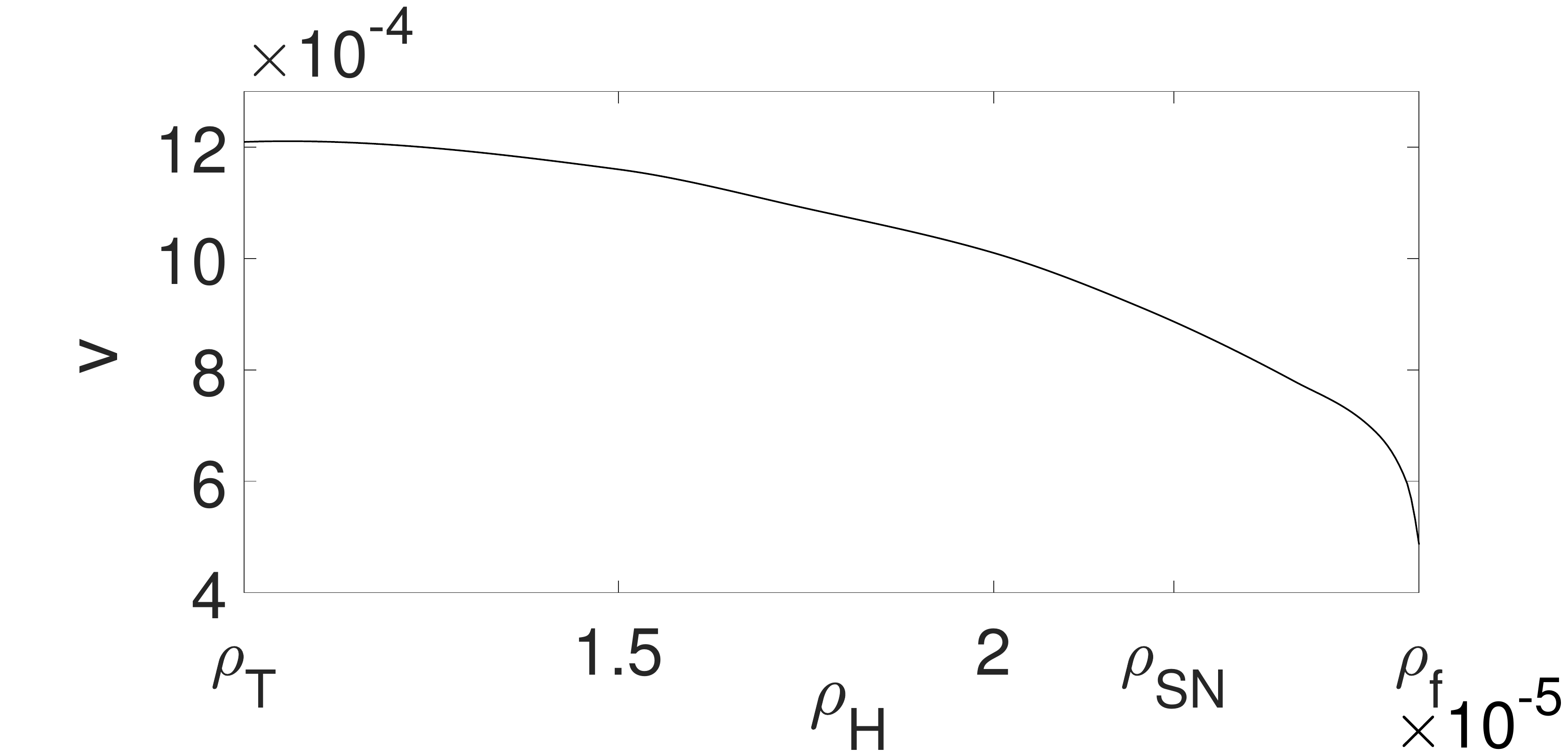}
    \caption{(a) Speed $v$ of steadily translating front solutions connecting $\vP_*$ to the trivial state $\vP_0$ computed with NBC far from the front and $D_{\text{Y}}=10^{-7}$. Stable front solutions are present below the red dot corresponding to the fold point $\rf\approx 2.6 \times 10^{-5}$. Above this point the solution branch spirals into a codimension-two point tentatively identified as a T-point. The inset shows the front profiles in the comoving frame at three color-coded locations along the spiral branch and reveals that each turn of the branch results in an additional oscillation in the front profile as it approaches $\vP_0$. (b) Same as (a) but showing the oscillating front present for $\rH<\rf$ in terms of its maximum amplitude $A_{\rm max}$ (left axis, solid blue line) together with its oscillation period $T$ (right axis, solid red). The shaded area indicates the foliated snaking region, $\rT<\rH<\rSN$ (for details, see Ref.~\cite{knobloch2021stationary}). Bullets show the front oscillation period $T$ obtained from direct numerical integration of Eqs.~(\ref{eq:AI}) while the solid red line is a fit to $1/\sqrt{\rf-\rH}$. The space-time plot in the inset represents a typical oscillating front over approximately three oscillation periods $T \approx 150$, computed at $\rH=2\times 10^{-5}$ (diamond symbols), and shows the $A$ field (top) together with the $Y$ field (bottom), the dark colors indicating higher values of the field. The computation in the inset was performed with $x\in [0,20]$ and $t\in[0,500]$. (c) Same as (a) but showing the {mean speed of the oscillating front in $\rH<\rf$, averaged over the period $T$,} obtained by direct numerical integration of Eqs.~\eqref{eq:AI}.}
\label{fig:speed}
\end{figure}

In Figure~\ref{fig:speed}(a), we show that stable steadily translating fronts exist for $\rH>\rf \approx 2.6 \times 10^{-5}$ and below the red dot in the figure. The speed of these fronts is of order $10^{-4}$ and slowly increases towards $v=v_f\approx 8.94\times 10^{-5}$ as one approaches the fold point at $\rH=\rf$ from above, where the stable fronts annihilate with unstable fronts. The front profile (red curve) at this point is shown in the inset and exhibits an overshoot at the location of the front. The leading spatial eigenvalues $\lambda$ of $\vP_*$ at ($\rH,v$)=($\rf,v_{\rm f}$) are $\lambda_s=-1.4163$ and $\lambda_u=1.4207$, consistent with the observed overshoot, i.e., the presence of a growing spatial instability of $\vP_*$ as $\xi$ increases from $-\infty$. The corresponding spatial eigenvalues of $\vP_0$, $\lambda_s=-1.4097$ and $\lambda_u=1.4164$, are also real, accounting for the monotonic approach of the profile to $\vP_0$ as $\xi\to\infty$. Both sets of eigenvalues remain real for larger values of $\rH$. Consequently, the behavior of the front profile as $\xi\to\pm\infty$ remains qualitatively the same as $\rH$ increases.

Beyond the fold point $\rH=\rf$ the branch of unstable front solutions organizes into a spiral, adding with each turn an additional oscillation to the front profile as it descends towards $\vP_0$, as shown by the {green} and black profiles in the inset in Fig.~\ref{fig:speed}(a). We speculate that this spiral form is associated with a codimension-two T-point,\cite{glendinning1986t,zimmermann1997pulse,sneyd2000traveling,or2001pulse,romeo2003stability,champneys2007shil,yochelis2022versatile,moreno2022bifurcation,raja2023} but could not continue the computations closer to such a point. Because stable, constant speed fronts do not exist below $\rf$ we think of $\rf$ as representing {\it propagation failure}. However, as shown in the following sections, this characterization is not completely accurate: the region $\rH<\rf$ does in fact contain stable propagating fronts, albeit not of constant form; it also contains unstable fronts of constant form generated by a complex mechanism involving the T-point. 

\begin{figure*}[tp]
\sbox{\measurebox}{%
\begin{minipage}[b]{0.62\textwidth}
  {(a)\includegraphics[width=\textwidth]{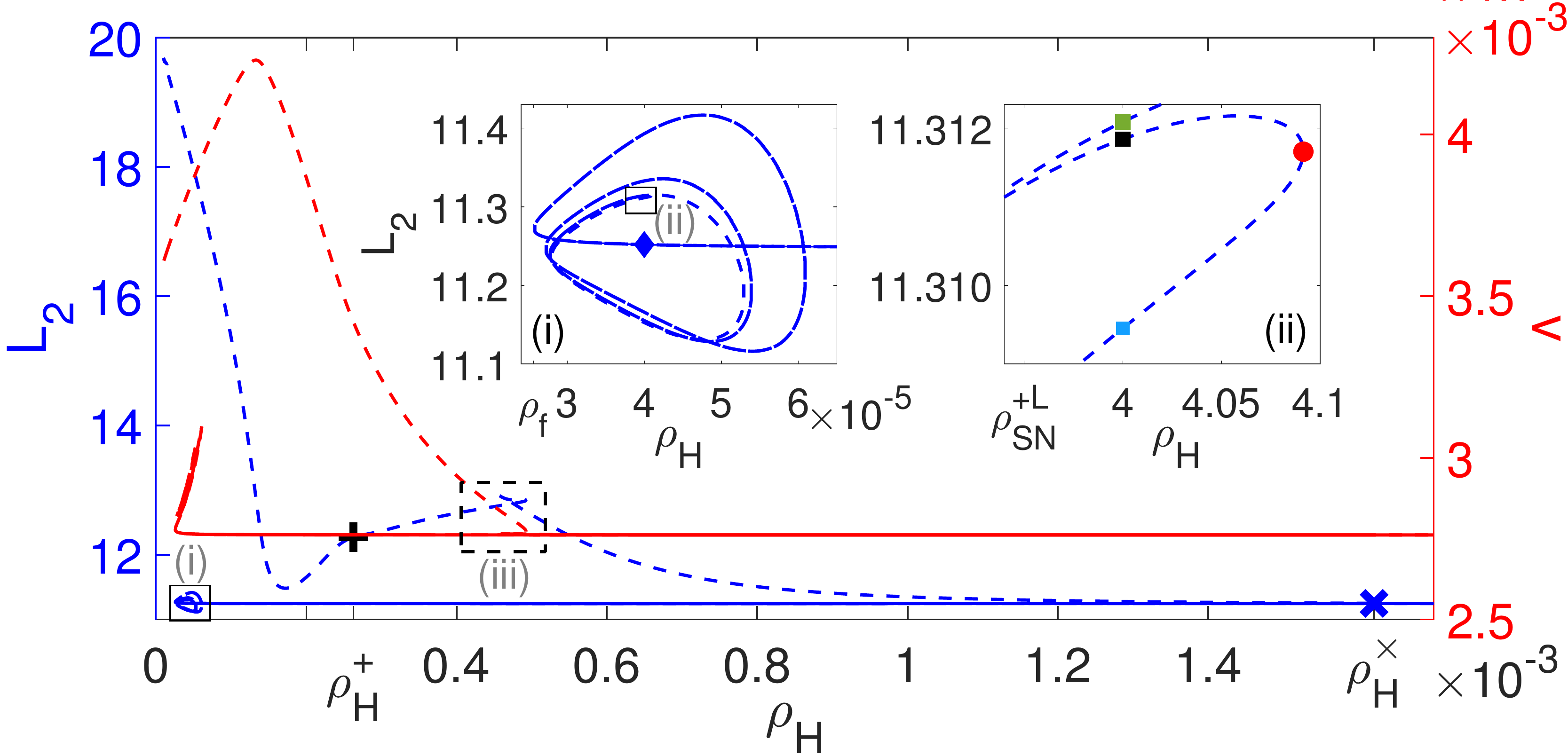}}
  {(e)\hskip 0.2in \includegraphics[width=0.9\textwidth]{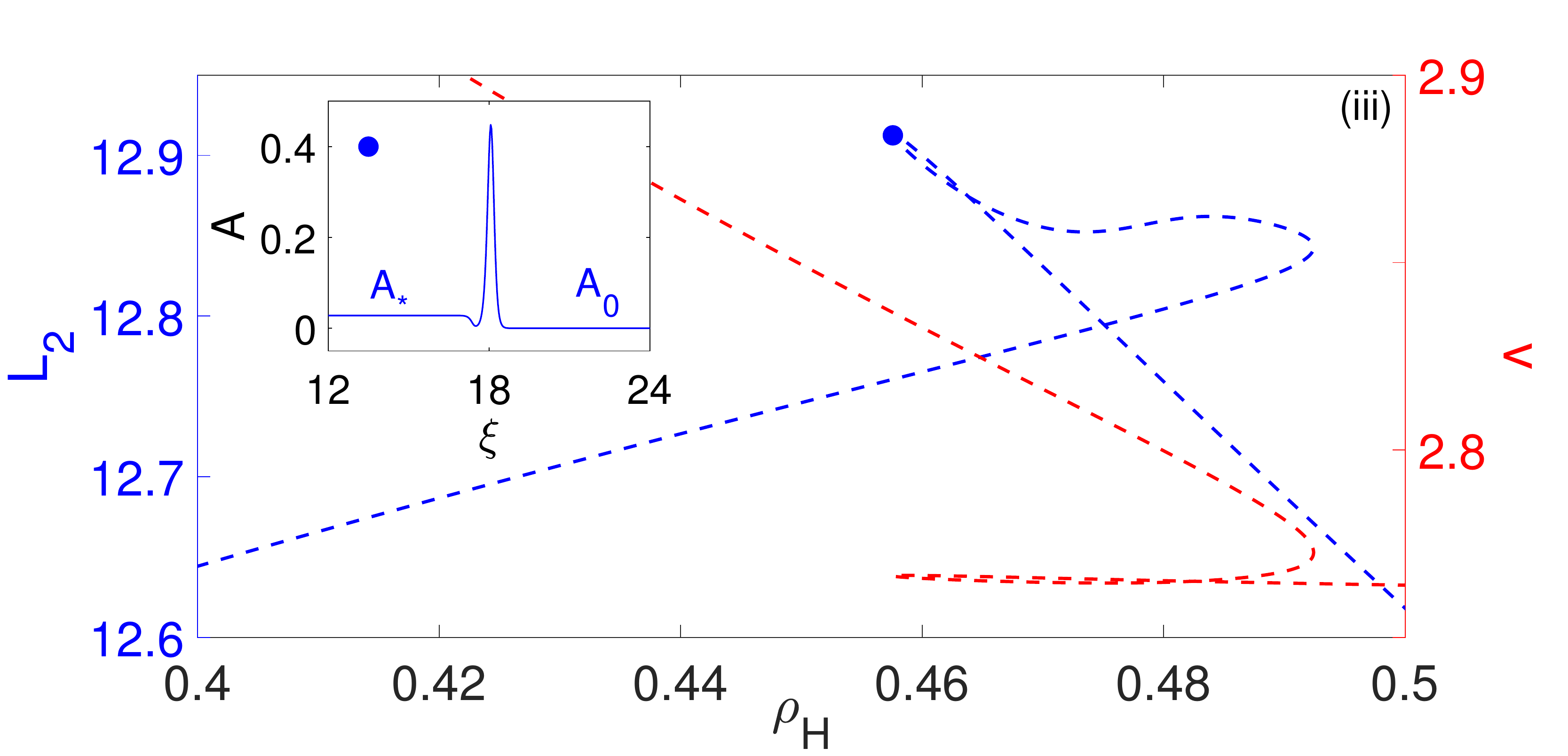}}
\end{minipage}}
\usebox{\measurebox}\qquad \qquad
\begin{minipage}[b][\ht\measurebox][s]{.25\textwidth}
  {(b)\includegraphics[width=\textwidth]{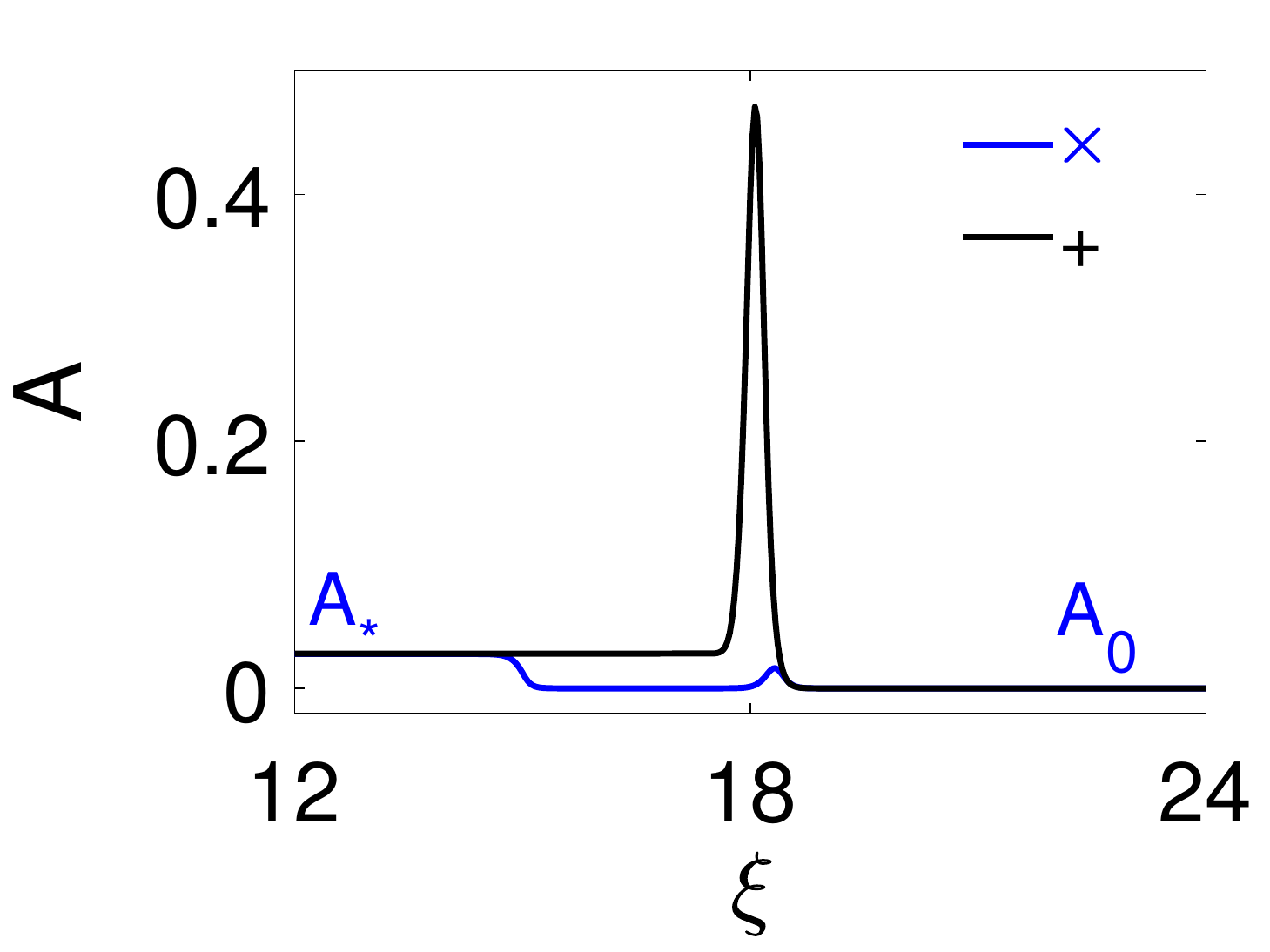}}
  {(c)\includegraphics[width=\textwidth]{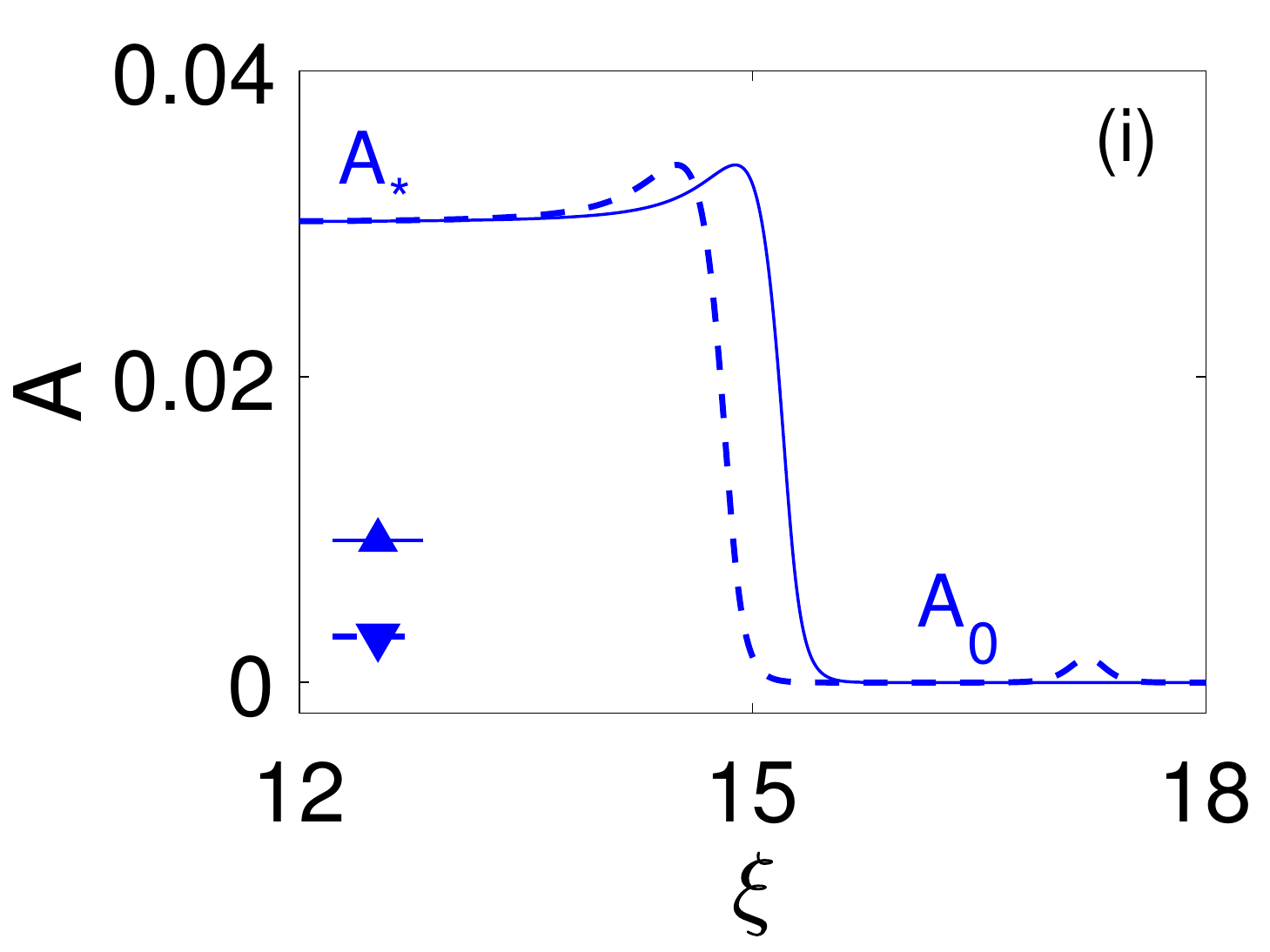}}
  {(d)\includegraphics[width=\textwidth]{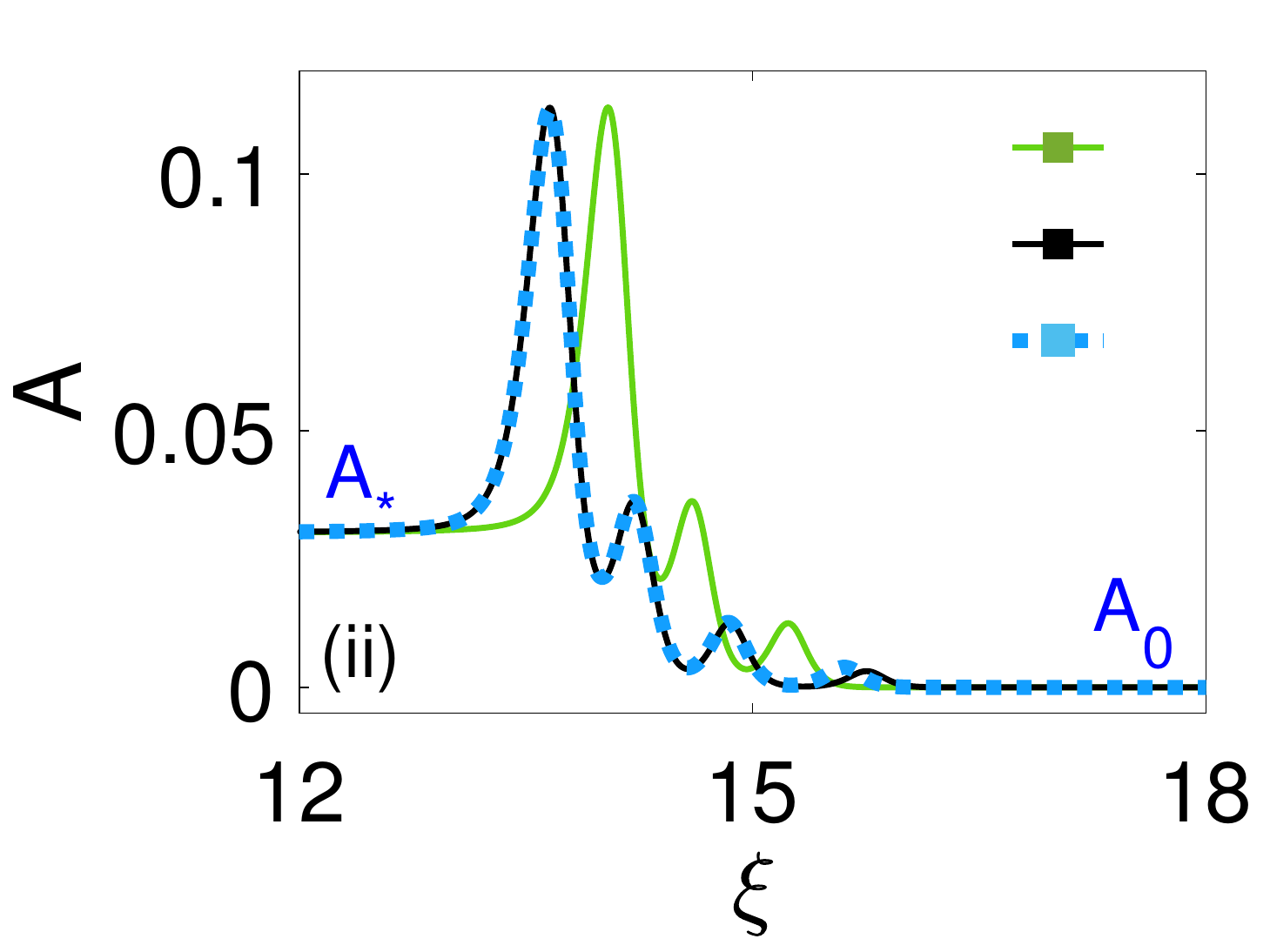}}
\end{minipage}
\caption{ {(a) Bifurcation diagram, showing constant velocity front and locked front-peak solutions in terms of the ${\rm L}_2$ norm (left axis, blue) and speed $v$ (right axis, red); $v>0$ indicates rightward propagation. Inset (i) zooms into the bottom left region while inset (ii) represents a further zoom into the same region (at L$_2\simeq 11.3$). The unstable fast fronts (upper branch, dashed blue) develop a double structure beyond $\rH=\rH^+\sim 0.26\times 10^{-3}$ (black $+$ symbol, profile in (b)) and undergo a fold at $\rH=\rH^\times\approx 1.62\times 10^{-3}$ (blue $\times$ symbol, profile in (b)). (c,d) Profiles at the color-coded locations on the blue branch in insets (i,ii), respectively, all at $\rH=4.0 \times 10^{-5}$, with the profile corresponding to the blue square in (ii) shown using small blue squares to distinguish it from the solid black profile in (d). The red dot in (ii) defines the transition from inward spiraling to outward spiraling. The slow fronts come in from the right and are linearly stable down to $\rH=\rf$ (lowest branch, solid blue) while all other front solutions are unstable (dashed lines). (e) Zoom of the upper fast front branch, inset (iii) in the main panel, with an inset showing the profile at the fold marked by the solid blue dot. Here $D_{\text{Y}}=10^{-4}$, while all other parameters remain the same as in Fig.~\ref{fig:speed}; all solutions are computed in a domain of size $L=30$ with NBC.}}
\label{fig:DYm4}
\end{figure*}

\begin{figure*}[tp]
\sbox{\measurebox}{%
\begin{minipage}[b]{0.67\textwidth}
  {(a)\includegraphics[width=\textwidth]{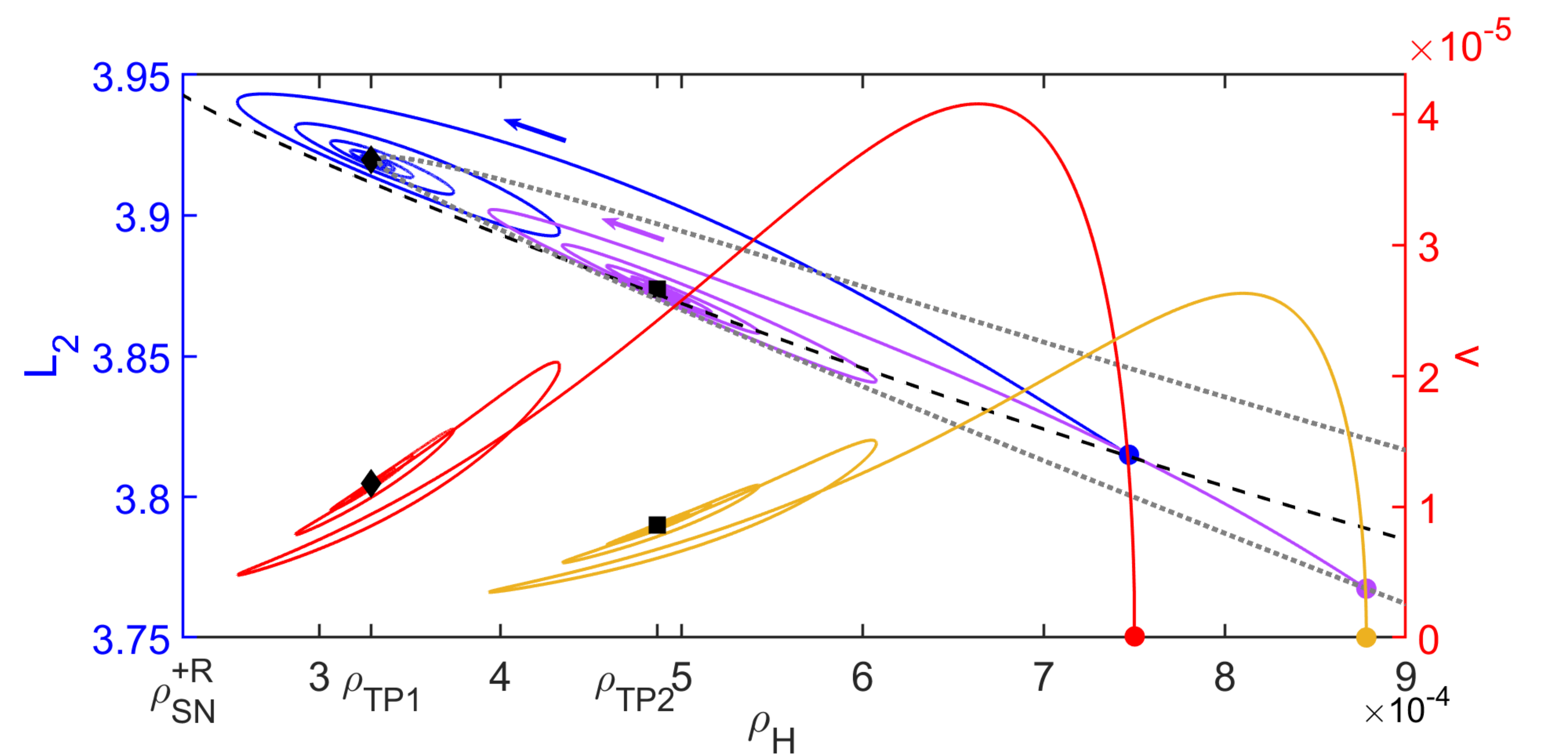}}
  {(d)\includegraphics[width=\textwidth]{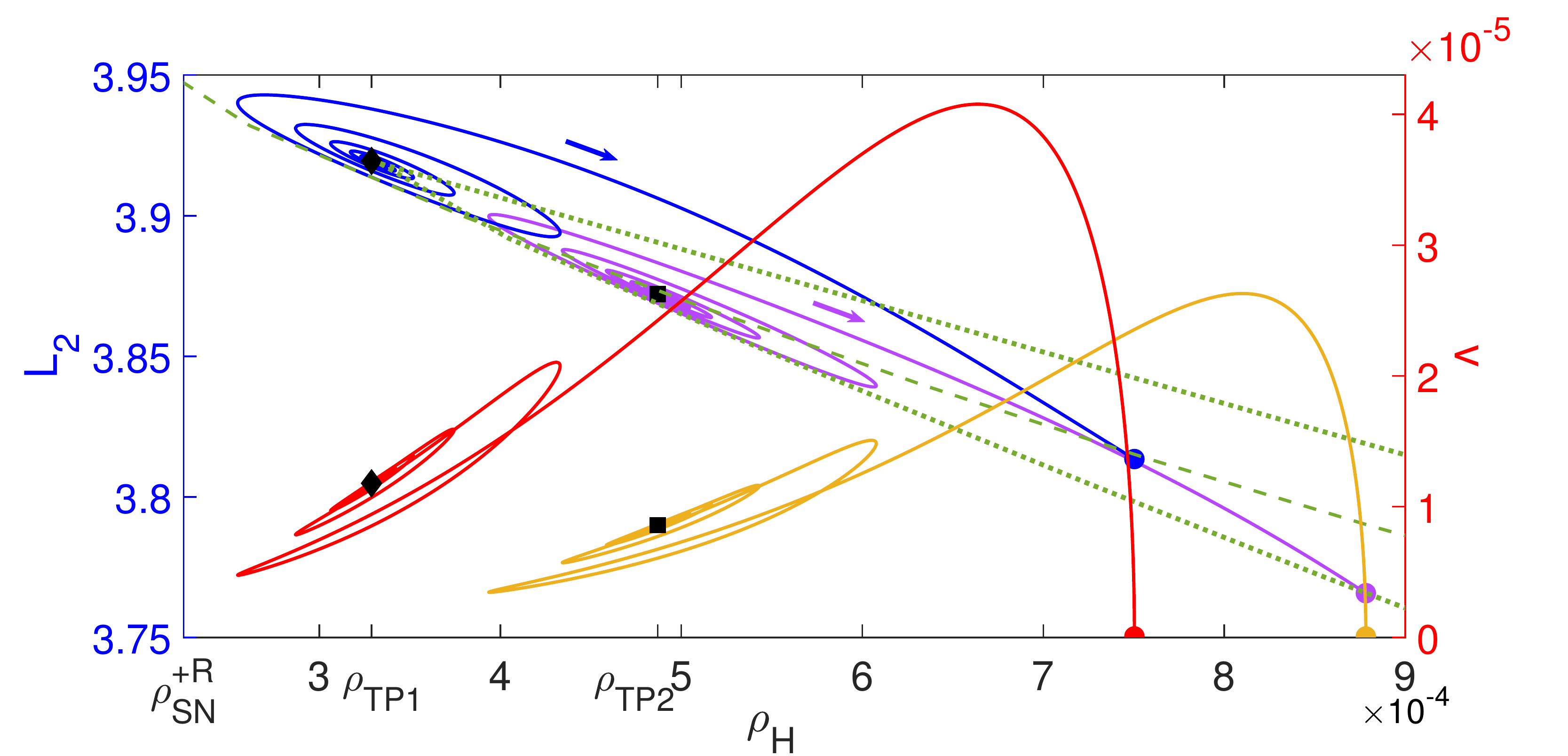}}
\end{minipage}}
\usebox{\measurebox}\qquad  \qquad
\begin{minipage}[b][\ht\measurebox][s]{.215\textwidth}
  {(b)\includegraphics[width=\textwidth]{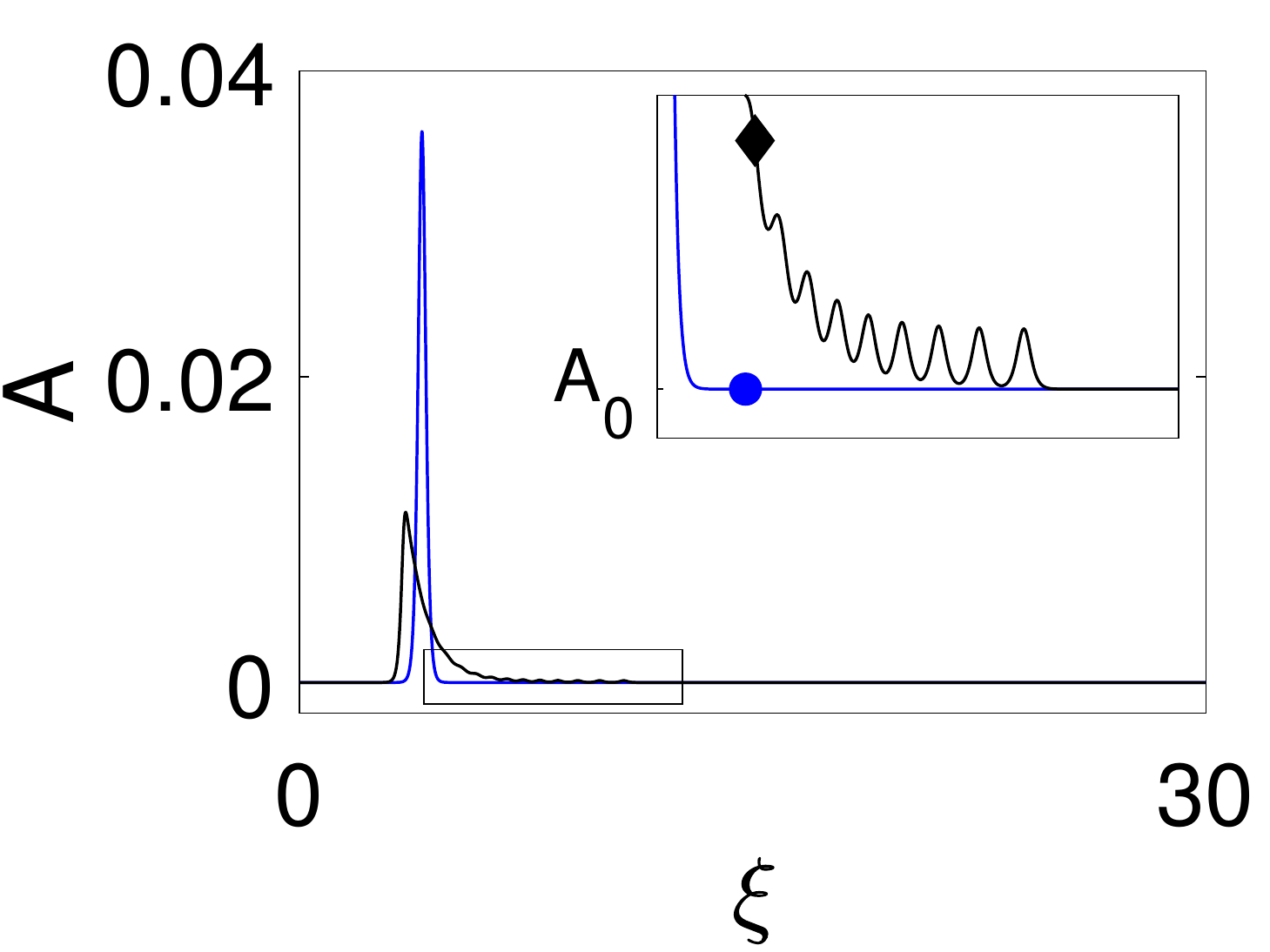}}
  {(c)\includegraphics[width=\textwidth]{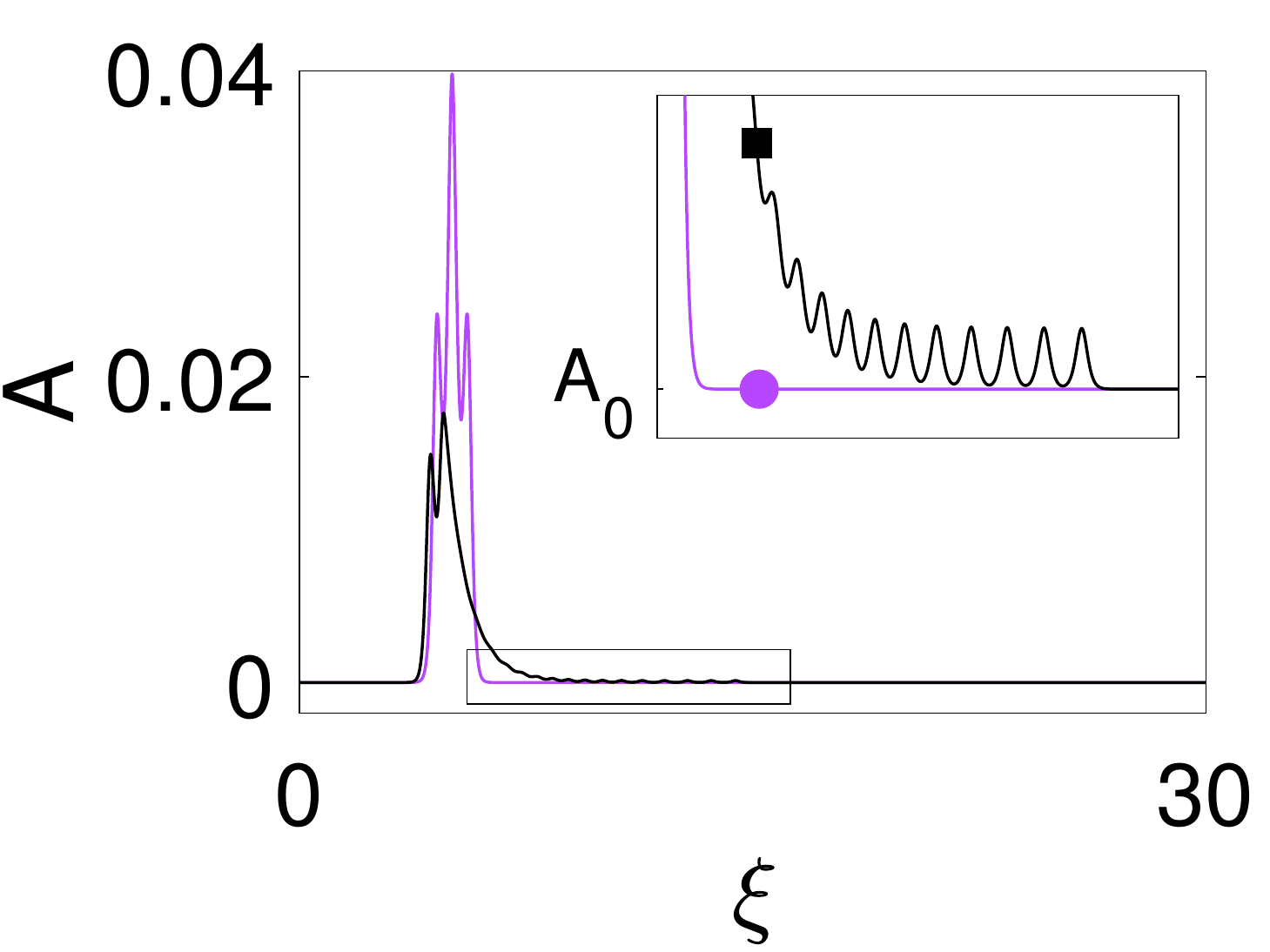}}
  {(e)\includegraphics[width=\textwidth]{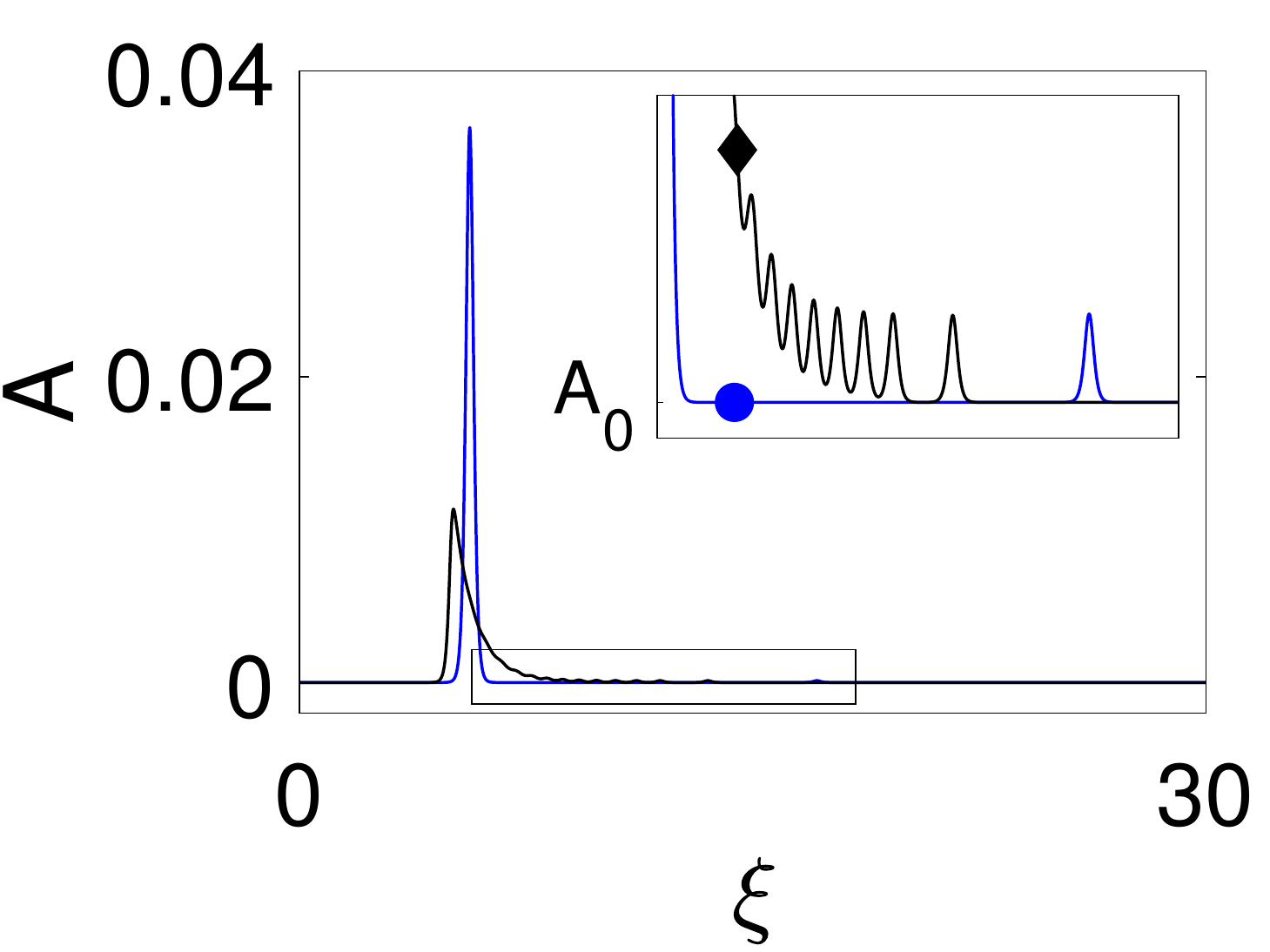}}
  {(f)\includegraphics[width=\textwidth]{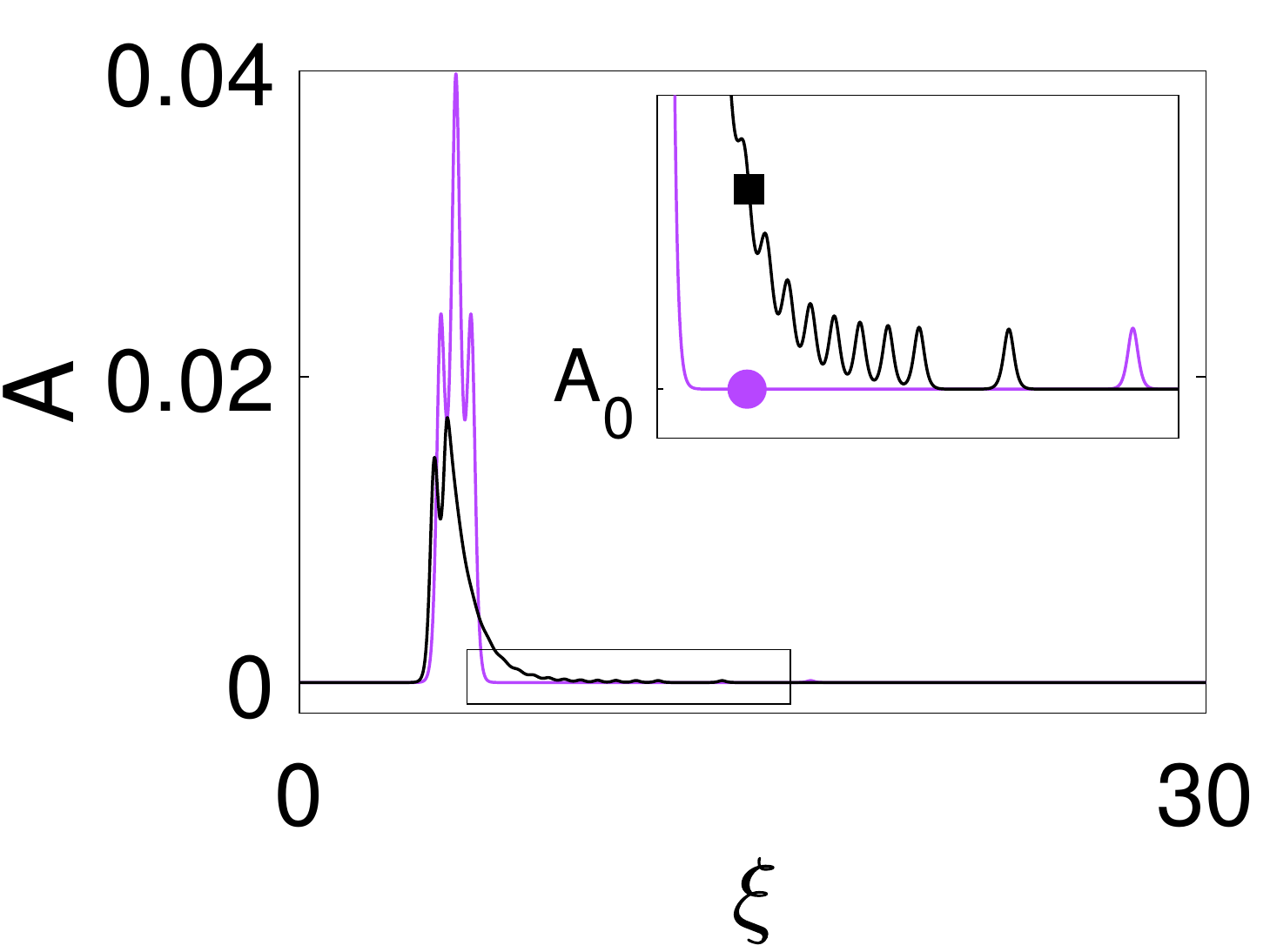}}
\end{minipage}
  \caption{ {(a) Traveling peaks without the small amplitude precursor peak on a periodic domain of length $L=30$, revealing the presence of two additional T-points (TP1: blue, TP2: purple); arrows indicate the direction of spiraling into the T-points. The right axis shows the corresponding velocities (TP1: red, TP2: orange). Associated branches of symmetric stationary states are shown in dashed (1P: black) and dotted (3P: gray) lines, respectively. (b,c) Color-coded solution profiles at the locations indicated in panel (a), corresponding to the parity-breaking bifurcation (solid dots, symmetric profiles), TP1 (black diamond, oscillatory profile) and TP2 (black square, oscillatory profile). Insets show profile details. (d) As for panel (a) but for traveling peaks upon their reemergence from the T-points, showing the presence of an extra small amplitude precursor peak in both cases. (e,f) As for (b,c) but showing the color-coded solution profiles at the locations indicated in panel (d). All profiles asymptote to $\vP_0$ as $\xi\to\pm\infty$. Parameter: $D_Y=10^{-4}$.}}
    \label{fig:DYm4_1P_rtrn_spiral2}
\end{figure*}

\begin{figure*}[tp]
    \includegraphics[width=\textwidth]{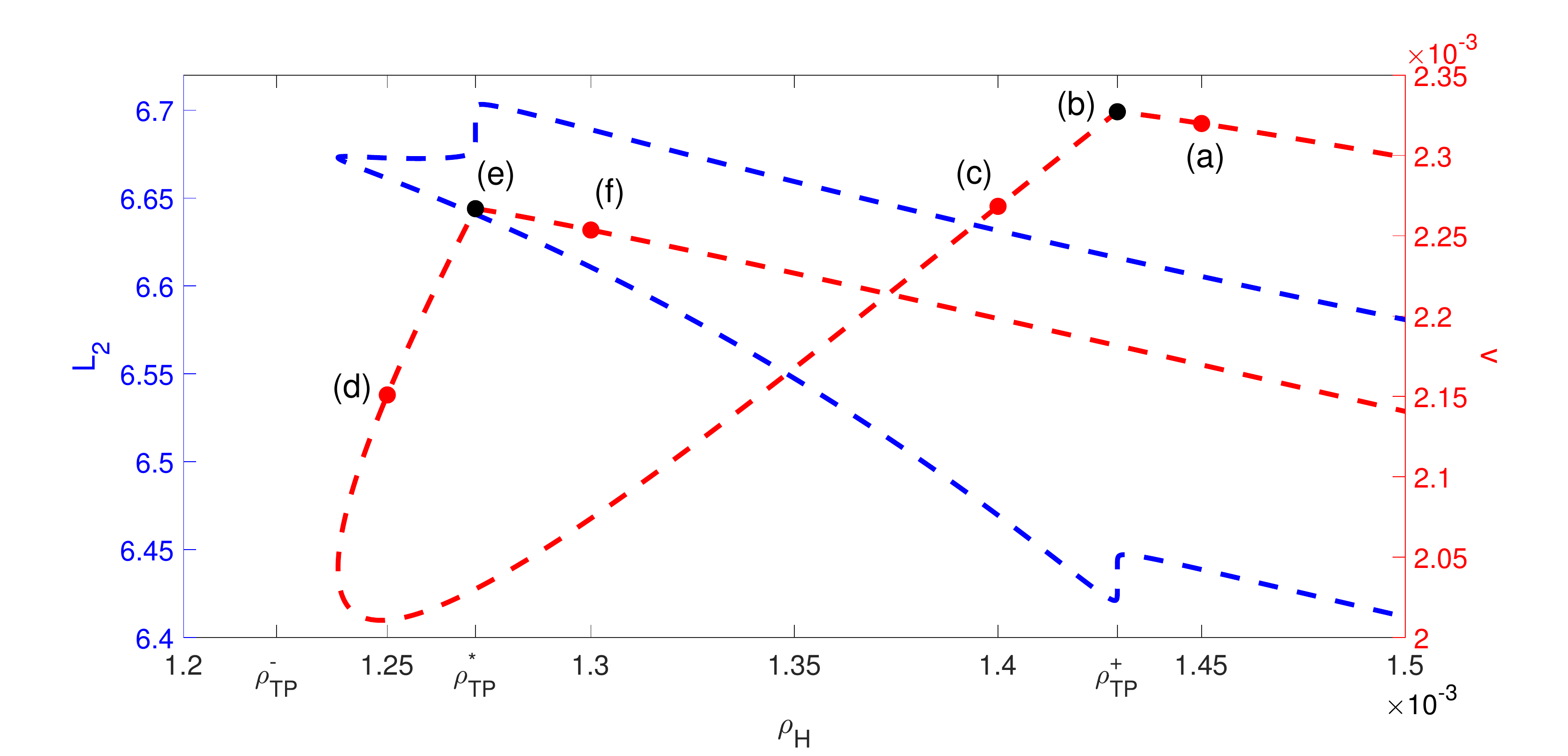}
    (a)\includegraphics[width=0.25\textwidth]{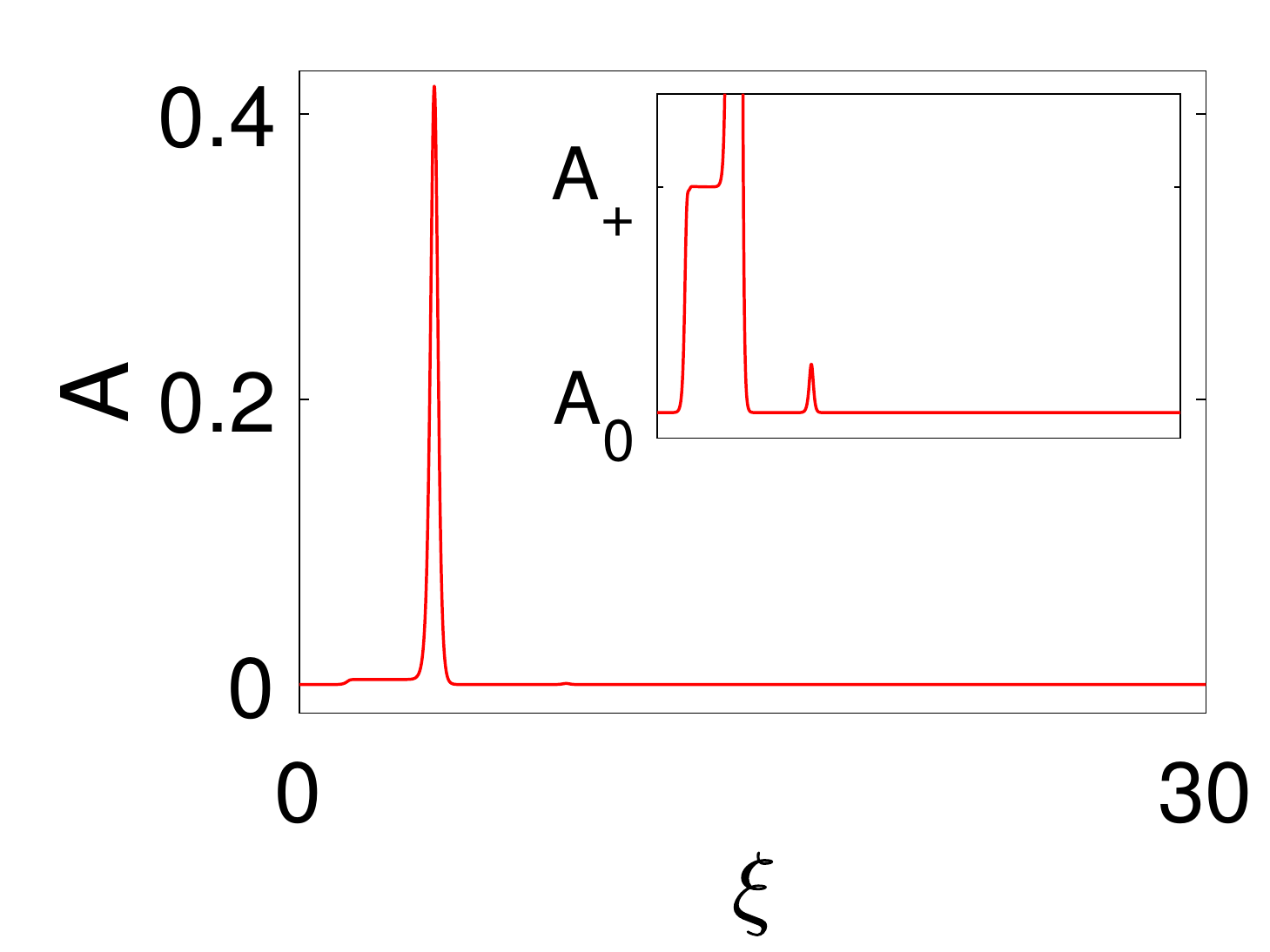}
    \quad \quad \quad (b)\includegraphics[width=0.25\textwidth]{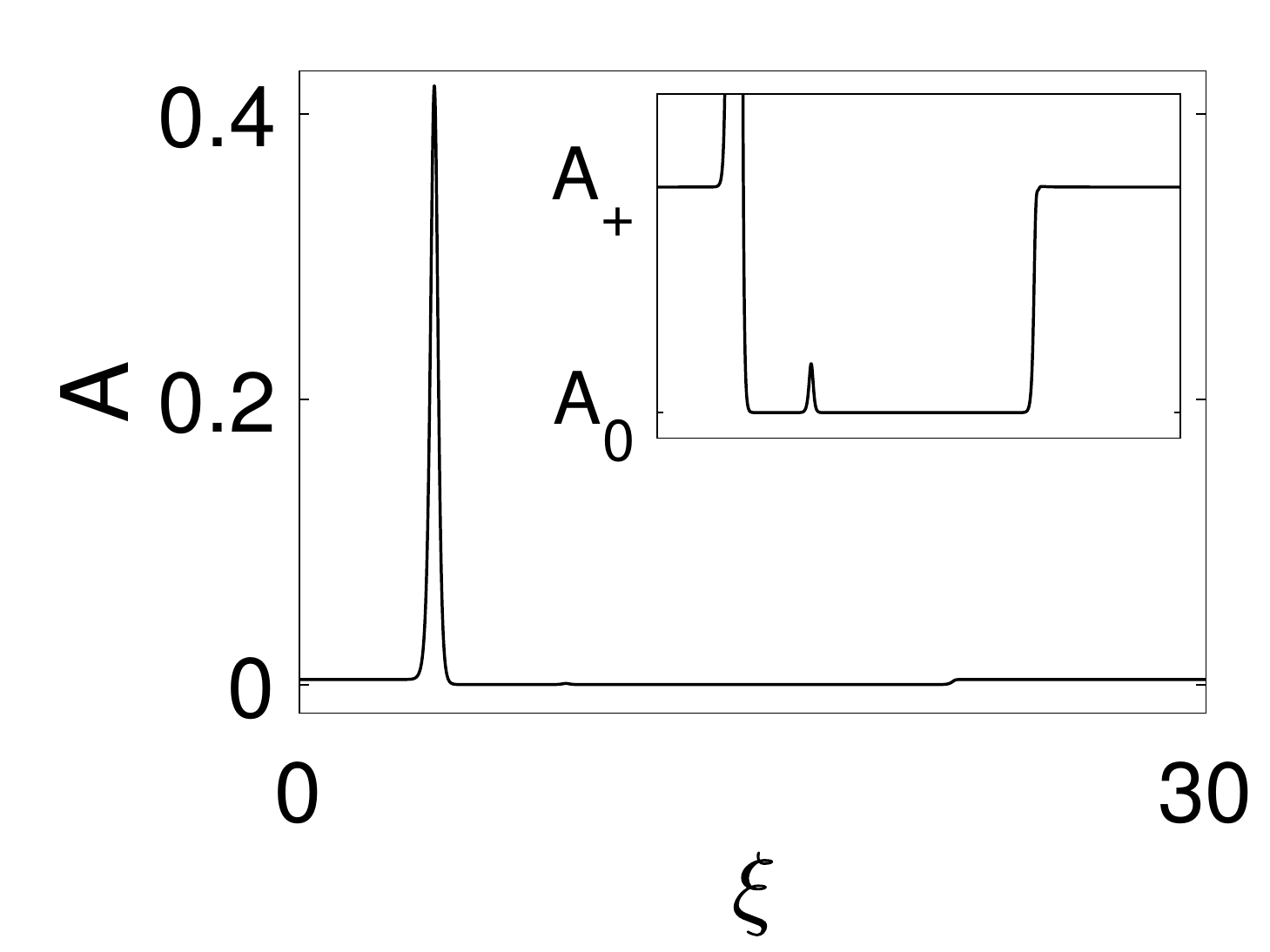}
    \quad \quad \quad (c)\includegraphics[width=0.25\textwidth]{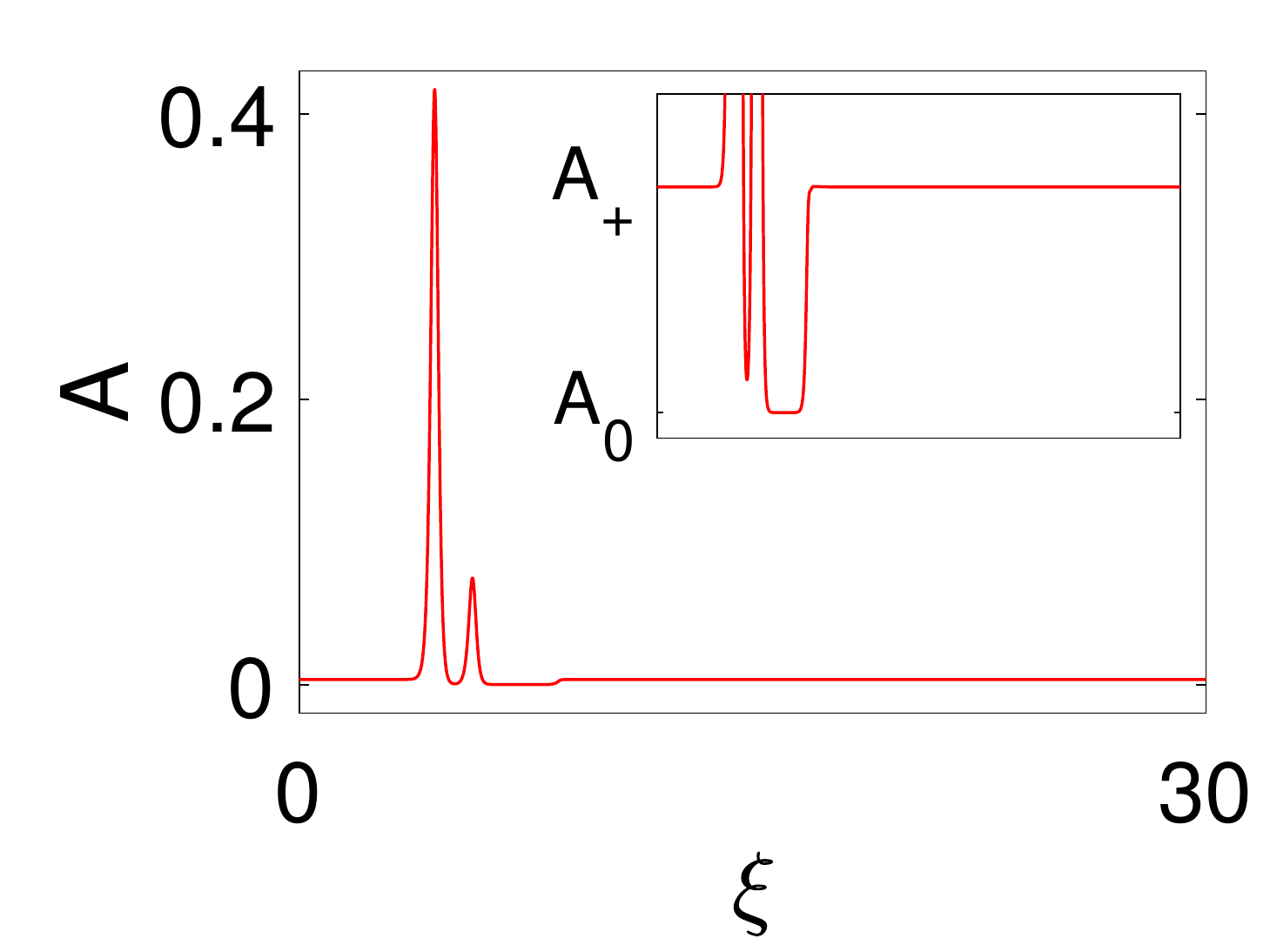}\\
    (d)\includegraphics[width=0.25\textwidth]{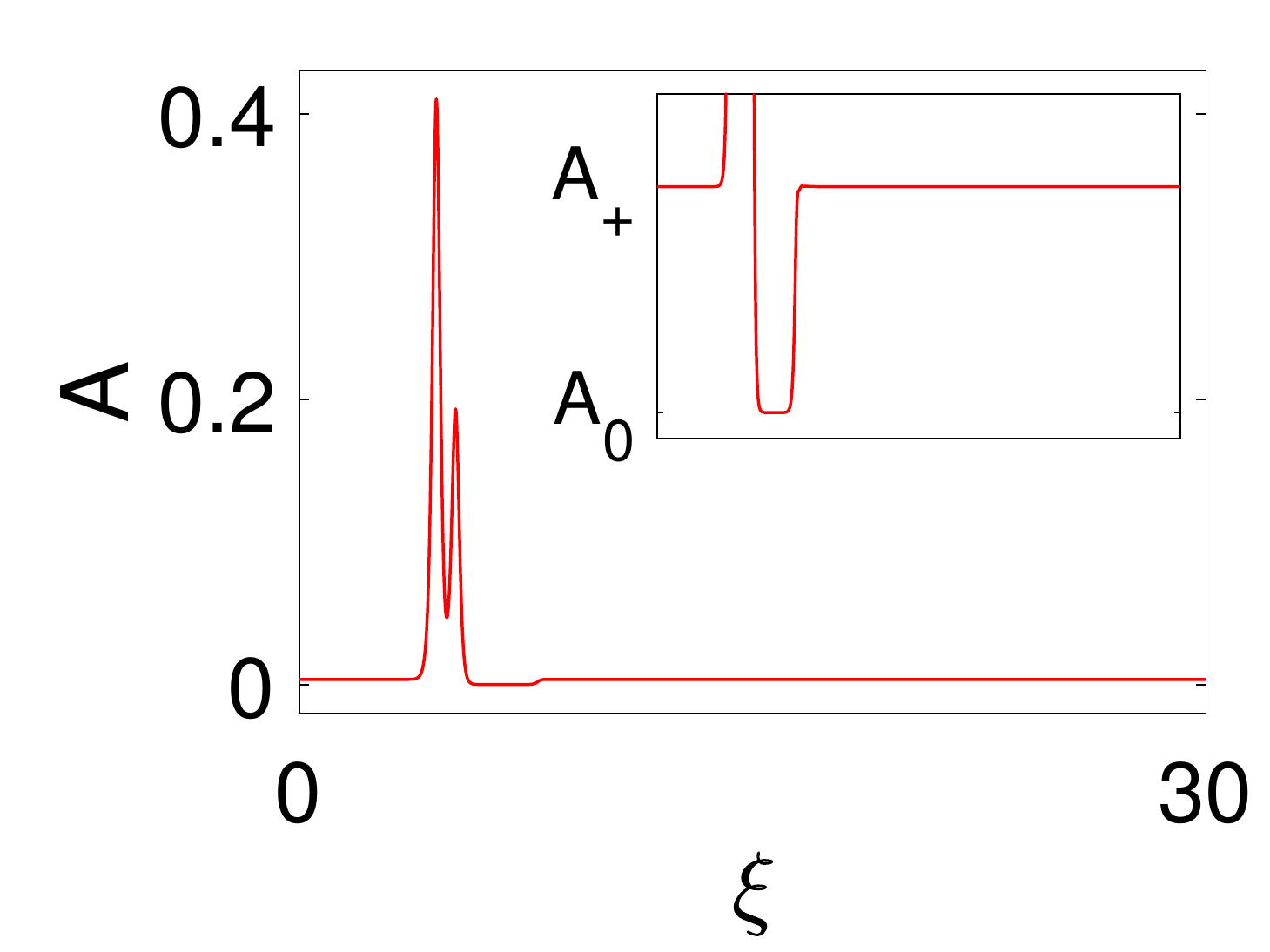}
    \quad \quad \quad (e)\includegraphics[width=0.25\textwidth]{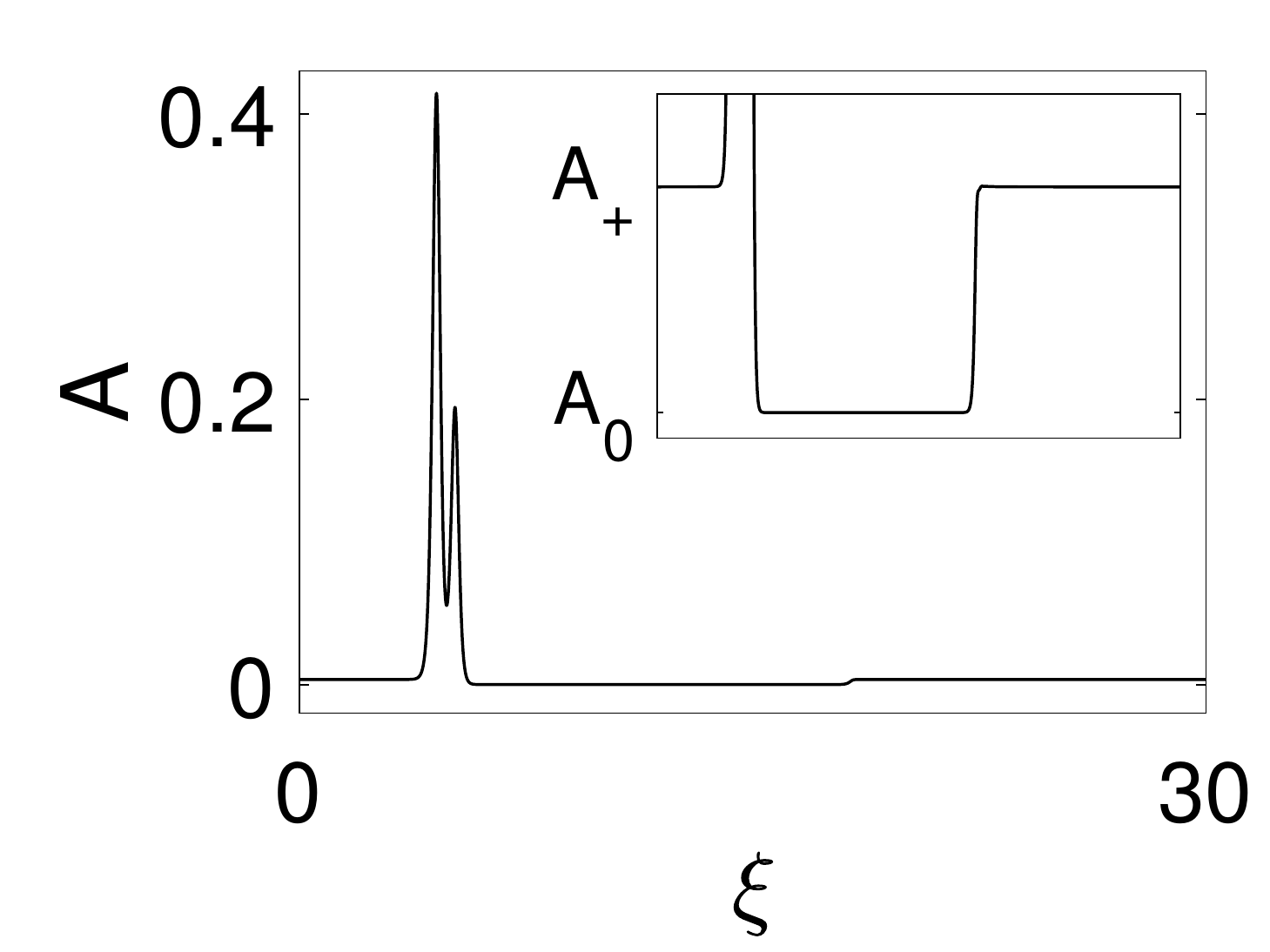}
    \quad \quad \quad (f)\includegraphics[width=0.25\textwidth]{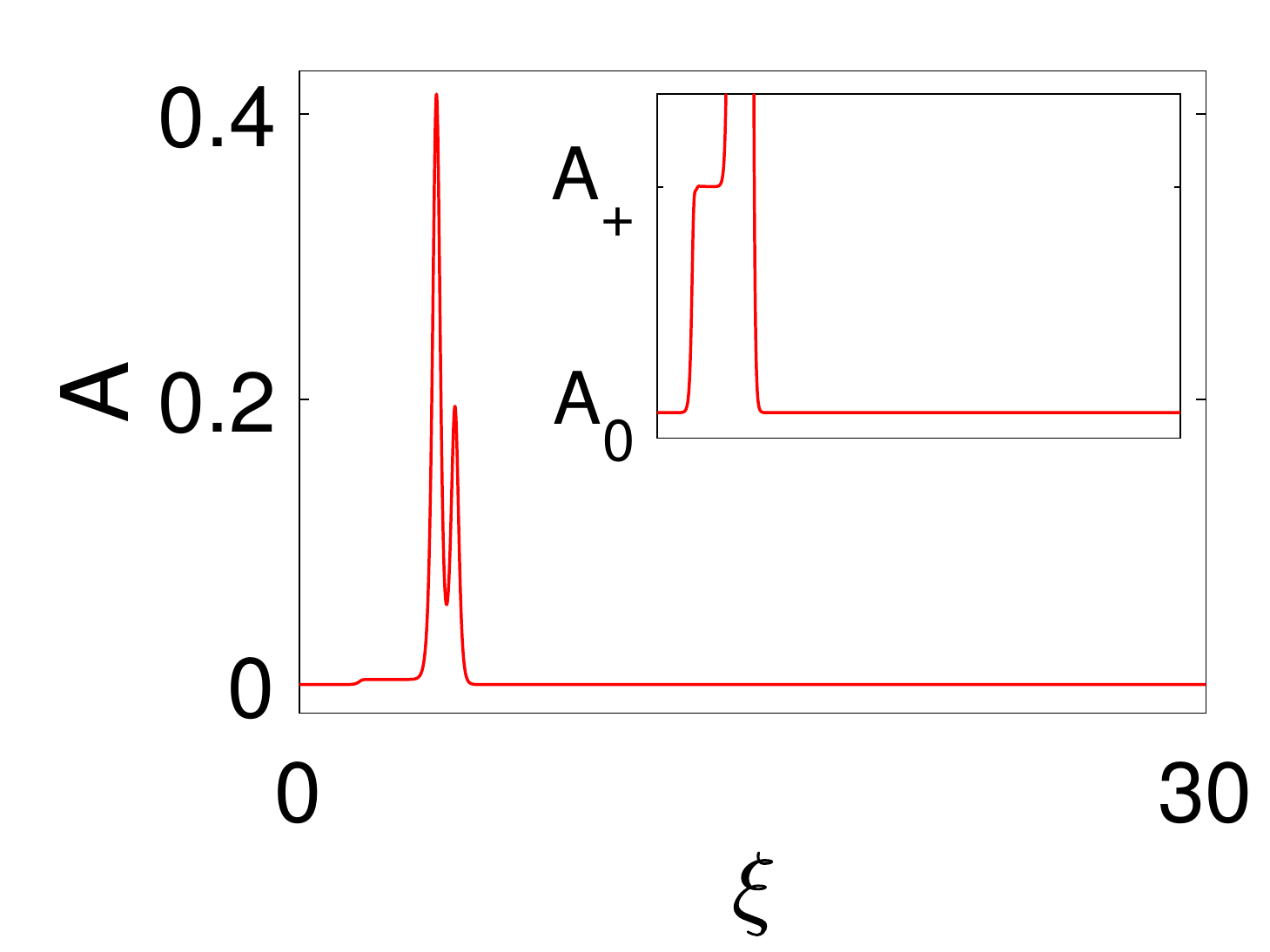}
    \caption{Top panel: Key detail of Fig.~\ref{fig:DYm4_1P} showing a branch of traveling peak states (blue) together with their speed $v$ (red) computed on a $L=30$ domain with PBC. The blue branch consists of 1P states with a precursor created in a parity-breaking bifurcation (blue dot in Fig.~\ref{fig:DYm4_1P}), passes through two successive T-points (first TP$^+$, then TP$^*$) connecting the $\vP_0$ and $\vP_+$ states, and eventually terminates in a parity-breaking bifurcation of symmetric 2P states without a precursor {(effectively at the purple diamond} in Fig.~\ref{fig:DYm4_1P}). Bottom panels: Color-coded profiles at the codimension-two TP points (b,e) indicated in the top panel by black dots and at locations on either side (red dots). For clarity these locations are indicated on the velocity plots only; in L$_2$ norm the T-points are associated with discontinuous behavior (see text). Parameter: $D_Y=10^{-4}$.}
    \label{fig:DYm4_TP_1P}
\end{figure*}
\begin{figure*}[tp]
    \includegraphics[width=\textwidth]{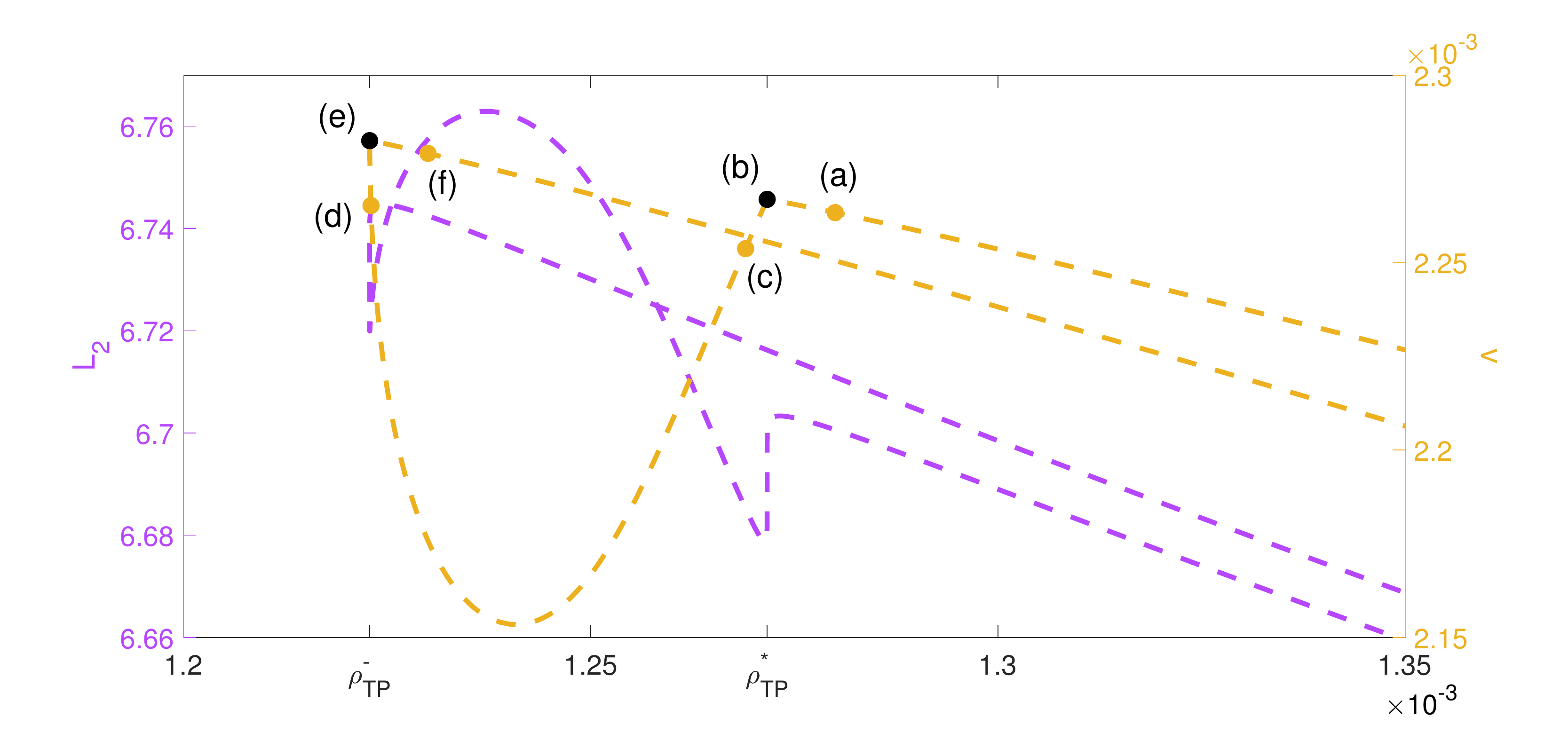}
    (a)\includegraphics[width=0.25\textwidth]{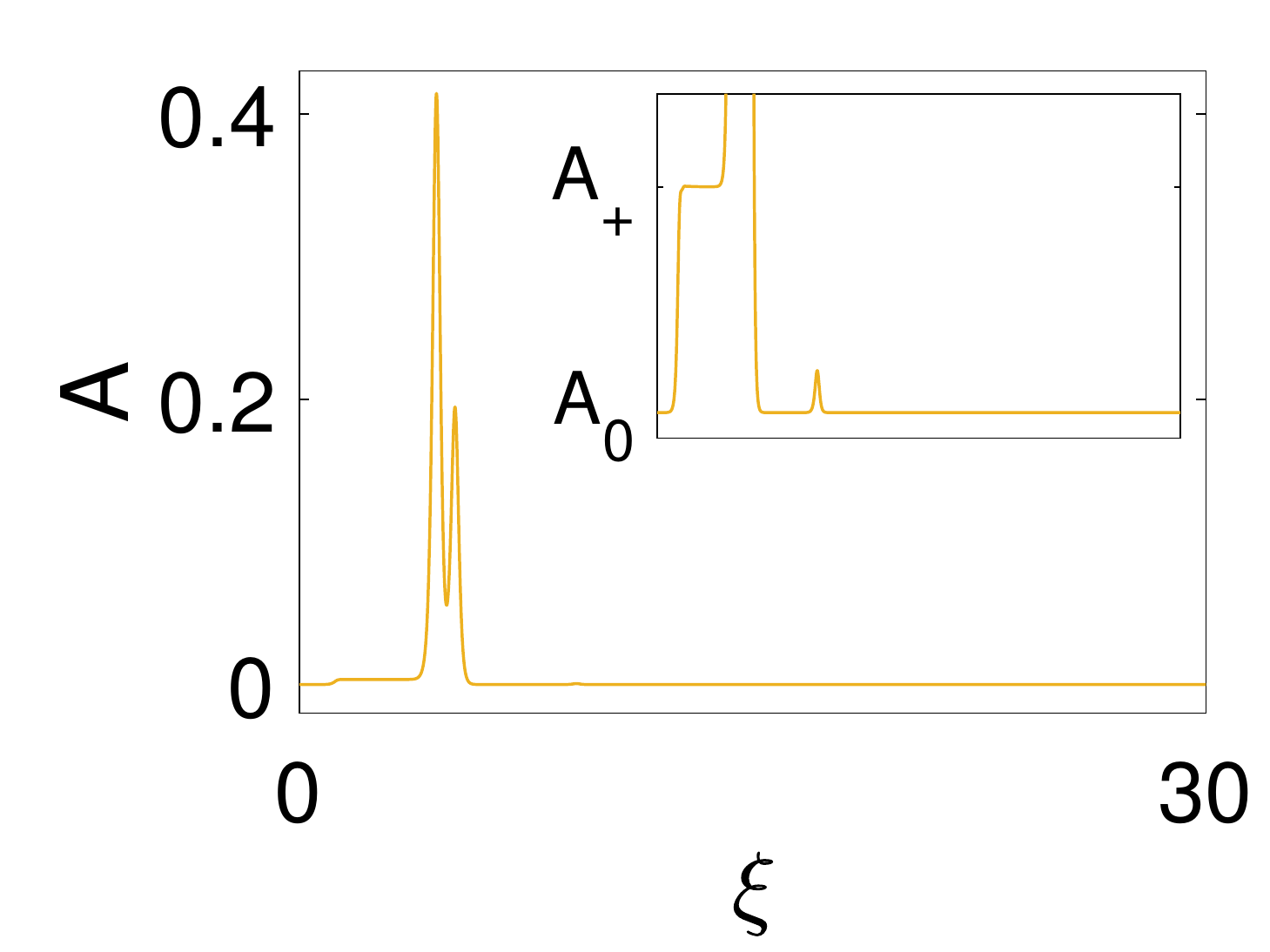}
    \quad \quad \quad (b)\includegraphics[width=0.25\textwidth]{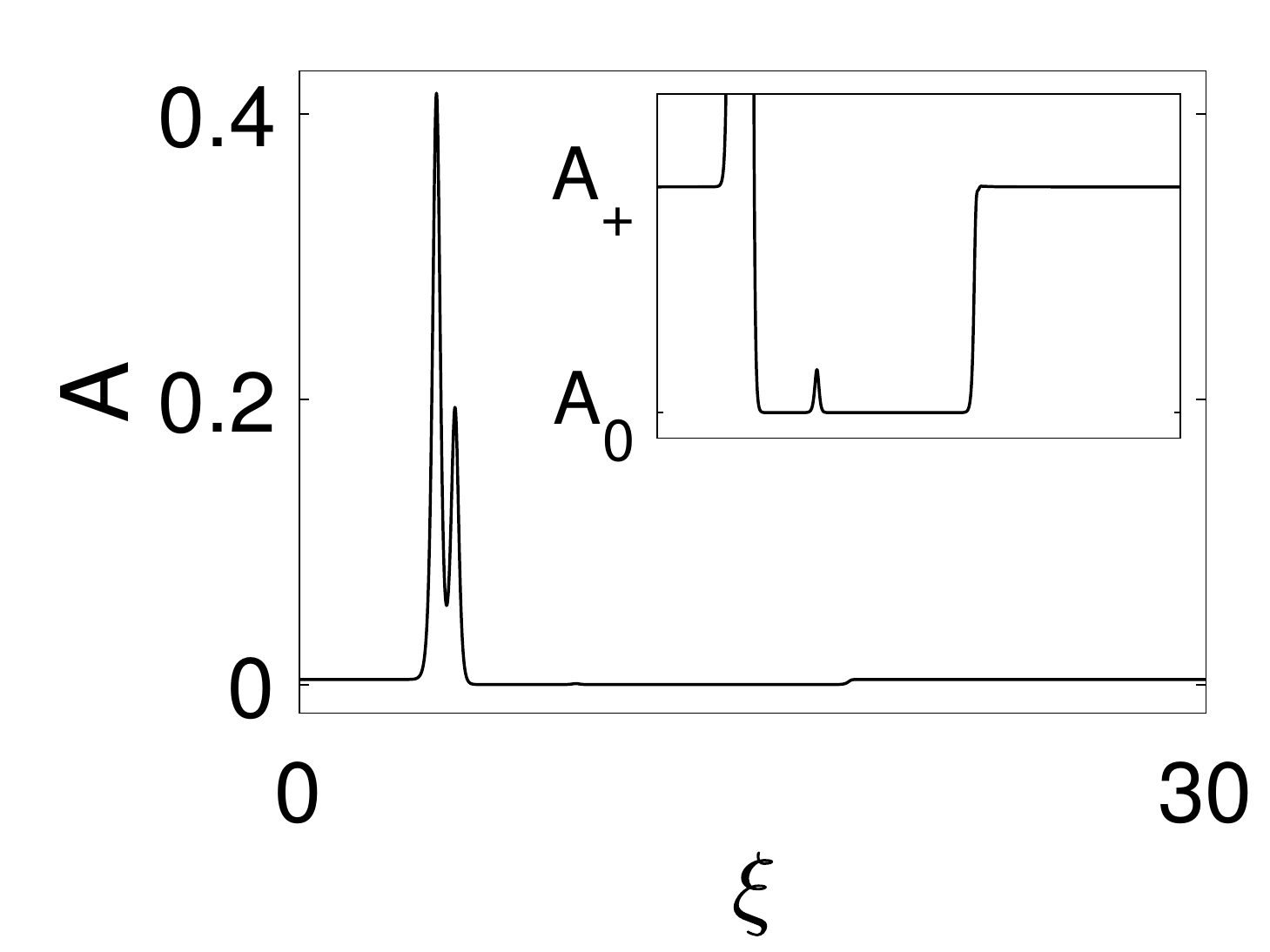}
    \quad \quad \quad (c)\includegraphics[width=0.25\textwidth]{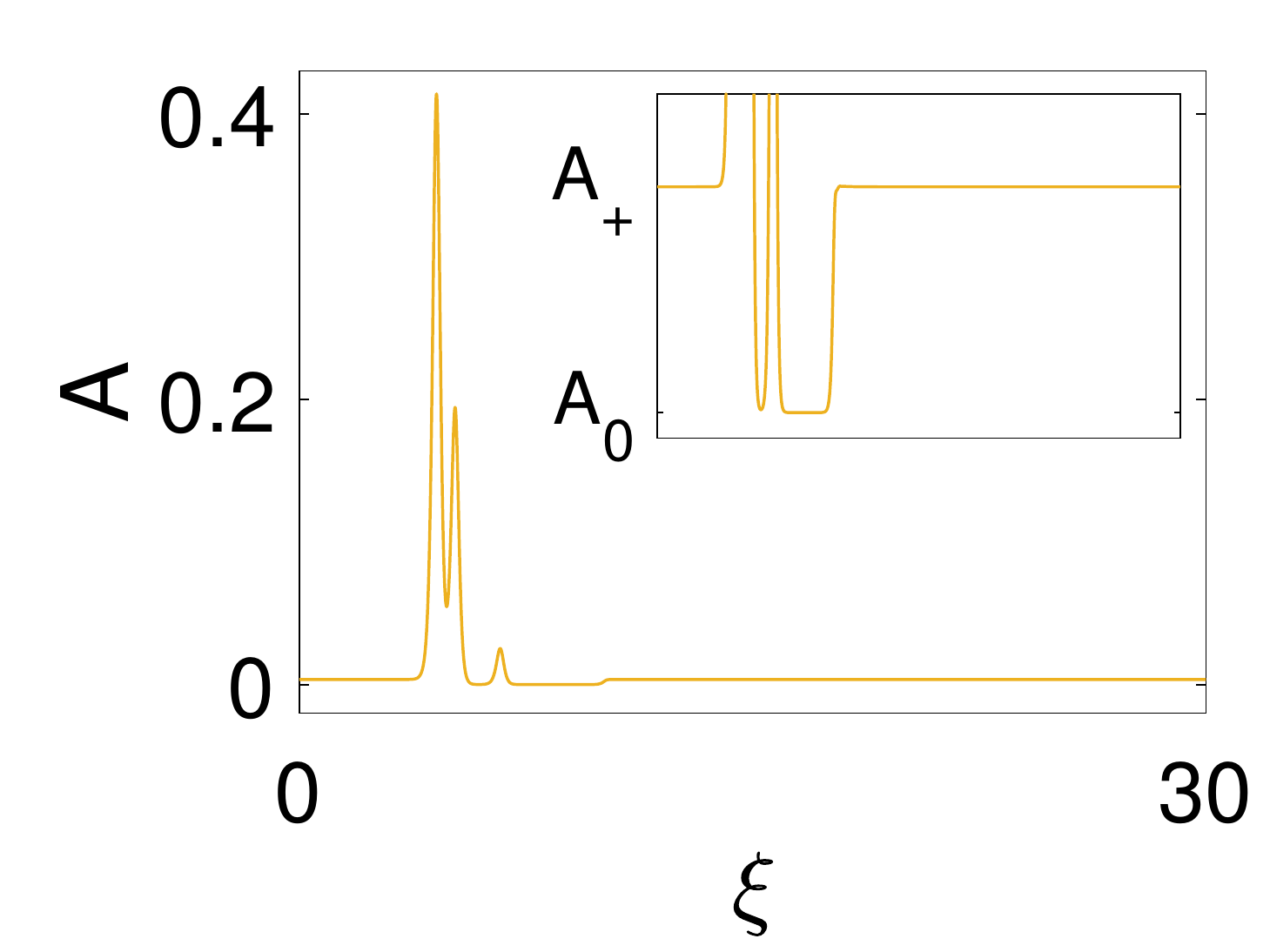}\\
    (d)\includegraphics[width=0.25\textwidth]{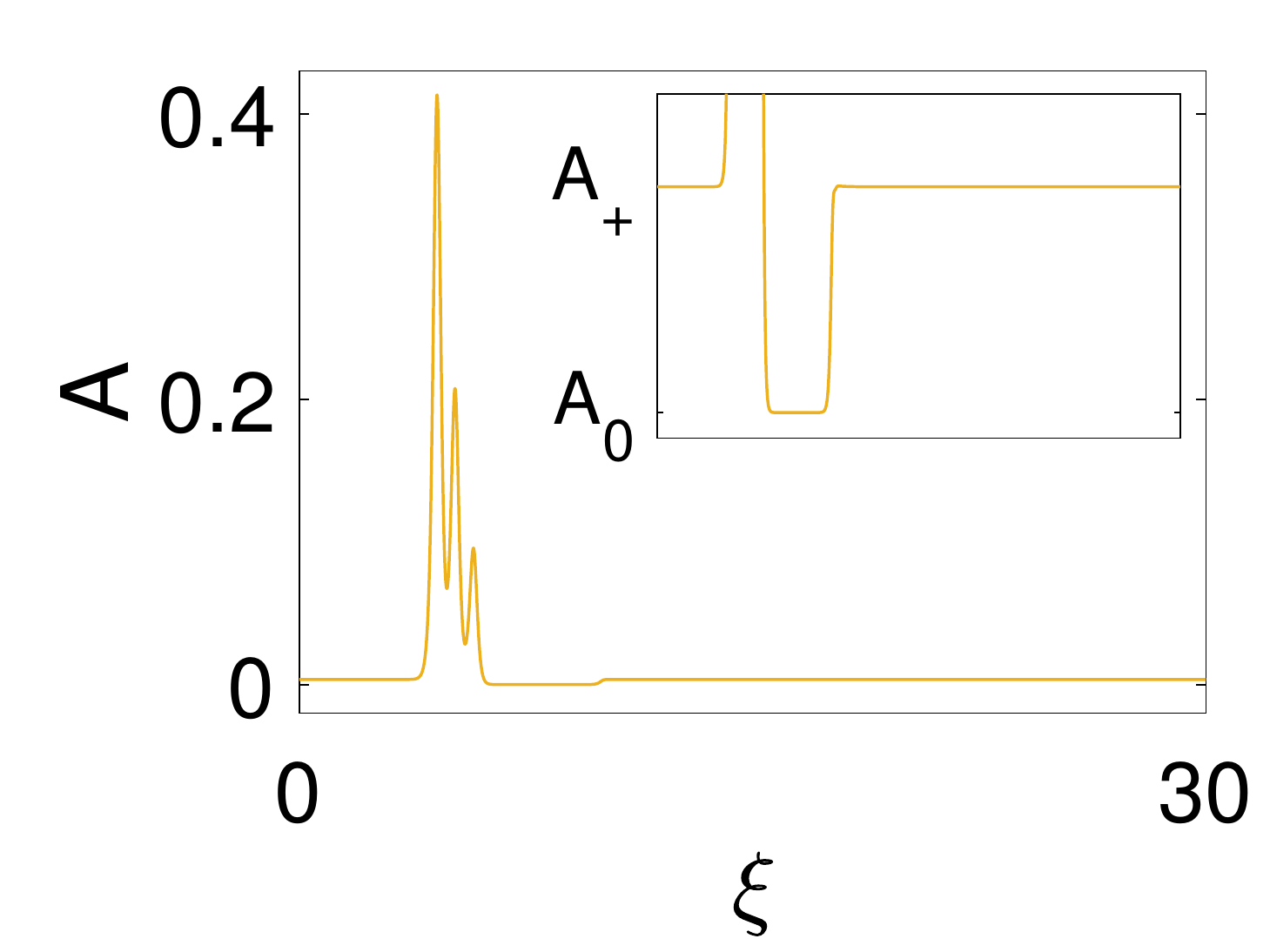}
    \quad \quad \quad (e)\includegraphics[width=0.25\textwidth]{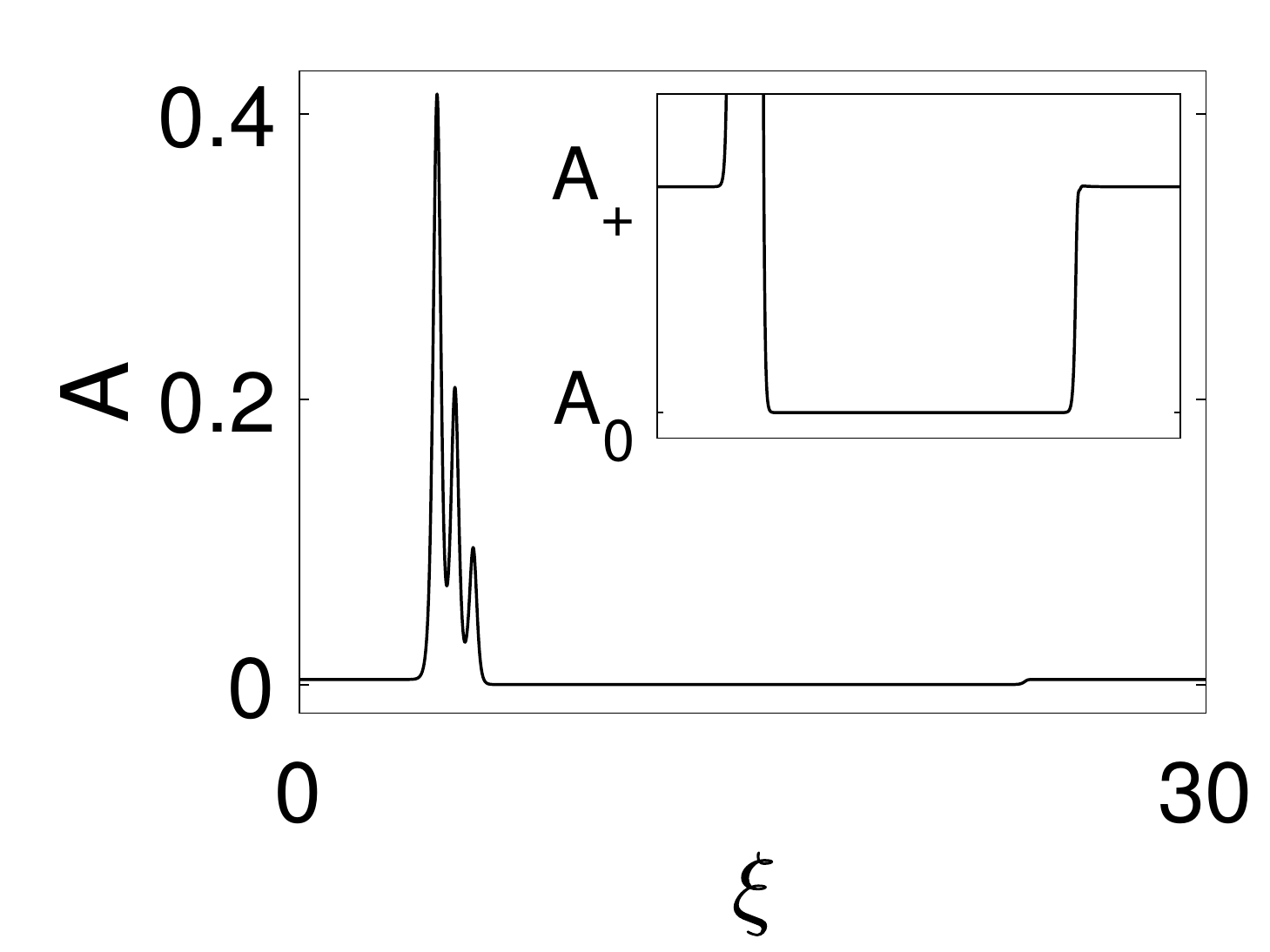}
    \quad \quad \quad (f)\includegraphics[width=0.25\textwidth]{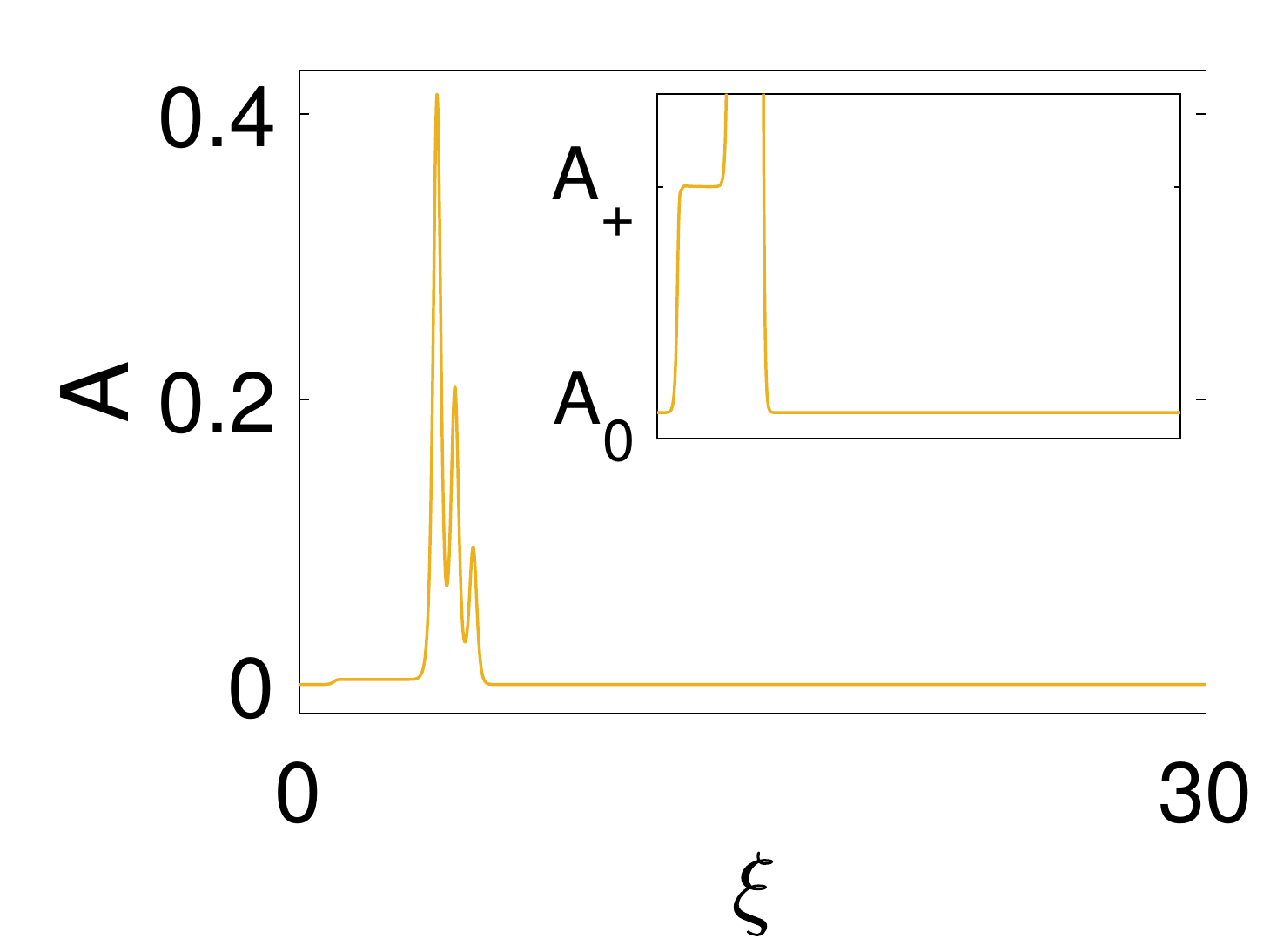}
    \caption{As in Fig.~\ref{fig:DYm4_TP_1P} but for traveling 2P states with a precursor (purple) created in a parity-breaking bifurcation of stationary 2P states {(effectively at the purple diamond} in Fig.~\ref{fig:DYm4_1P}) and terminating in a parity-breaking bifurcation of a stationary 3P state without a precursor (purple square in Fig.~\ref{fig:DYm4_1P}). The branch passes TP$^*$ first and then TP$^-$ but the overall behavior is similar to that in Fig.~\ref{fig:DYm4_TP_1P}. Parameter: $D_Y=10^{-4}$.}
    \label{fig:DYm4_2P_TP}
\end{figure*}

\subsection{SNIPER bifurcation of fronts}\label{sec:sniper}

In Ref.~\cite{knobloch2021stationary} we have shown that the interval $\rT<\rH<\rSN$ is occupied by unstable stationary multipeak states organized in a foliated snaking structure (see Fig.~\ref{fig:speed}(b)). In addition, we have seen that for $\rH>\rf$ one finds steadily propagating fronts connecting $\vP_*$ to the trivial state $\vP_0$. The latter cannot be continued into $\rH<\rf$ because of the fold at $\rf$. Since $\rSN<\rf$ it is of both theoretical and practical interest to determine what happens in the interval $\rSN<\rH<\rf$.

Direct numerical integration of Eqs.~(\ref{eq:AI}) starting from an initial condition of $\tanh$ type shows that in this interval the system generally evolves into a stable propagating {\it and} oscillating front. Such fronts are no longer steady states in an appropriately moving frame, and indeed the translation speed of this state oscillates along with its amplitude although it remains positive at all times. The simulations show that the oscillation period lengthens as $\rH$ increases and appears to diverge at $\rf$ as $1/\sqrt{\rf-\rH}$ (Fig.~\ref{fig:speed}(b), red dots and solid line), suggesting that the fold at $\rH=\rf$ in fact corresponds to a SNIPER bifurcation~\cite{strogatz2018nonlinear} of traveling fronts. Since both $\vP_*$ and $\vP_0$ are linearly stable, the emerging oscillations are localized at the location of the front and so correspond to an {\it oscillating front}. We show such an oscillating front solution at $\rH=1.6\times 10^{-5}$ in a space-time plot in the inset in Fig.~\ref{fig:speed}(b), shown over approximately three temporal periods of the oscillation period $T\approx150$. Moreover, the average propagation speed of these fronts is an order of magnitude larger than that of the nonoscillating fronts that are present for $\rH>\rf$, as shown in Fig.~\ref{fig:speed}(c).

\section{Coexistence of multiple fronts}\label{sec:fronts}

 {In order to understand the origin of the behavior reported in Fig.~\ref{fig:speed}, we have recomputed some of our results for larger values of the parameter $D_{\text{Y}}$ and over larger domains ($L=30$). Figure \ref{fig:DYm4} shows the results for $D_{\text{Y}}=10^{-4}$, this time shown in terms of the L$_2$ norm (\ref{eq:L2}). The top panel of Fig.~\ref{fig:DYm4}(a) shows that the spiral structure shown in Fig.~\ref{fig:speed}(a) remains, although it is highly compressed owing to the large range of $\rH$ values shown. However, the results shown in the figure reveal much additional and helpful detail. In particular, they reveal that there are in fact {\it two} branches of distinct steadily moving fronts. The first comes in from large $\rH$ and consists of stable fronts (solid blue line) with an almost constant, $\rH$-independent speed $v$ (solid red line). These fronts resemble those depicted in Fig.~\ref{fig:speed}(a) and exhibit a characteristic overshoot in amplitude that coincides with the location of the front. Moreover, they remain stable until the fold at $\rH=\rf$ beyond which they again appear to spiral into a T-point, resulting in the addition of more and more oscillations to the front profile with each turn (see profiles in Fig.~\ref{fig:DYm4}(d) corresponding to the three locations indicated in inset (ii) of panel (a)). However, there is a second branch of fronts as well, with larger L$_2$ norm and higher speed $v$ (dashed blue/red lines in Fig.~\ref{fig:DYm4}(a)). As shown in Fig.~\ref{fig:DYm4}(b), profile labeled with a black $+$ symbol, these fronts also connect $\vP_*$ to the trivial state $\vP_0$, albeit with a much larger overshoot, but are unstable.}

 {The existence of these fast (but unstable) fronts is unexpected but essential for understanding the behavior of our system. This is because the two branches are in fact connected (red dot in inset (ii) of Fig.~\ref{fig:DYm4}(a)). To see how this happens let us follow the fast fronts as $\rH$ increases from the top left of panel (a). Between $\rH\approx 4.6\times 10^{-4}$ and $\rH\approx 4.9\times 10^{-4}$ the solution branch passes through a pair of folds (see Fig.~\ref{fig:DYm4}(e), which zooms in on the dashed rectangle labeled (iii) in the main panel of Fig.~\ref{fig:DYm4}(a)). The inset in Fig.~\ref{fig:DYm4}(e) shows the front profile at the location indicated by the solid blue dot and shows that the profile has started to develop a small dip between the $\vP_*$ asymptote and the large overshoot peak associated with the fast fronts. As $\rH$ increases further this dip broadens into a plateau while the overshoot peak decreases in amplitude, becoming a small precursor peak by the time the fast fronts reach the right fold at $\rH=\rH^{\times}\approx 1.62\times 10^{-3}$. This location is indicated by a blue $\times$ symbol in panel (a) and the corresponding profile is shown in Fig.~\ref{fig:DYm4}(b), blue profile. At this point the L$_2$ norm of the fast fronts is almost indistinguishable from that of the slow fronts and this continues to be so as the fast fronts turn towards smaller $\rH$ and follow the slow fronts all the way into the spiral structure at the far left where the two branches finally connect, as shown in inset (ii) of Fig.~\ref{fig:DYm4}(a), and the precursor peak is assimilated into the oscillatory front profile. Prior to this reconnection both branches undergo spiraling behavior similar to that identified in Fig.~\ref{fig:speed} for $D_{\text{Y}}=10^{-7}$, adding oscillations as they spiral in. Figure~\ref{fig:DYm4}(c) compares the stable profile (solid blue line, no precursor) with the corresponding unstable profile (dashed blue line, with precursor), both at $\rH=4\times 10^{-5}$ (inset (i) of Fig.~\ref{fig:DYm4}(a)). Thus in the spiralling region these two branches differ only in the presence of the small precursor peak on the unstable branch which turns into an extra oscillation in the front profile as the putative T-point is approached.}

 {A T-point is a codimension-two point corresponding to the simultaneous presence of a double heteroclinic orbit.\cite{glendinning1986t,zimmermann1997pulse,sneyd2000traveling,or2001pulse,romeo2003stability,champneys2007shil,yochelis2022versatile,moreno2022bifurcation,raja2023} The behavior shown in Fig.~\ref{fig:DYm4}(a) is strongly suggestive of the proximity to such a point, which we tentatively identify with simultaneous heteroclinic connections between $\vP_*$ and the intermediate homogeneous state $\vP_+$ and between $\vP_+$ and $\vP_0$ (Fig.~\ref{fig:bif_uni}). As evidence we point to the value of $\rho_{\rm SN}^{+L}$ at which $\vP_+$  undergoes a fold (Fig.~\ref{fig:bif_uni}). This location is indicated in Fig.~\ref{fig:DYm4}(a), inset (ii), and lies just to left of the putative T-point while the next fold, at $\rho_{\rm SN}^{+R}$, lies to its right and farther out. In addition, the pair of folds in Fig.~\ref{fig:DYm4}(e) and the associated appearance of the dip between the overshoot and the background state $\vP_*$ in the inset profile are both a consequence of a close approach to the state $\vP_+$. This approach just misses $\vP_+$, however, and past the pair of folds results in a monotonic approach to $\vP_0$, albeit accompanied by a precursor peak on top of $\vP_0$. This question is explored in greater detail in the subsections that follow.}

\begin{figure*}[tp]
    \includegraphics[width=\textwidth]{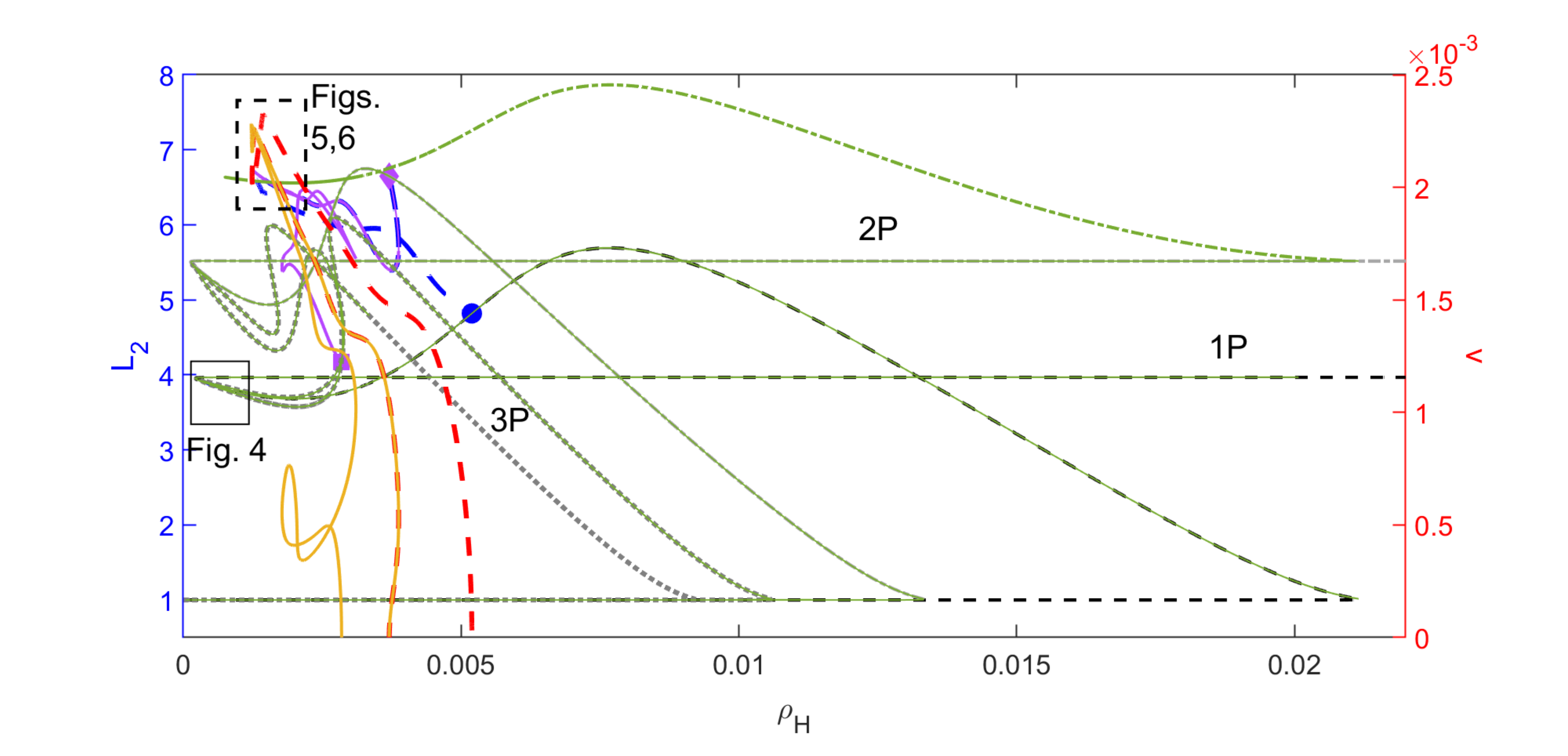}
    \caption{The bigger picture, showing branches of stationary peak states with different numbers of peaks (1P, 2P, 3P) computed on a $L=30$ domain with PBC (green and black curves), together with the traveling states shown in Figs.~\ref{fig:DYm4_TP_1P} and \ref{fig:DYm4_2P_TP} (region bounded by the dashed rectangle) and their velocity $v$ {(red and yellow curves)}. Both the blue and purple branches originate in parity-breaking bifurcations where the red and yellow curves drop to zero (right axis), corresponding to the blue dot and purple square, respectively, and terminate together on the 2P branch (purple diamond). The superposed black dashed curves correspond to stationary peak solutions without the small precursor peak while the green curves correspond to stationary profiles with a precursor. The region bounded by the solid rectangle contains the spiral T-points discussed in Fig.~\ref{fig:DYm4_1P_rtrn_spiral2}, but these are not included in this plot. The figures that follow describe the behavior of stationary 1P, 2P, and 3P branches in more detail. Parameter: $D_Y=10^{-4}$.}
          \label{fig:DYm4_1P}
\end{figure*}
\begin{figure*}[tp]
\sbox{\measurebox}{%
\begin{minipage}[b]{0.67\textwidth}
  {(a)\,\,\,\includegraphics[width=\textwidth]{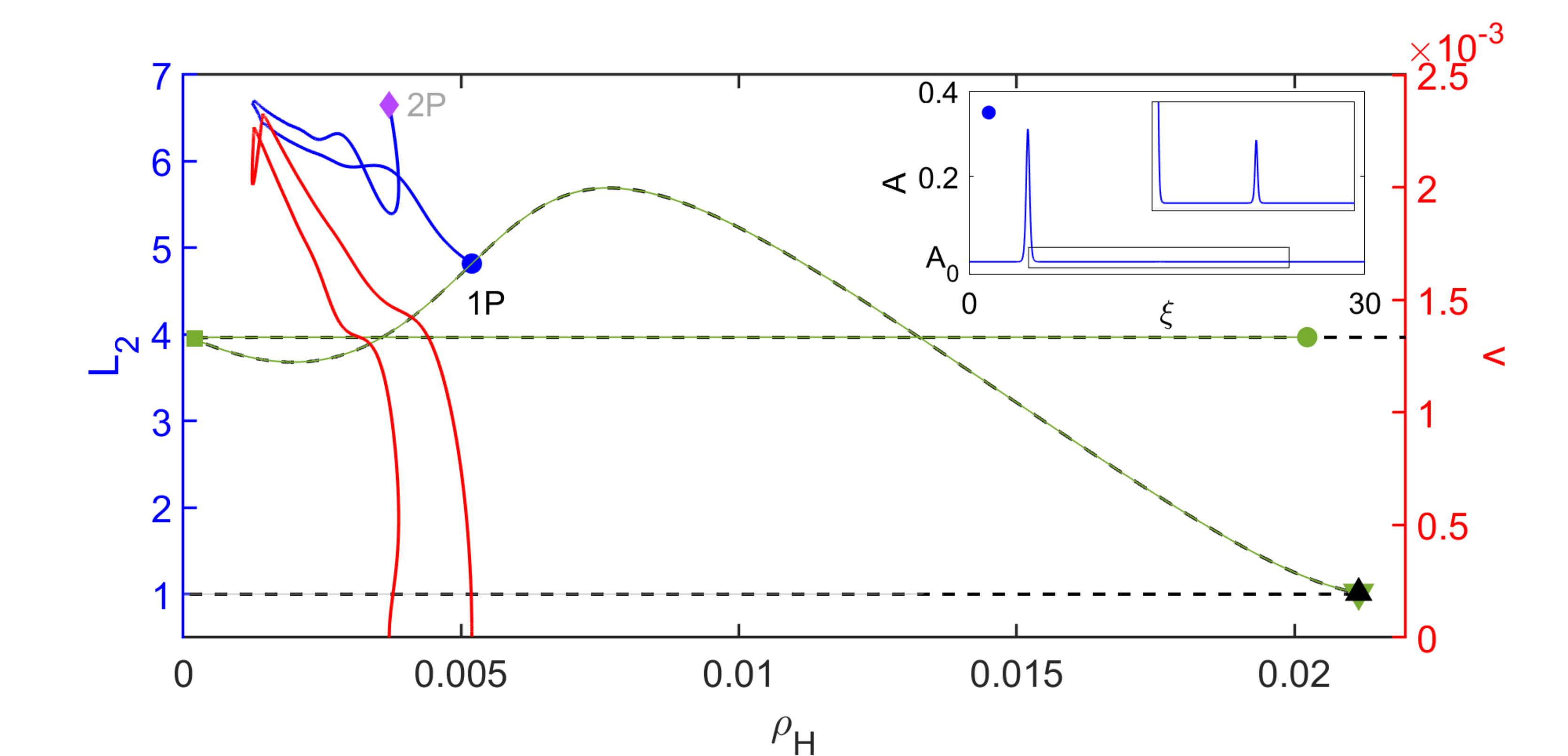}}
  {(b)\includegraphics[width=\textwidth]{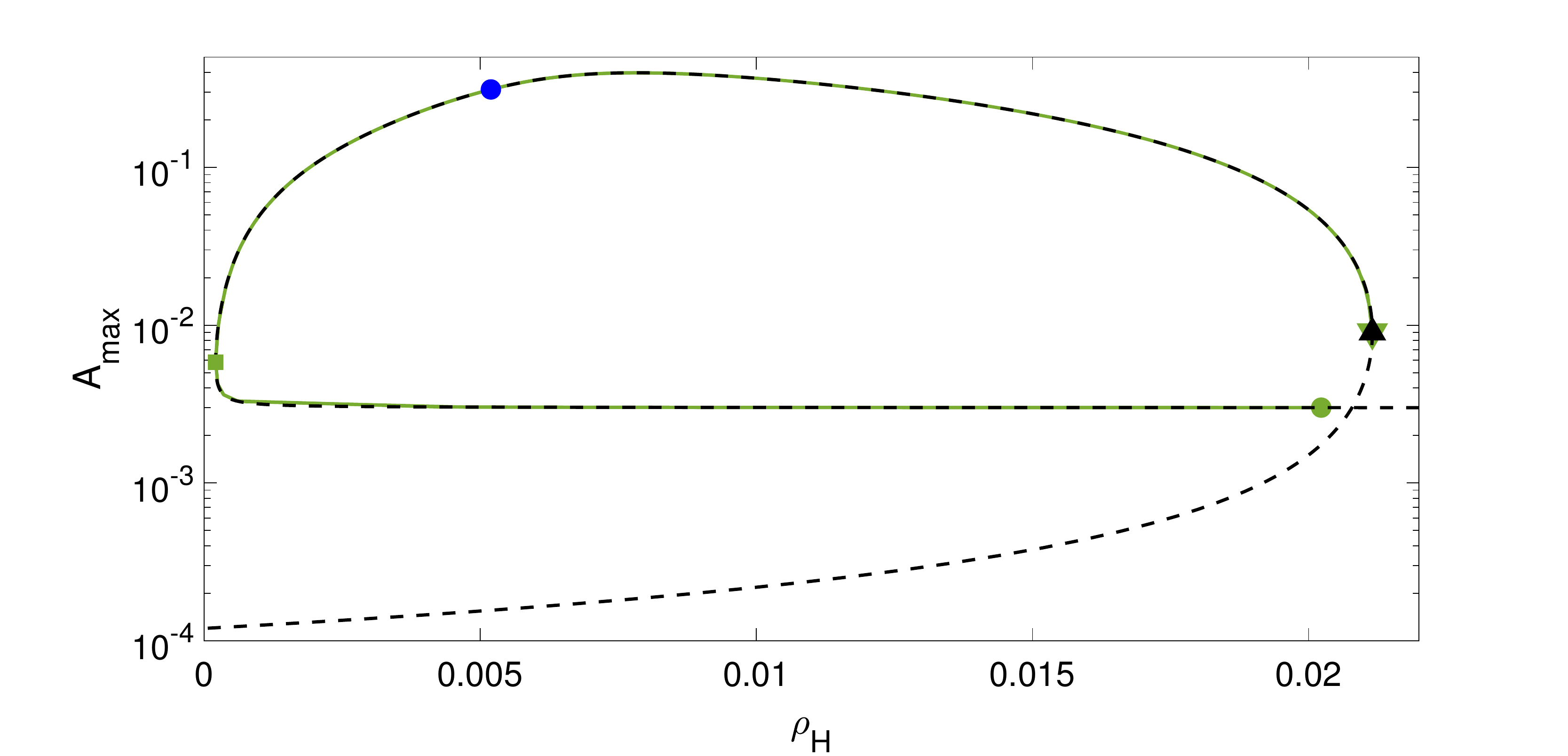}}
\end{minipage}}
\usebox{\measurebox}\qquad  \qquad
\begin{minipage}[b][\ht\measurebox][s]{.215\textwidth}
  {\textcolor{white}{(c)}\includegraphics[width=\textwidth]{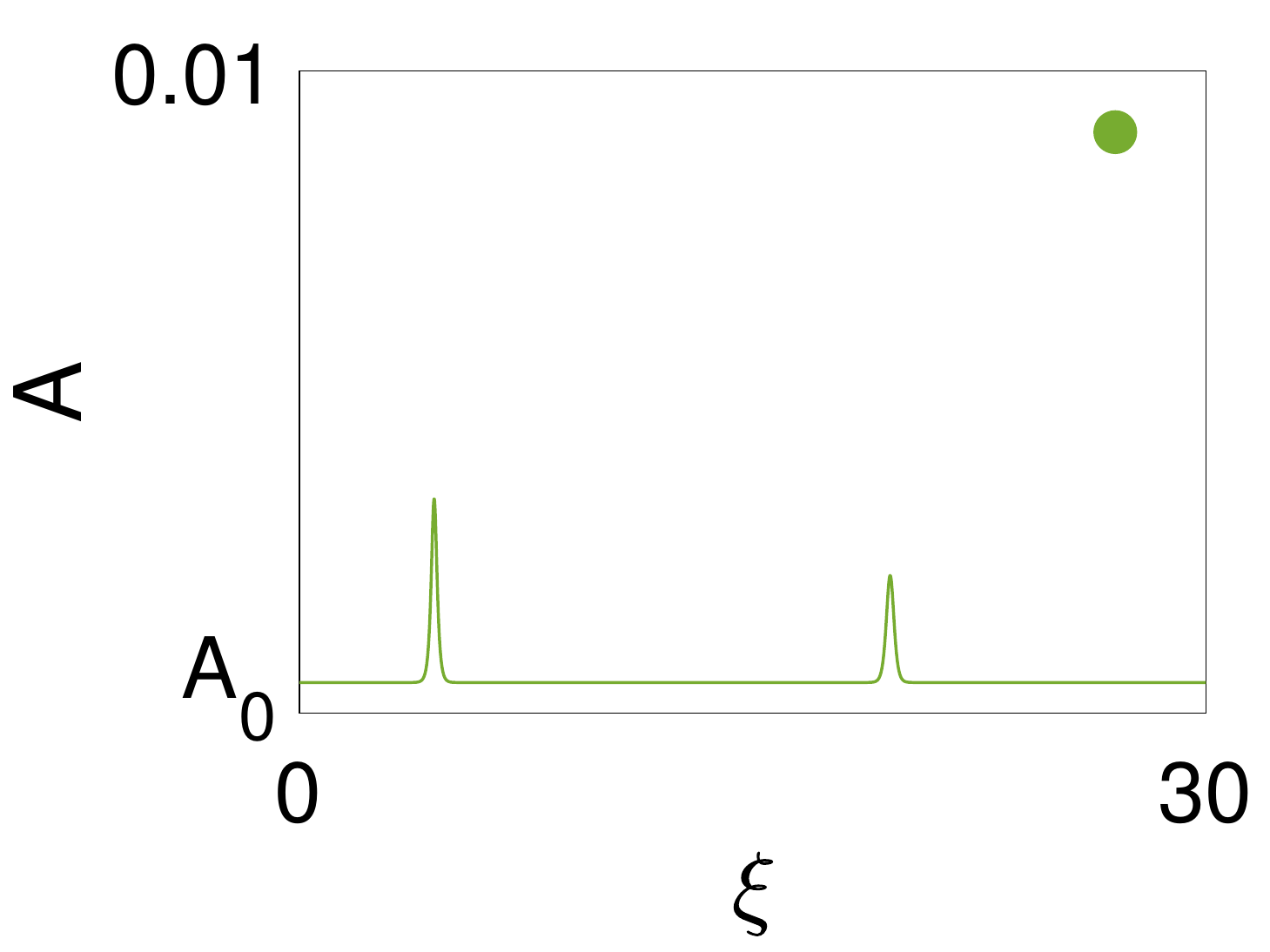}}
  {\textcolor{white}{(c)}\includegraphics[width=\textwidth]{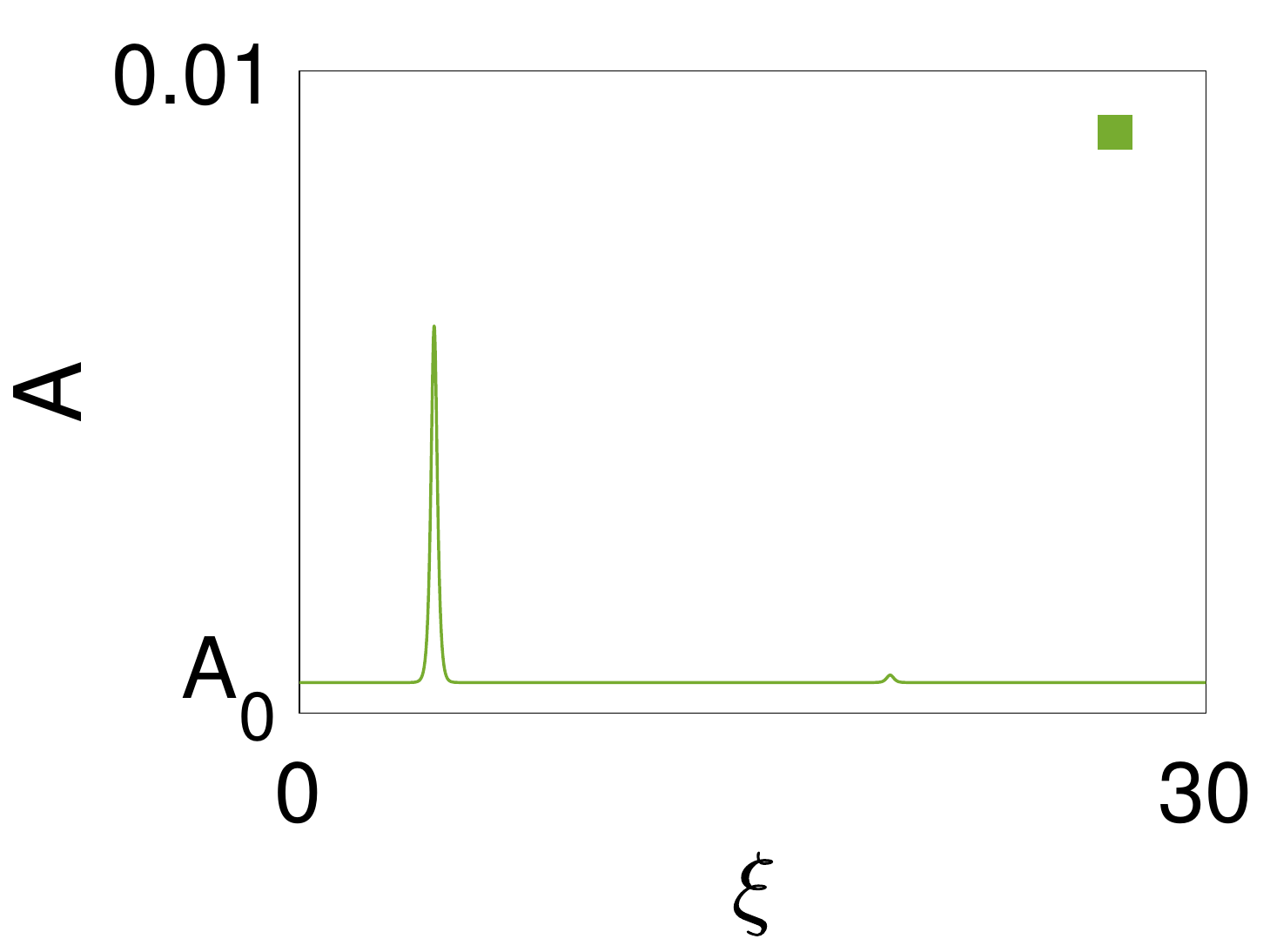}}
  {\textcolor{white}{(c)}\includegraphics[width=\textwidth]{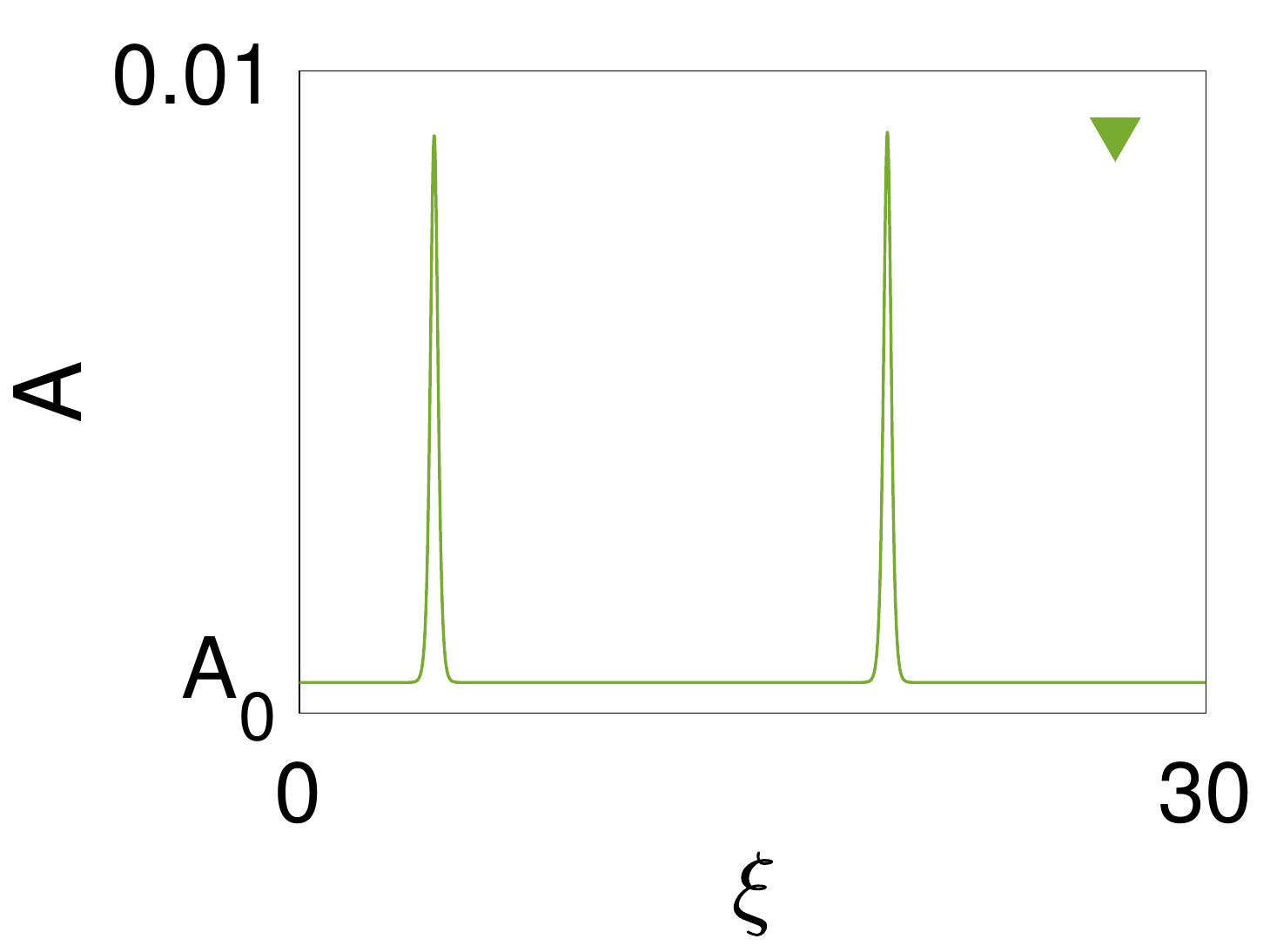}}
  {(c)\includegraphics[width=\textwidth]{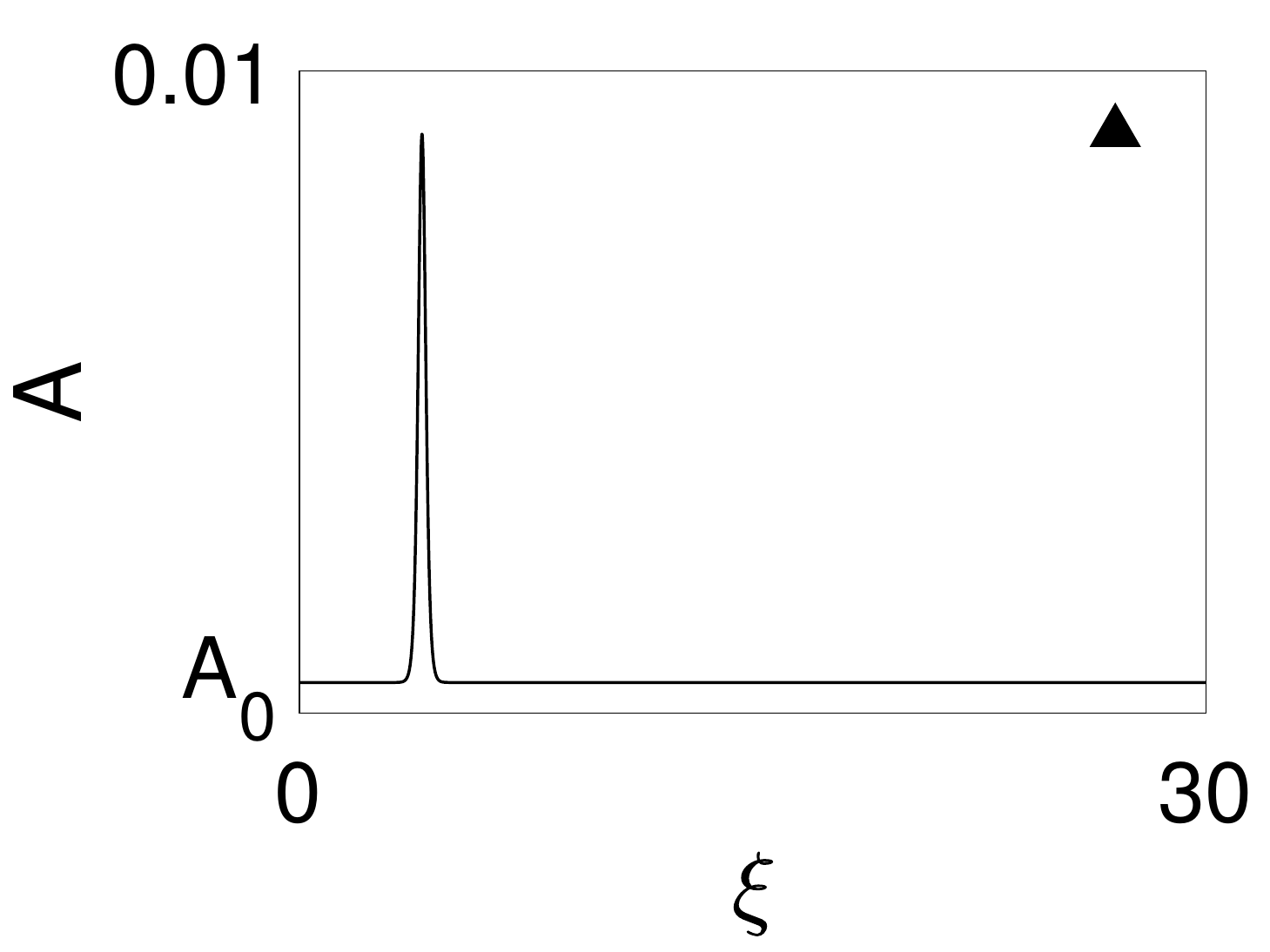}}
\end{minipage}
  \caption{(a) Deconstruction of Fig.~\ref{fig:DYm4_1P} showing the one-peak traveling states (blue) and their velocity $v$ (red) together with the stationary 1P and 2P branches: the black curve shows the stationary states with no precursor peak, while the superposed green curve shows the corresponding branch with a precursor peak. The parity-breaking bifurcation to traveling states is indicated by a blue dot on the green branch; the corresponding profile is shown in the inset. The traveling states terminate on the 2P branch (not shown) in another parity-breaking bifurcation (purple diamond). (b) The evolution of the stationary 1P branch with a precursor peak from the location of the green dot in (a) to its termination at the green down-triangle in (a) in terms of the maximum peak amplitude A$_{\rm max}$ on a log scale, showing that the profile transitions from a 1P precursor profile (blue dot) to a 2P state with two identical equispaced peaks (green down-triangle). (c) Sample 1P (black up-triangle) and 2P (green down-triangle) profiles at the same location. The corresponding locations in the L$_2$ norm are indicated in panel (a).}
    \label{fig:DYm4_1P_stat}
\end{figure*}

\begin{figure*}[tp]
\sbox{\measurebox}{%
\begin{minipage}[b]{0.67\textwidth}
  {(a)\,\,\,\includegraphics[width=\textwidth]{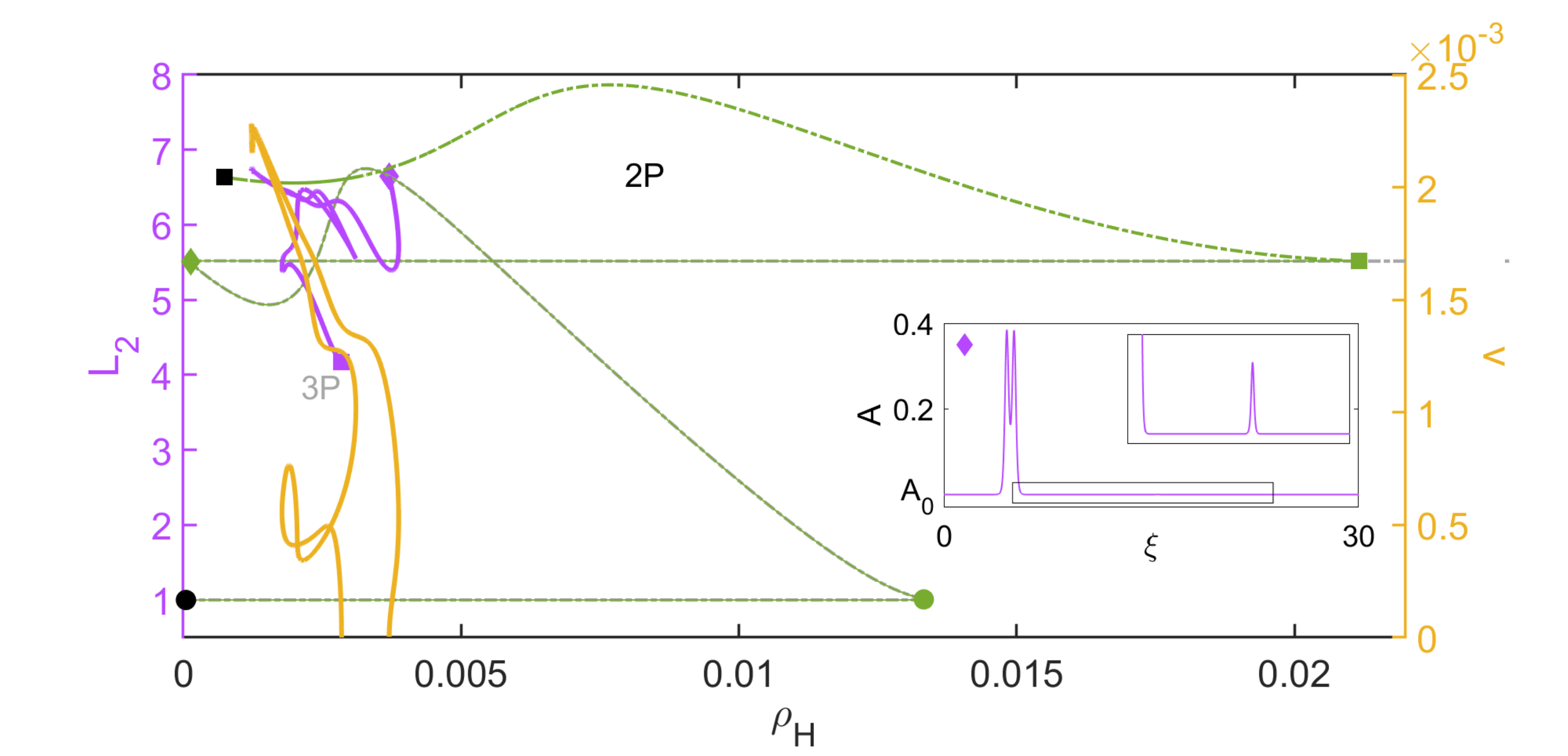}}
  {(b)\includegraphics[width=\textwidth]{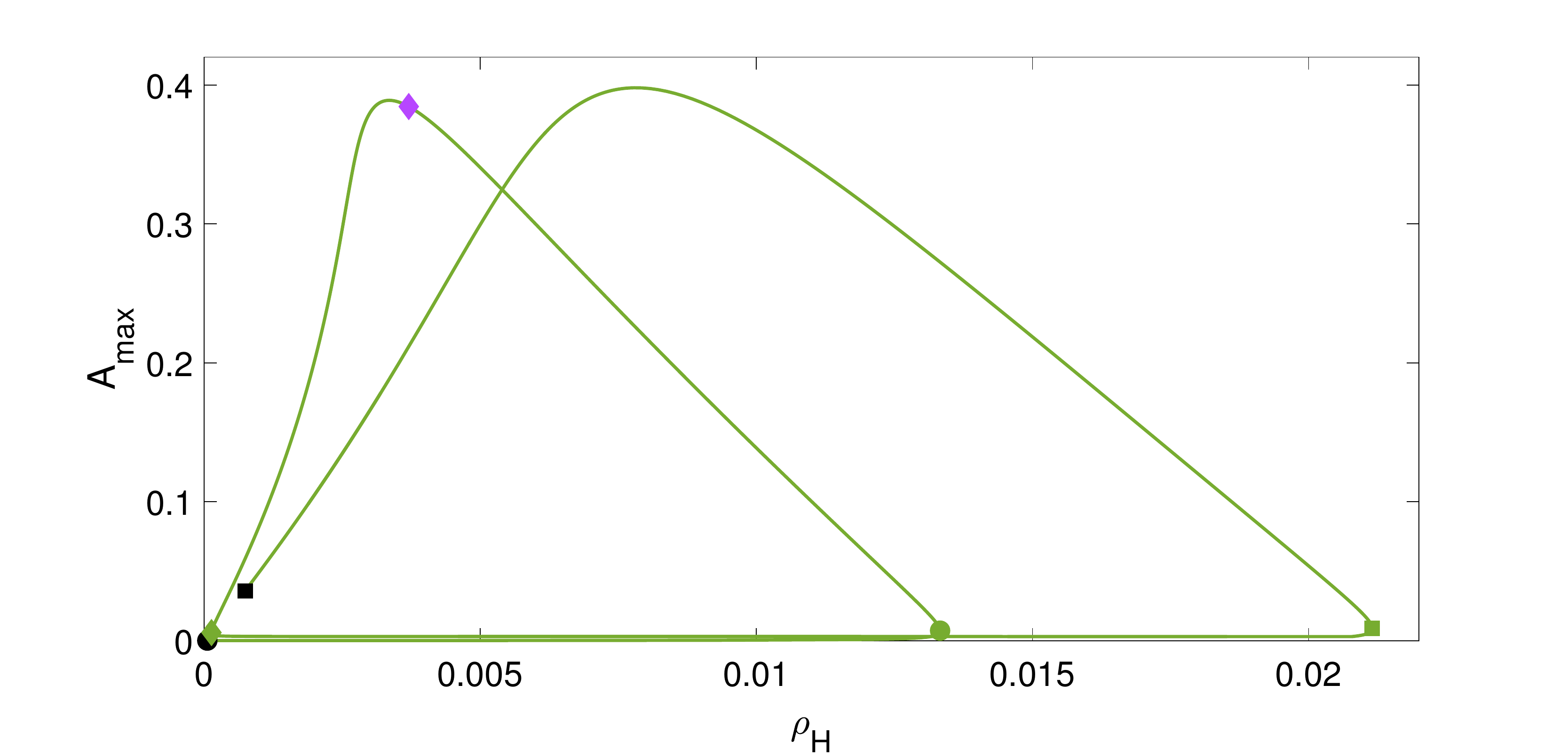}}
\end{minipage}}
\usebox{\measurebox}\qquad  \qquad
\begin{minipage}[b][\ht\measurebox][s]{.215\textwidth}
  {\textcolor{white}{(c)}\includegraphics[width=\textwidth]{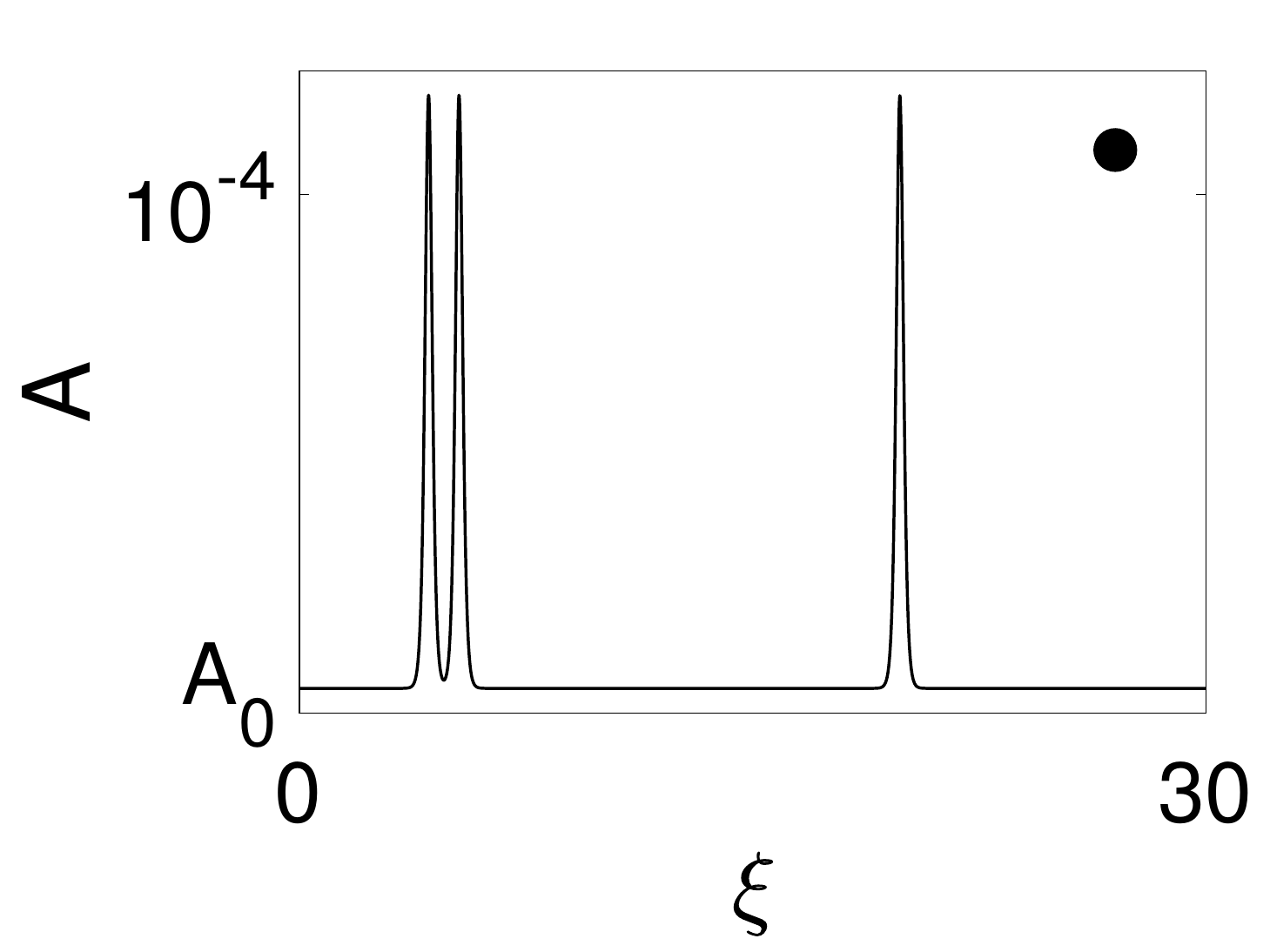}}
  {\textcolor{white}{(c)}\includegraphics[width=\textwidth]{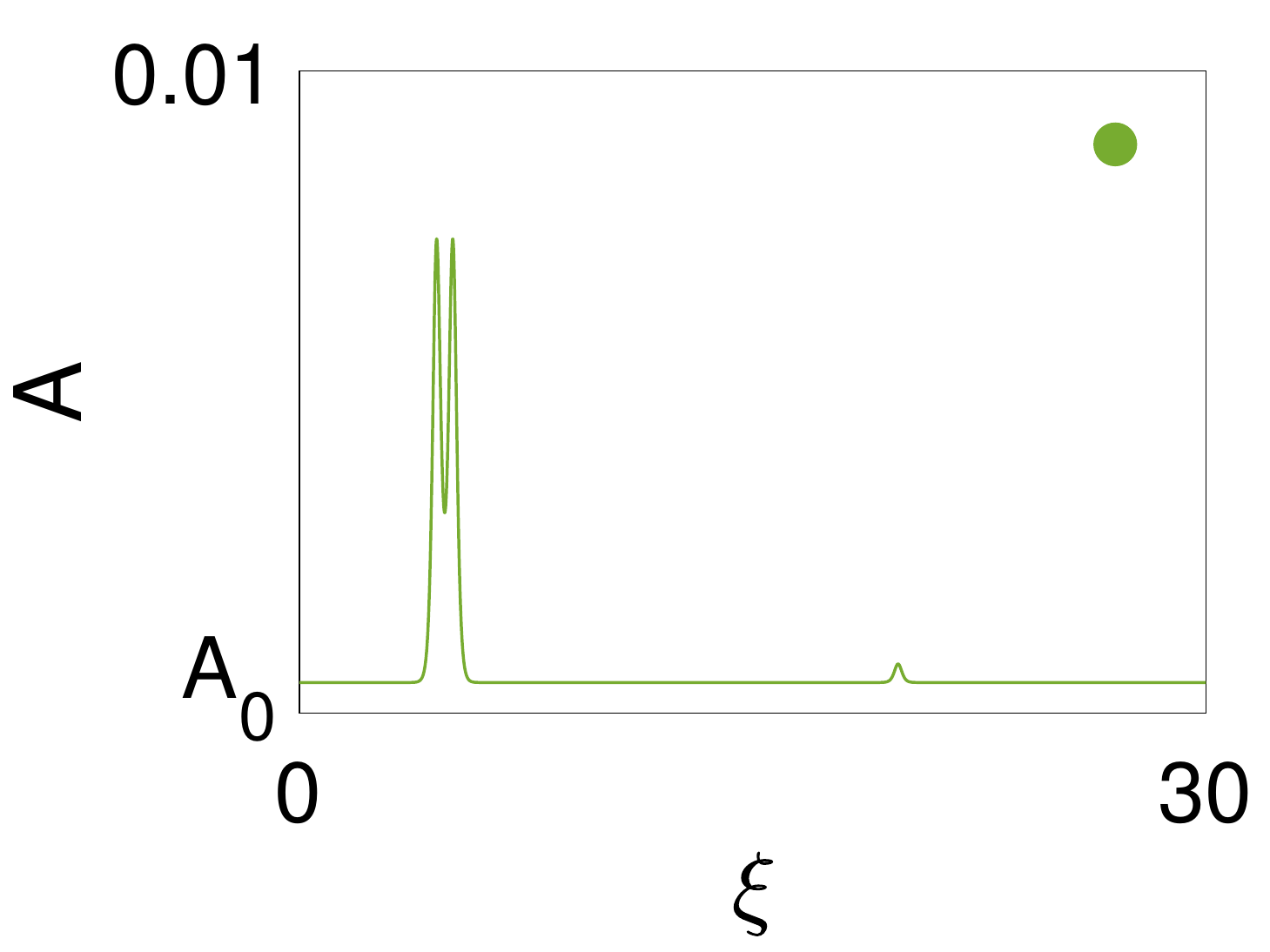}}
  {\textcolor{white}{(c)}\includegraphics[width=\textwidth]{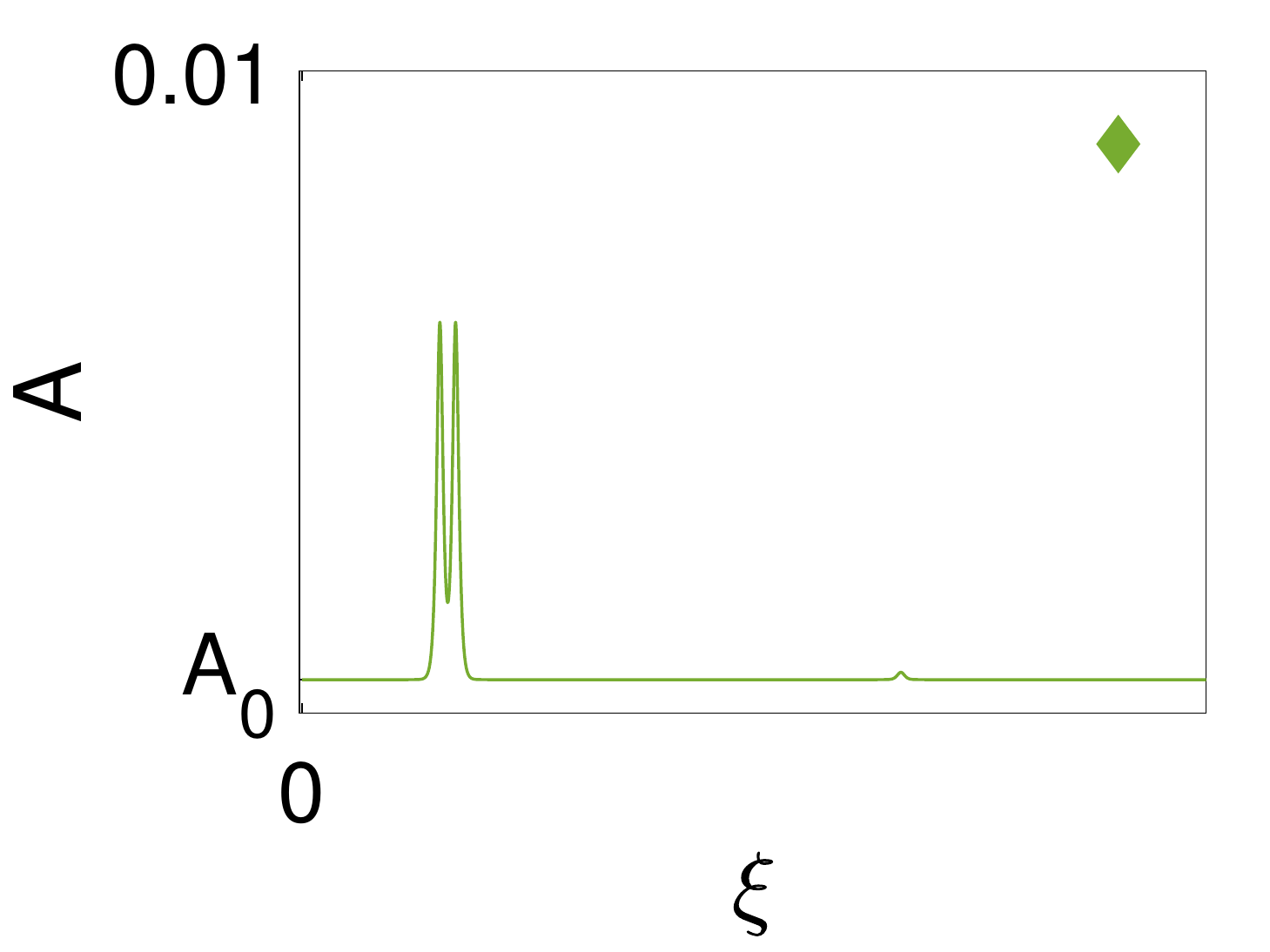}}
  {(c)\includegraphics[width=\textwidth]{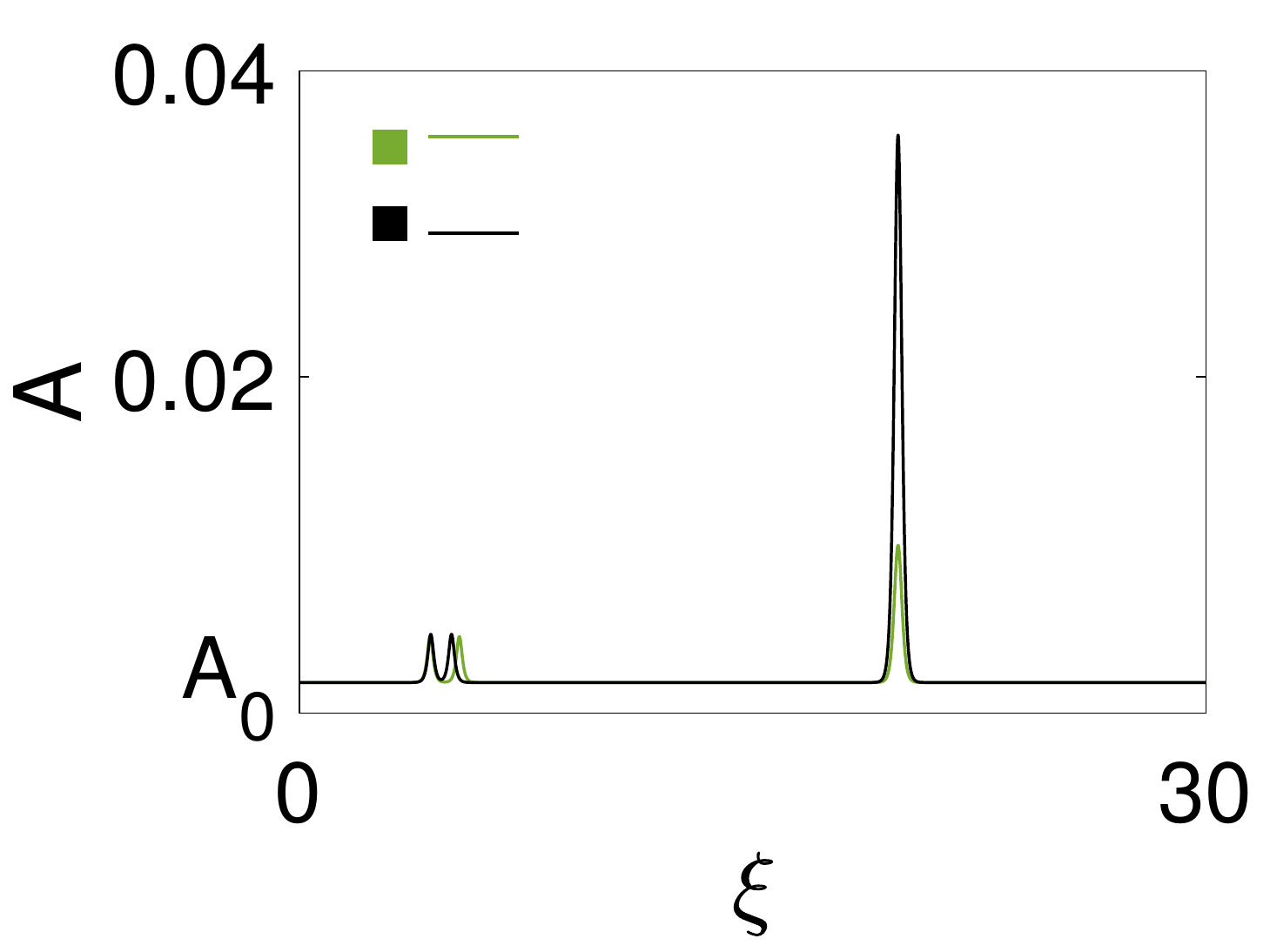}}
\end{minipage}
  \caption{(a) Deconstruction of Fig.~\ref{fig:DYm4_1P} showing the two-peak traveling states (purple) and their velocity $v$ (yellow) together with the stationary 2P and 3P branches showing the transition from a 2P state with precursor to a 3P state without one. The black curve shows the stationary states with no precursor, while the superposed green curve shows the corresponding branch with a precursor. The traveling states originate in a parity-breaking bifurcation on the branch of stationary 2P states with precursor (purple diamond on the green branch) and terminate on the black branch of stationary 3P states without precursor in another parity-breaking bifurcation. The inset shows the profile corresponding to the purple diamond: the precursor is exactly half way between the 2P profile and its periodic translate and so is symmetric. (b) The evolution of the stationary 2P branch with a precursor peak in terms of the maximum peak amplitude A$_{\rm max}$. (c) Four sample profiles, where the symbols indicate corresponding locations in (a) and (b). The green branch spirals counterclockwise from the black dot (bound state of identical 2P and 1P states) to the green dot (2P profile with a precursor) to the green square (1P profile with a single precursor) and finally to the black square (1P profile with a double precursor). The corresponding locations in the L$_2$ norm are indicated in panel (a).}
    \label{fig:DYm4_2P_stat}
\end{figure*}

\begin{figure*}[tp]
\sbox{\measurebox}{%
\begin{minipage}[b]{0.67\textwidth}
  {(a)\,\,\,\includegraphics[width=\textwidth]{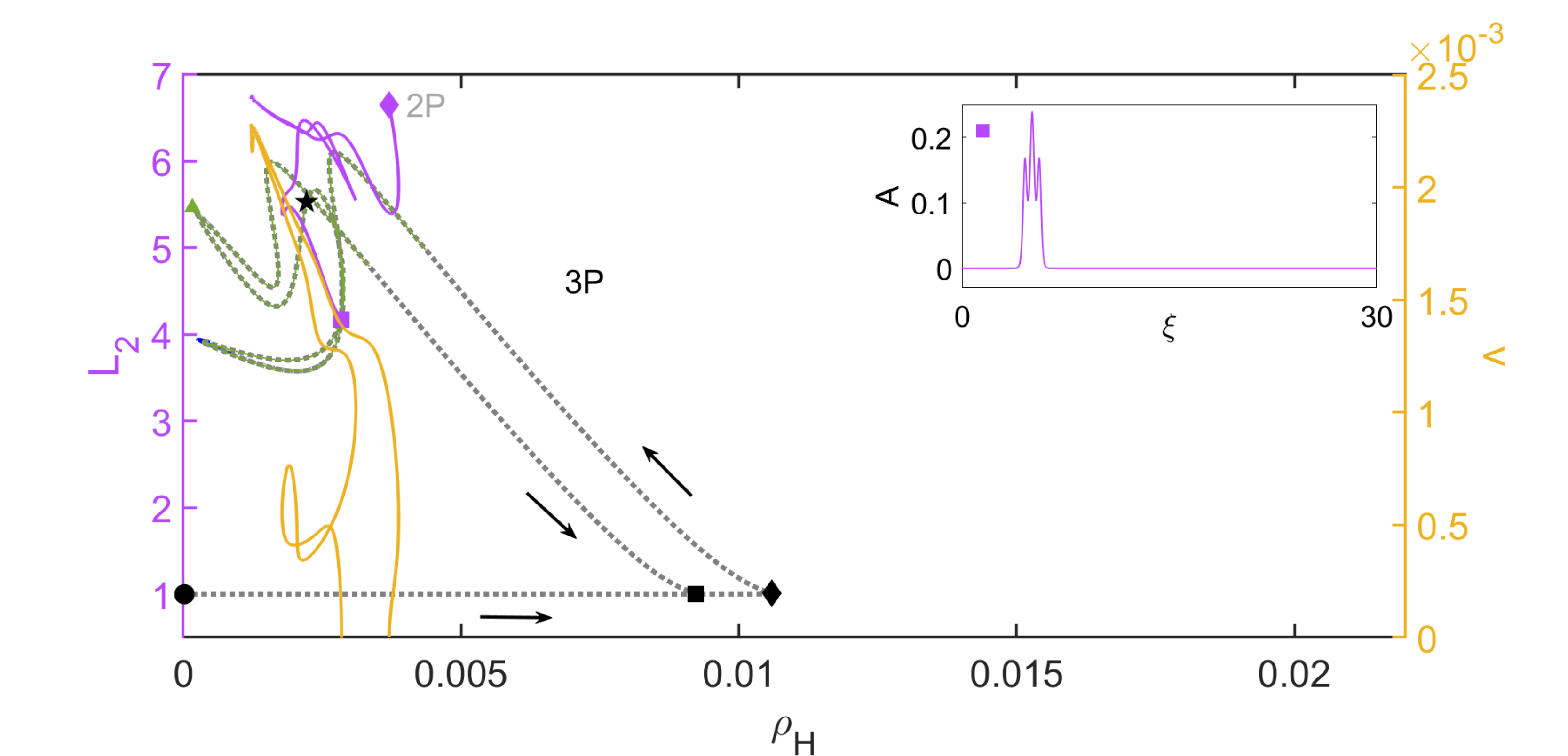}}
  {(b)\includegraphics[width=\textwidth]{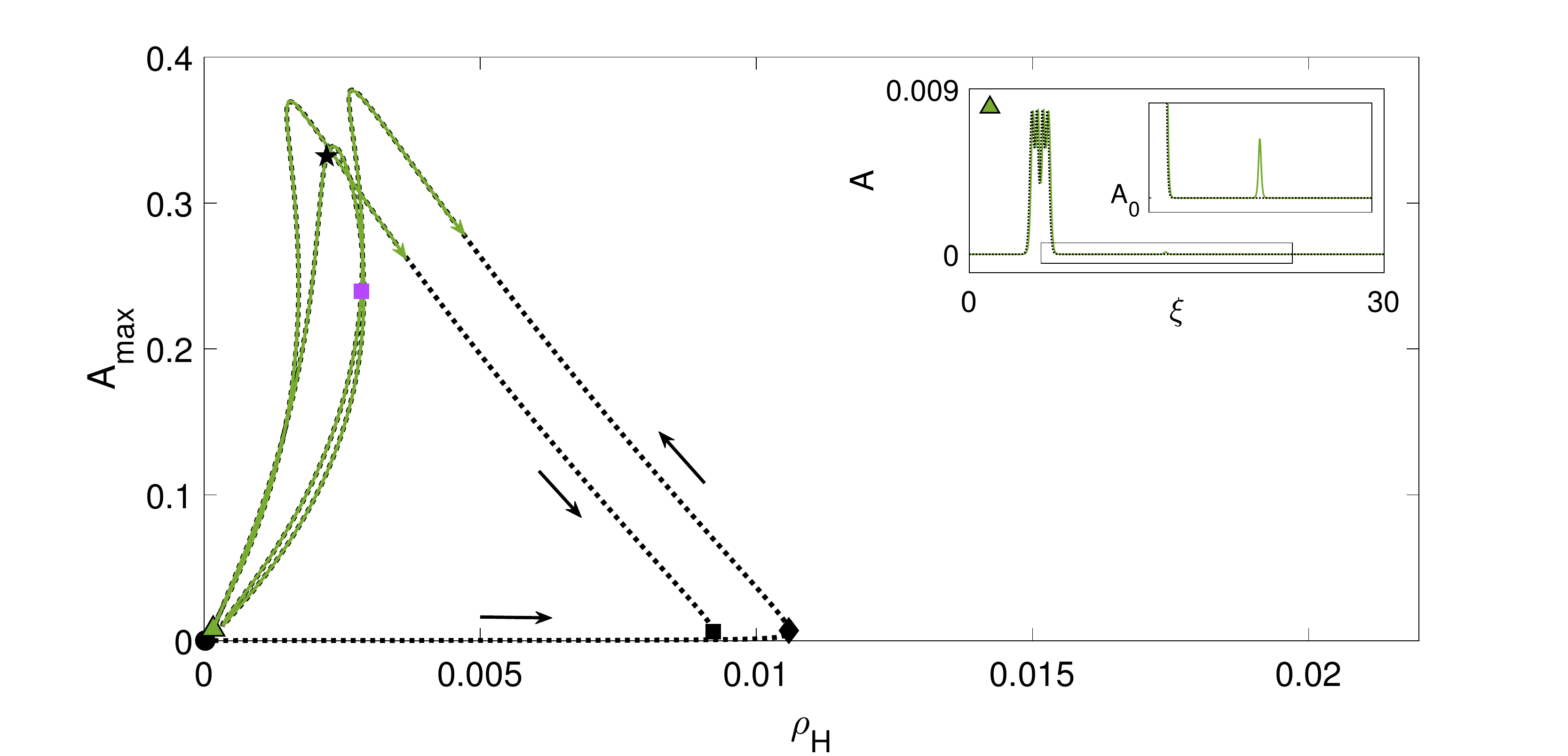}}
\end{minipage}}
\usebox{\measurebox}\qquad  \qquad
\begin{minipage}[b][\ht\measurebox][s]{.215\textwidth}
  {\textcolor{white}{(c)}\includegraphics[width=\textwidth]{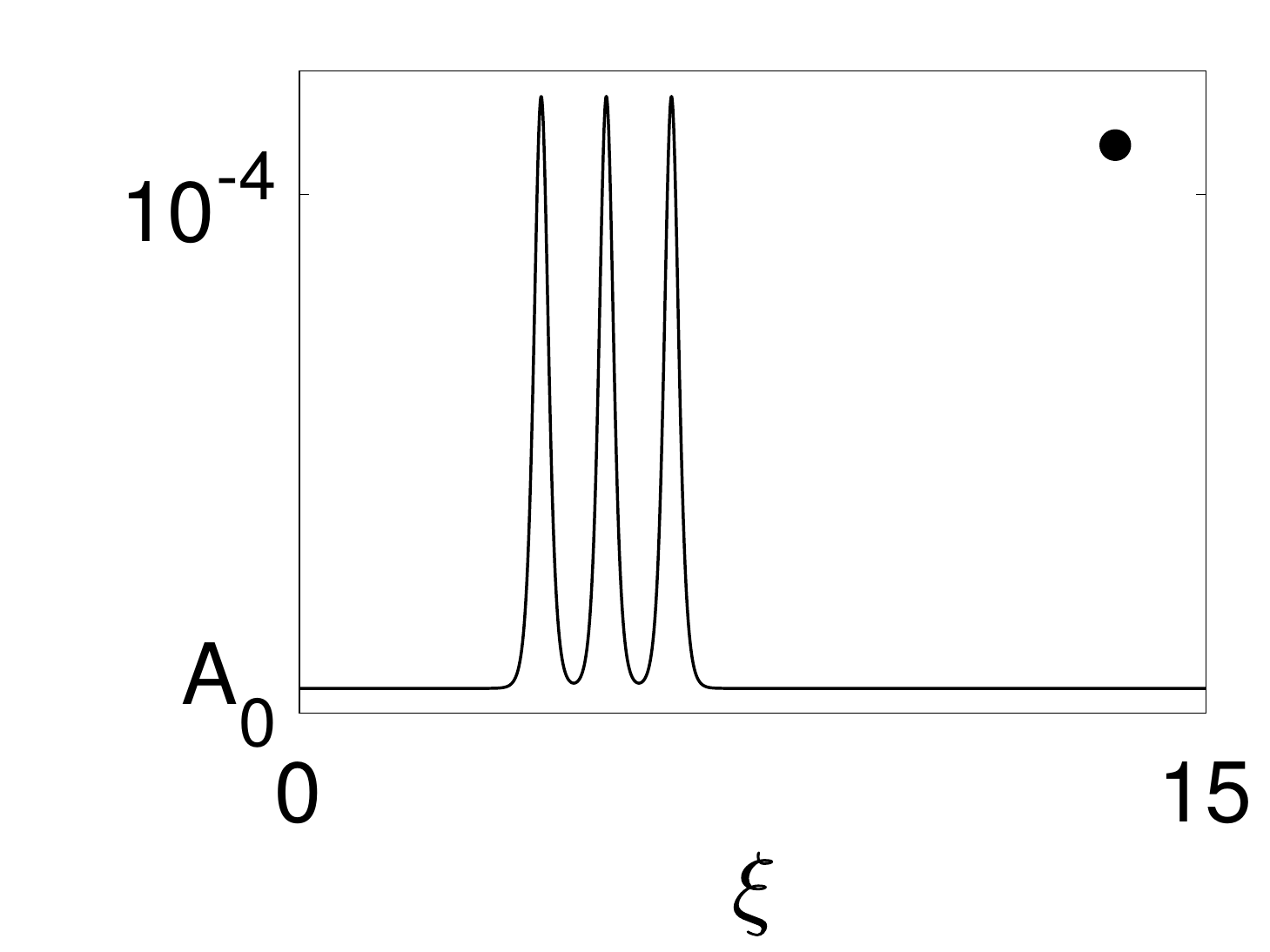}}
  {\textcolor{white}{(c)}\includegraphics[width=\textwidth]{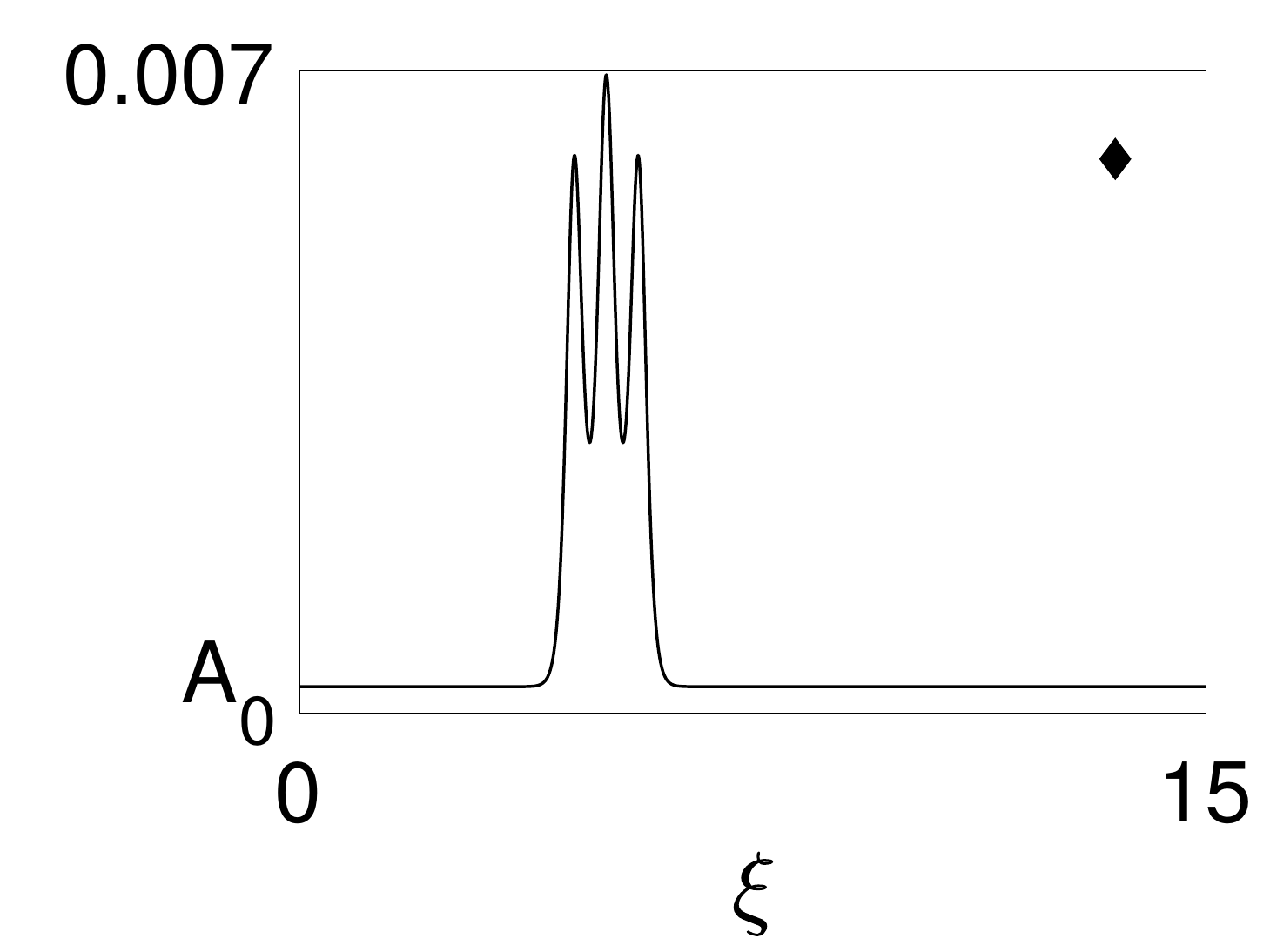}}
  {\textcolor{white}{(c)}\includegraphics[width=\textwidth]{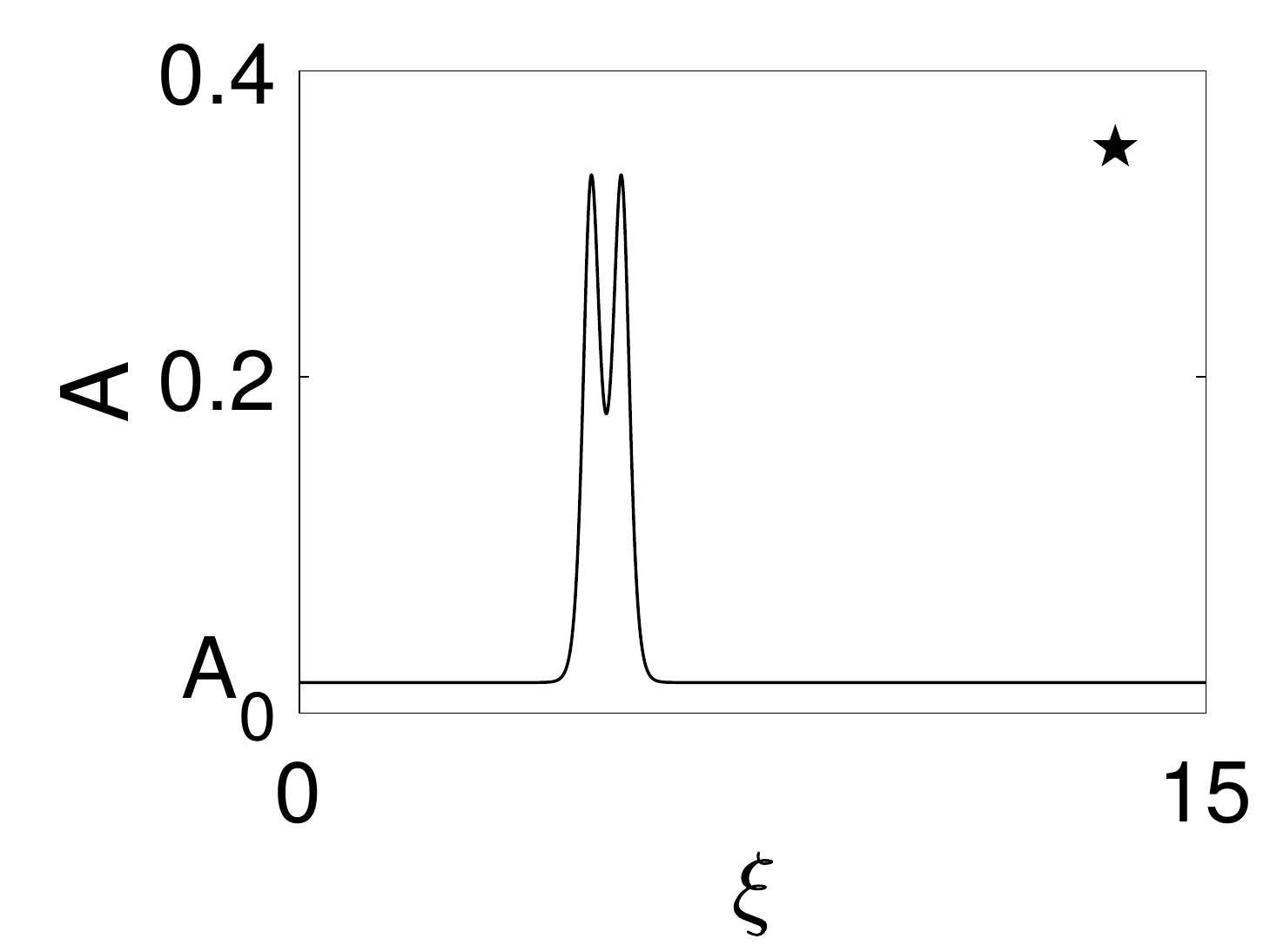}}
  {(c)\includegraphics[width=\textwidth]{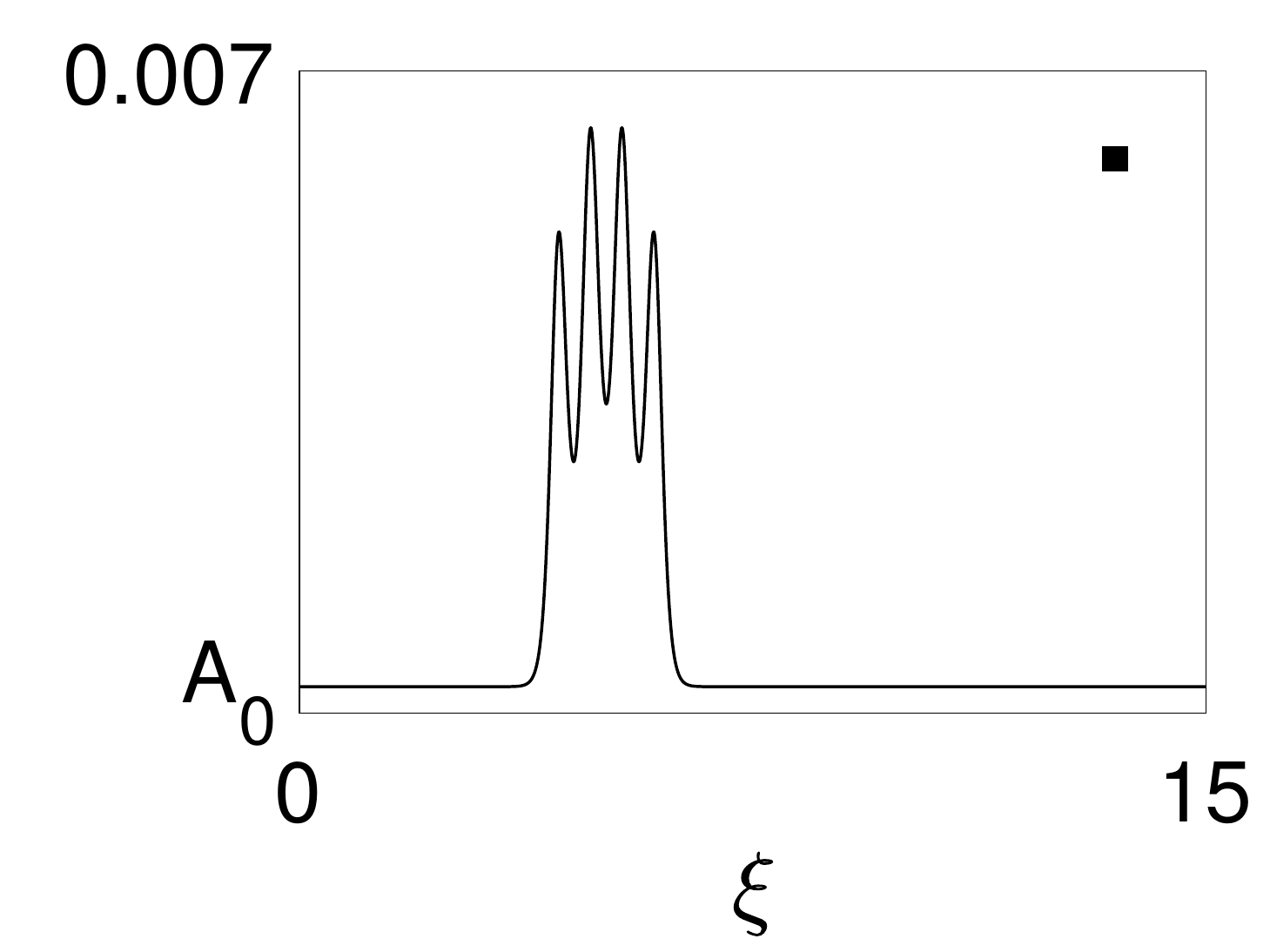}}
\end{minipage}
    \caption{Stationary 3P branch without (black) and with the extra small peak (partial branch, green) in terms of (a) the L$_2$ norm and (b) the maximum amplitude A$_{\rm max}$. The color-coded insets show the profiles at the locations in the main panels: 3P state at the termination of the traveling states (magenta) and a 4P state with precursor (green), respectively. (c) The solution profiles at the four locations indicated in panels (a) and (b) by the solid black symbols, {shown on the half domain.} The arrows indicate the direction along the black branch starting with the leftmost profile in panel (b).}
    \label{fig:DYm4_3P_stat}
\end{figure*}

\subsection{Fronts, peaks, and T-points}

 {In the following we refer to the T-point in Fig.~\ref{fig:DYm4}(a) as TP0. This point is located at $\rH \approx 4.0 \times 10^{-5}$, i.e. in between the fold point $\rho_{SN}^{+L}\approx 3.95\times 10^{-5}$ and the fold point $\rho_{SN}^{+R}\approx 2.25\times 10^{-4}$ (Fig.~\ref{fig:bif_uni}). However, this T-point is not the only one. Figure~\ref{fig:DYm4_1P_rtrn_spiral2}(a) shows two {\it additional} T-points, labeled TP1 (blue curve) and TP2 (purple curve), both with spiral structure similar to that associated with TP0, but computed with PBC. These points are located, respectively, at $\rH \approx 3.3 \times 10^{-4}$ (black diamond) and $\rH \approx 4.9\times10^{-4}$ (black square); both points are located to the right of $\rho_{SN}^{+R}\approx 2.25\times 10^{-4}$. Because of the use of PBC we can no longer interpret the associated behavior in terms of traveling fronts but instead do so in terms of traveling and stationary peak states. Figure~\ref{fig:DYm4_1P_rtrn_spiral2}(a) shows that the traveling peak states responsible for TP1 are no longer part of an extended branch of traveling fronts but are created via a parity-breaking bifurcation of a stationary one-peak state (black dashed line) that takes place at $\rH \approx 7.5\times 10^{-4}$ (Fig.~\ref{fig:DYm4_1P_rtrn_spiral2}(a), blue dot). The velocity $v$ of these states vanishes at this point (red dot). Thus with PBC propagation sets in via a spontaneous parity-breaking bifurcation instead of being due to asymmetry imposed by different states at either end of the domain as in Fig.~\ref{fig:DYm4}.}

 {Despite this difference the behavior near TP1 resembles that near TP0 as do the corresponding solution profiles. Figure~\ref{fig:DYm4_1P_rtrn_spiral2}(b) shows the change in the solution profile in passing from the parity-breaking bifurcation to the spiral center: as in the case of TP0 the profile picks up a new oscillation with each turn of the spiral. Moreover, as for TP0, the solution branch turns around near the spiral center and spirals back out, emerging with a precursor peak (Figs.~\ref{fig:DYm4_1P_rtrn_spiral2}(d-f)). The resulting traveling bound state terminates with increasing $\rH$ in a parity-restoring bifurcation on a different steady state branch, this time consisting of a bound state of a large peak and a small precursor (Fig.~\ref{fig:DYm4_1P_rtrn_spiral2}(d), green dashed line). Owing to the very small size of the precursor peak, the green dashed line is almost identical to the black dashed line, and the blue dots indicating the parity-breaking bifurcations are located at almost identical values of $\rH$.}

The point TP2 behaves very similarly. This time the traveling front is created in a parity-breaking bifurcation on a branch of stationary 3-peak states (black dotted line) that takes place at the purple dot ($\rH \approx 8.8\times 10^{-4}$). Once again the traveling state spirals into TP2, picking up oscillations in the front profile with each turn of the spiral, and then reemerges with a small precursor peak, terminating on a branch of stationary bound states of a 3-peak state and a small precursor peak (Fig.~\ref{fig:DYm4_1P_rtrn_spiral2}(d), green dotted line) at almost the same value of $\rH$, $\rH \approx 8.8\times 10^{-4}$. Neither TP1 nor TP2 is shown in Fig.~\ref{fig:DYm4}.

 {Despite appearances, there is no spatially periodic state at the center of the TP1 and TP2 spirals: our computations show that the solution branch spirals in for a certain number of turns and then spirals back out. We believe that this is a consequence of the finite domain size used in the computations, and expect that on larger domains additional turns would be uncovered as the oscillation wavelength at the leading edge of the profile and its speed $v$ simultaneously adjust towards the T-point, cf.~\cite{raja2023} No connection to a spatially periodic state therefore takes place and indeed no periodic state has been identified at these parameter values. Moreover, because of the PBC used, both the main peak and the precursor peak sit on an $\vP_0$ background. In addition, the leading stable and unstable spatial eigenvalues of the precursor state are real and almost equal in magnitude: the precursor peak profile is therefore monotonic and almost symmetric.}

 {When $\rH$ is increased further, additional T-points are encountered. Figure \ref{fig:DYm4_TP_1P} shows the corresponding bifurcation diagram, also with PBC. Two T-points are found on a branch of traveling peaks (dashed blue line) and are associated with vertical segments or discontinuities in the solution branch. The corresponding locations on the plots of the velocity $v$ are indicated by black dots.} Figure \ref{fig:DYm4_TP_1P}(b) shows that the traveling solution at $\rho^+_{\rm TP}$ (black dot in the top panel) consists of a large peak with a small precursor peak. This part of the profile thus resembles the traveling front state in Fig.~\ref{fig:DYm4} but because of the PBC the $\vP_0$ profile ahead of the precursor peak connects back to $\vP_+$. Thus the solution actually takes the form of a heteroclinic {\it cycle}, connecting $\vP_+$ to $\vP_0$ via the large peak, followed by a homoclinic $\vP_0\to \vP_0$ connection via the precursor peak and then a monotonic $\vP_0\to \vP_+$ connection. The leading stable and unstable spatial eigenvalues of both $\vP_+$ and $\vP_0$ are real at these parameter values accounting for the observed monotonic profiles. The profiles in Figs.~\ref{fig:DYm4_TP_1P}(a,c) explain what happens as $\rH$ passes through $\rho^+_{\rm TP}$. In (a) the solution takes the form of a overall $\vP_0$ to $\vP_0$ connection, while in (c) it takes the form of an overall $\vP_+$ to $\vP_+$ connection, thereby accounting for the discontinuity in L$_2$ norm across $\rho^+_{\rm TP}$. Thus TP$^+$ corresponds to the simultaneous presence of an $\vP_+\to \vP_0$ connection and a (distinct) $\vP_0\to \vP_+$ connnection, as shown in panel (b). Moreover, panel (c) shows clearly how the small precursor peak is assimilated in the large peak once $\rH<\rho^+_{\rm TP}$ forming a double-peak structure in a process that is similar to that observed earlier in Fig.~\ref{fig:DYm4_1P_rtrn_spiral2}. Evidently the discontinuous amplitude behavior at TP$^{+}$ in Fig.~\ref{fig:DYm4_TP_1P} is a consequence of a switch between an overall homoclinic connection from $\vP_0\to \vP_0$ (smaller amplitude) to an overall homoclinic connection from $\vP_+\to \vP_+$ (larger amplitude) that occurs at this point and is thus a consequence of the use of PBC.

The profile at $\rho^*_{\rm TP}$ (also indicated by a black dot in the top panel) is quite different: by TP$^*$ the original large peak has turned into a double structure consisting of two unequal peaks and no precursor (Fig.~\ref{fig:DYm4_TP_1P}(e)). Figures \ref{fig:DYm4_TP_1P}(d,f) show that this point corresponds to a transition from an overall $\vP_+$ to $\vP_+$ connection to an overall $\vP_0$ to $\vP_0$ connection, once again accounting for the observed discontinuity in the L$_2$ norm of the solution.

Figure \ref{fig:DYm4_2P_TP} shows similar behavior associated with traveling 2P states. These states are created in a parity-breaking bifurcation on a branch of stationary 2P states and terminate in another parity-breaking bifurcation on a branch of stationary 3P states. In between this state passes through two T-points, TP$^*$ and TP$^-$ (black dots). The profile at TP$^*$ is shown in Fig.~\ref{fig:DYm4_2P_TP}(b) and consists of back-to-back connections between $\vP_+$ and $\vP_0$ and $\vP_0$ and $\vP_+$ with a small precursor peak in between. Panels (a) and (c) show that TP$^*$ corresponds to a transition from an overall $\vP_0$ to $\vP_0$ connection to an overall $\vP_+$ to $\vP_+$ connection; (c) in particular captures the process whereby the precursor peak is assimilated in the double-peak structure forming a 3P state with no precursor. The transition at TP$^-$ corresponds to a switch from a $\vP_+$ to $\vP_+$ connection to one from $\vP_0$ to $\vP_0$ but now based on the 3P structure with no precursor. 

Past $\rho^*_{\rm TP}$ and $\rho^-_{\rm TP}$, the double- and triple-peak states created by this process track one another and travel with almost identical speeds. Neither structure contains a precursor. Following this part of the branch one finds that the multipeak structures in the double- and triple-peak states become more and more symmetric, becoming reflection-symmetric and hence stationary at the parity-breaking bifurcations with which these branches terminate. At the same time the location of the $\vP_0\to \vP_+$ front moves increasingly to the right, gradually eliminating the $\vP_+$ state. Thus at the parity-breaking bifurcation the two- and three-peak states sit on an $\vP_0$ background. In the following the corresponding states are shown using a broken black line.

We remark that the presence of the T-points is a consequence of using PBC: on an infinite domain the branch of $\vP_+\to \vP_0$ traveling states is not required to connect up with $\vP_0\to \vP_+$ traveling states.

We show a larger view in Fig.~\ref{fig:DYm4_1P}. This figure shows the continuation of the blue branch in Fig.~\ref{fig:DYm4_TP_1P} and reveals its origin at larger $\rH$ in a parity-breaking bifurcation from a 1P branch of stationary states {with a precursor} (green curve) and its termination on a 2P branch of stationary two-peak states with no precursor (black curve), albeit in the PBC setting. Figures \ref{fig:DYm4_TP_1P}(a-f) show how this transformation takes place. In particular, this figure shows that the stationary peak states at either end of the blue branch sit on a $\vP_0$ background, and explains how the $\vP_+$ pedestal is eliminated along the branch of traveling states.

Given the complexity of Fig.~\ref{fig:DYm4_1P} we find it useful to separate out certain features. In Figs.~\ref{fig:DYm4_1P_stat}(a,b) we show the branch of stationary 1P states with a precursor and explain how these transition to a 2P state with identical equispaced peaks (see profiles in Fig.~\ref{fig:DYm4_1P_stat}(c)), all on a $\vP_0$ background; $n$P states with $n>2$ identical equispaced peaks are generated by a similar process and all coexist at essentially one value of $\rH$. The parity-breaking bifurcation to traveling states is indicated by a blue dot on the green 1P branch; the corresponding profile is shown in the inset. The traveling states terminate on a 2P branch (not shown) in another parity-breaking bifurcation (purple diamond). Along the 1P branch (green curve) the profile changes from one with a small precursor peak (profile shown at the location of the green dot) to a 2P state with two identical equidistant peaks (profile shown at the location of the green down-triangle). The 1P state with no precursor is superposed (black dashed line). The right panel in Fig.~\ref{fig:DYm4_1P_stat}(c) shows the 1P solution with no precursor at the same location as the green down-triangle. Since the black 1P state is strongly exponentially localized it can be used to construct a 2P state consisting of two identical equispaced peaks in the domain at the same parameter location. Thus the black dashed branch below the black up-triangle in fact represents the continuation of the green branch, and the green triangle represents a spatial period-doubling bifurcation at which the green precursor branch terminates on the 2P branch of identical equispaced peaks. We mention that we have been unable to extend the solution corresponding to the green dot farther to the right and thus cannot determine whether it ever terminates on the 1P black branch consisting of single peak states and no precursor.

Figures~\ref{fig:DYm4_2P_stat}(a,b) show the corresponding behavior of the 2P states with a precursor (green curve) that generate the two-peak traveling states in a parity-breaking bifurcation (purple diamond) already indicated in Fig.~\ref{fig:DYm4_1P_stat}. These traveling states in turn terminate on the black branch of stationary 3P states without precursor (purple square, branch not shown) in another parity-breaking bifurcation. The inset shows the profile corresponding to the purple diamond: the precursor is exactly half way between the 2P profile and its periodic translate and so is symmetric. The symbols indicate corresponding locations in (a). The green branch spirals counterclockwise from the black dot (bound state of identical 2P and 1P states) via the green dot (2P profile with a precursor) to the green square (1P profile with a single precursor) and finally to the black square (1P profile with a double precursor). To describe these transitions in greater detail note that in going backward from the green dot, the dip in the 2P structure dips down to A$_0$ while its overall amplitude approaches that of the precursor; going in the opposite direction, the amplitude of the 2P structure decreases forming the double precursor while the amplitude of the original precursor grows, thereby forming the green profile in the right panel of Fig.~\ref{fig:DYm4_2P_stat}(c). The corresponding locations in the L$_2$ norm are indicated in panel (a). This behavior recapitulates that already seen in Fig.~\ref{fig:DYm4_1P_stat}. In particular, all these states sit on a $\vP_0$ background. In fact this spiraling behavior is a general feature of this type of transition (Fig.~\ref{fig:DYm4_3P_stat} shows a further example showing the transition from a 3P stationary symmetric state with three identical peaks (black dot) to a 4P stationary symmetric state (black square) via a precursor state (green triangle)) and evidently represents a general peak-adding mechanism.
\begin{figure*}[tp]
    \quad\includegraphics[width=0.95\textwidth]{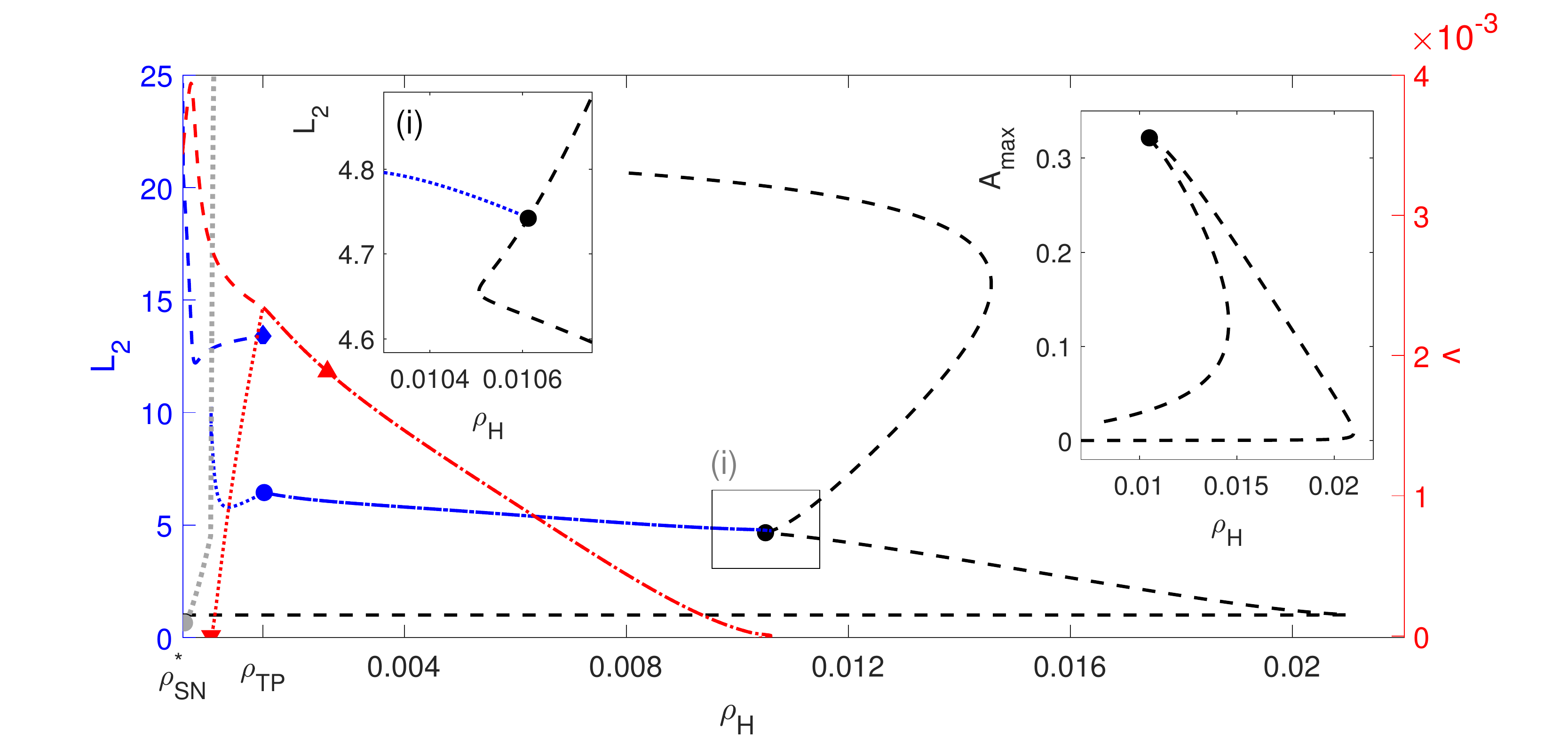}\\
    (a)\includegraphics[width=0.32\textwidth]{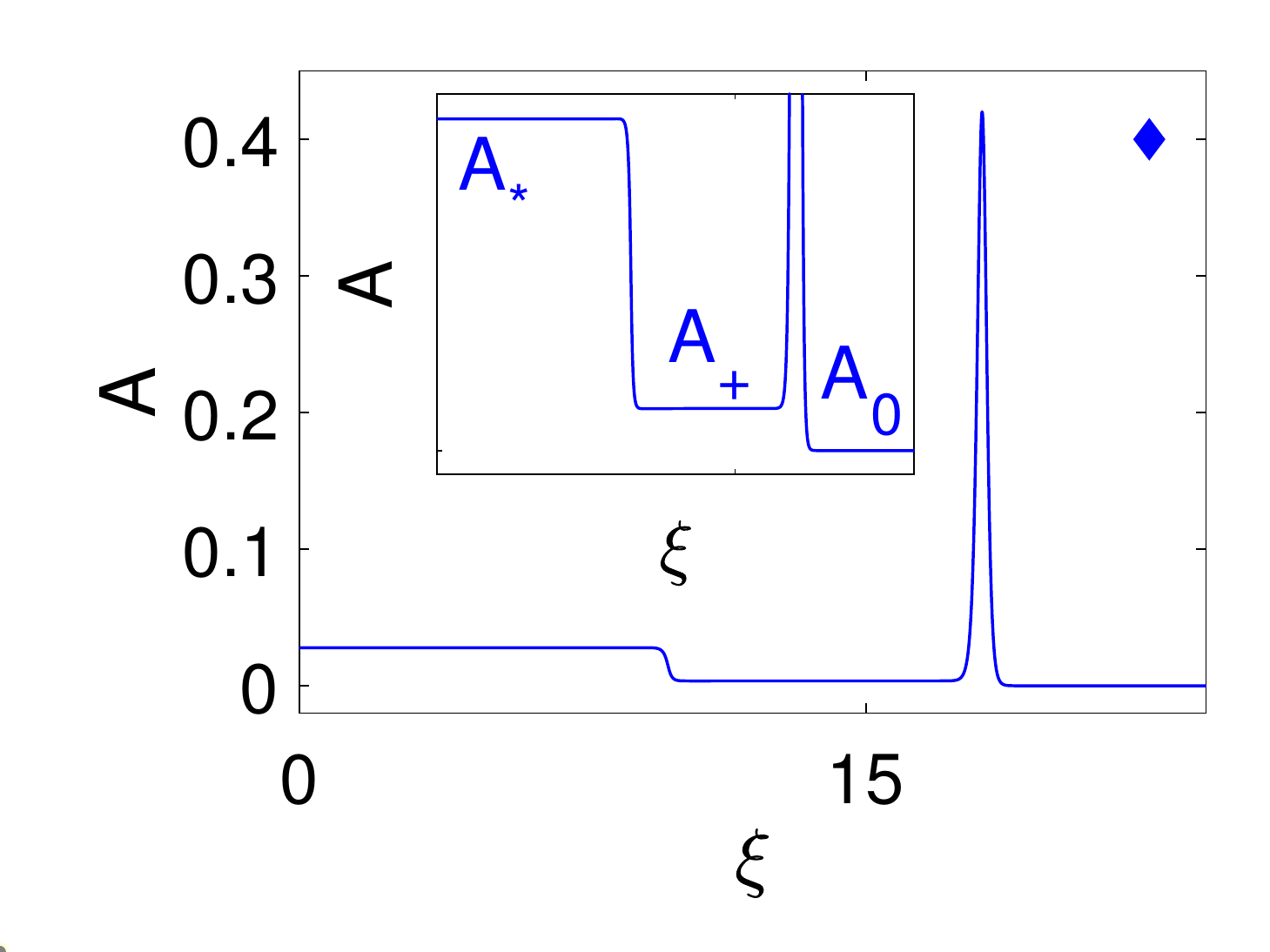}
    \includegraphics[width=0.32\textwidth]{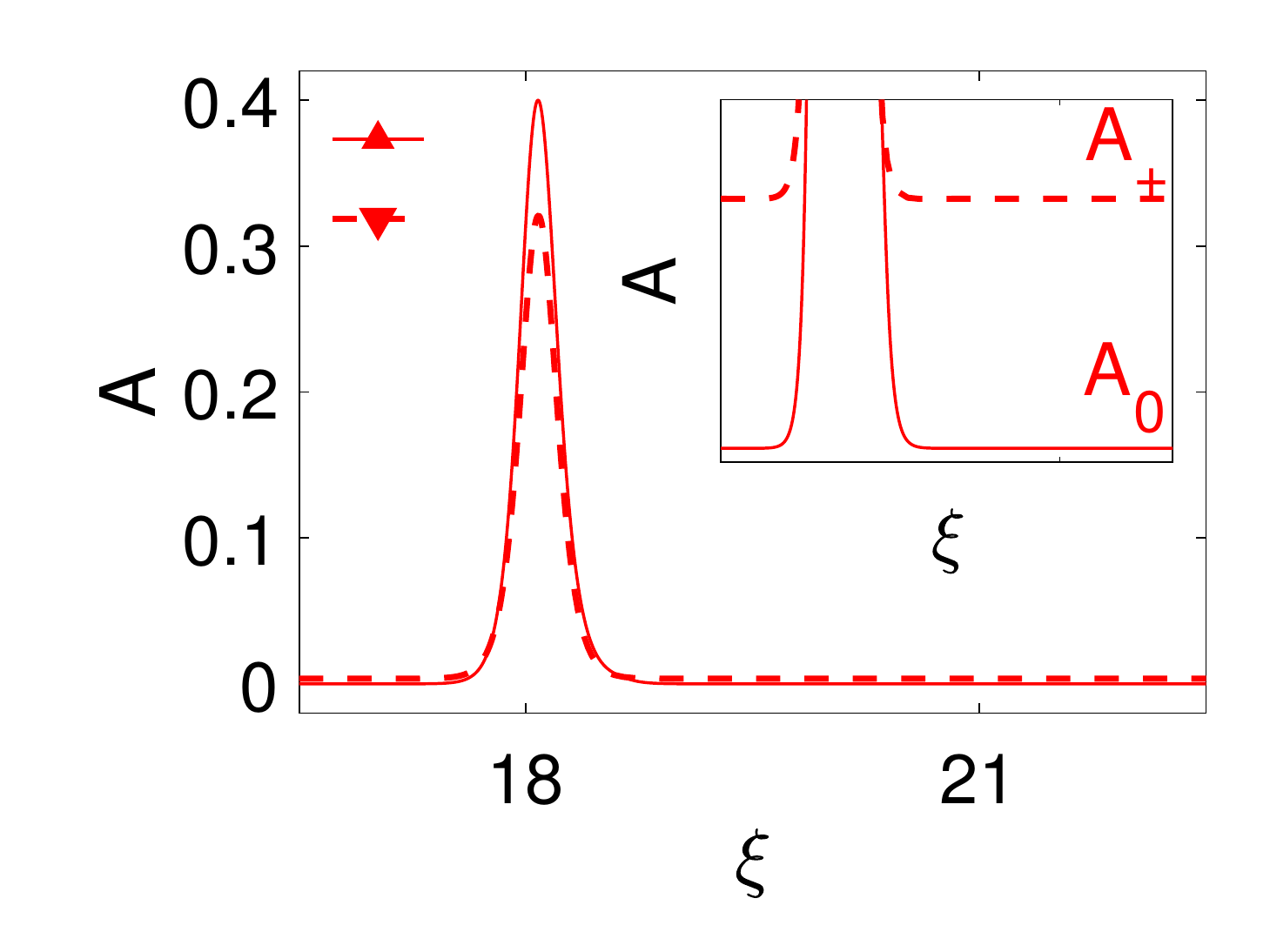}
    \includegraphics[width=0.32\textwidth]{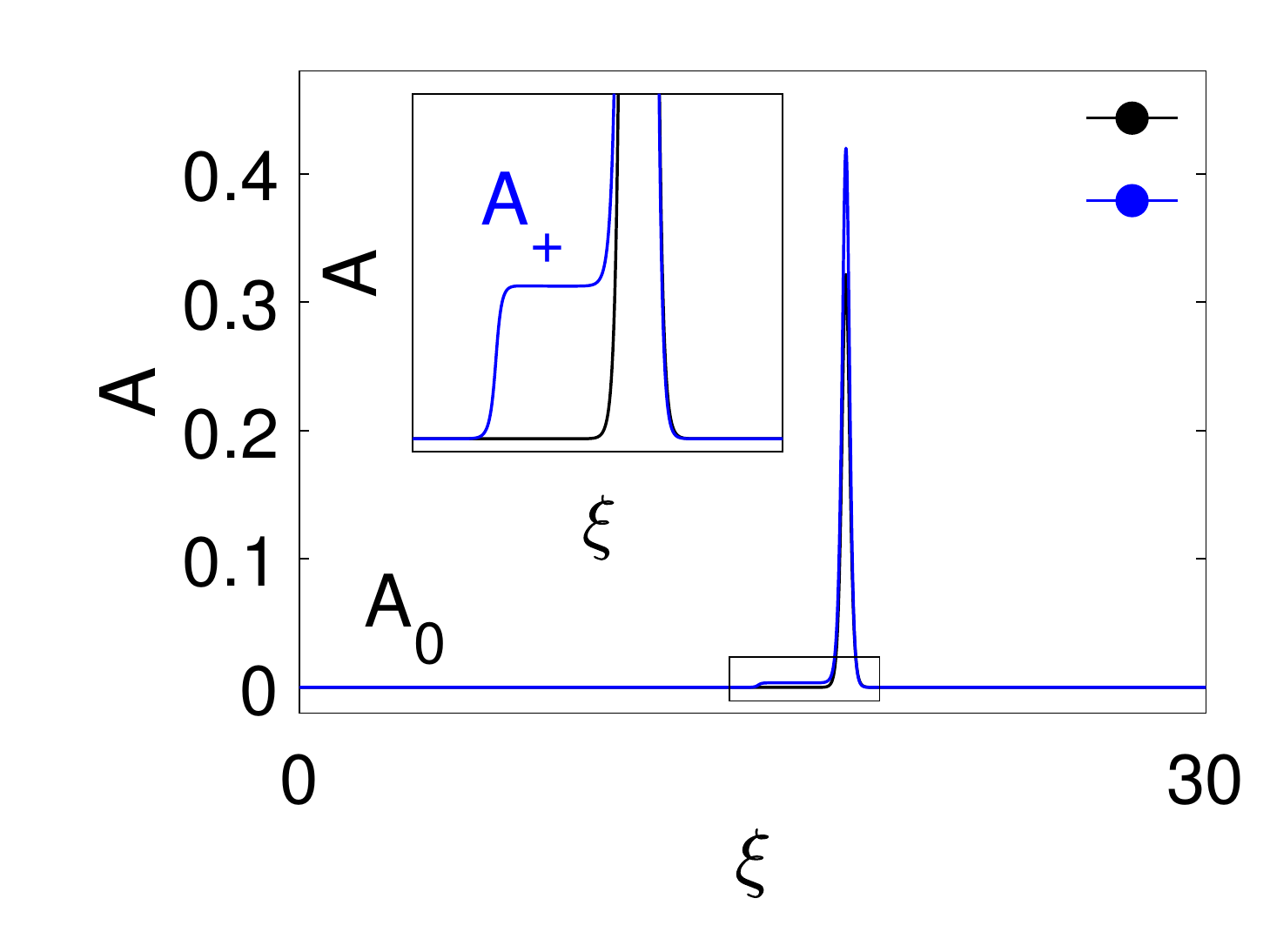} (b)\includegraphics[width=0.95\textwidth]{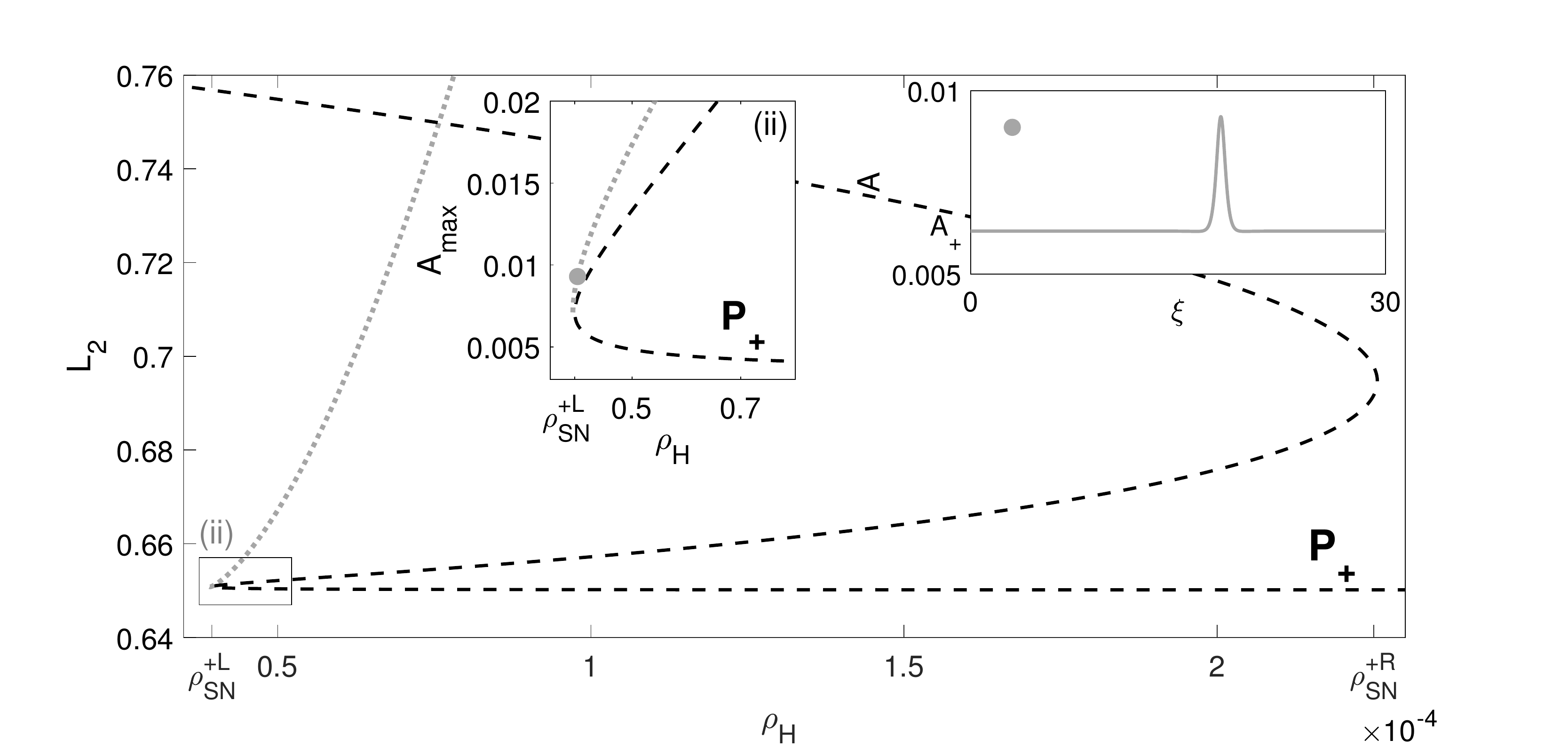}
    \caption{(a) Bifurcation diagram showing the unstable fast fronts (dashed, blue) that exist for $\rho_{\rm SN}^*<\rH<\rho_{\rm TP}\approx 1.45 \times 10^{-3}$ and two families of single peak solutions with finite speed (dotted, blue) and stationary (dashed black and dotted gray) on either side of the TP point (solid blue diamond/dot), computed with NBC. The bottom panels show the profiles at the marked locations: left of TP point (blue diamond), right of TP point (blue dot). Profiles further from the TP point are shown in the middle panel (red). The right panel also shows the profile (black) corresponding to the termination of the traveling peak branch (black dot); the associated branch of stationary states is shown in the main panel (dashed, black). Insets show two details of the diagram. (b) The termination of the 1P branch (dotted, gray) of stationary peaks on the fold of the uniform states $\vP_+$. The insets show details of the behavior near the fold together with a sample solution profile. Parameter: $D_Y=10^{-7}$.}
    \label{fig:DYm7}
\end{figure*}

\subsection{Fast fronts and negligible diffusion of the differentiation field}

Based on the above results, we were able to identify similar behavior in the case negligible diffusion, $D_{\text{Y}}=10^{-7}$, i.e., the case shown in Fig.~\ref{fig:speed}. The results of this effort are reported in Fig.~\ref{fig:DYm7}, computed with NBC, and reveal the formation of double front structure analogous to that seen in Fig.~\ref{fig:DYm4}(b). In this case, however, the double front structure involves all three uniform states that exist at these parameter values: these fronts consist of a connection between $\vP_*$ and the unstable uniform state $\vP_+$ and a second connection between $\vP_+$ and $\vP_0$ (Fig.~\ref{fig:DYm7}(a), lower left inset). The former may be viewed as a stable state invading an unstable state and its speed is determined in general by the marginal stability criterion. The latter may also be viewed as the invasion of $\vP_+$ by a stable state, this time $\vP_0$, and so its speed is determined by the same criterion. Of course for the double front these two speeds are connected since one front sees the other as an inhomogeneity and vice versa. Such an interaction may lead to locking of the two fronts and a steadily propagating state of the type computed here.

This double front state represents a transition between the fast fronts in the upper left of the figure (dashed blue line) and the slower (but still fast) fronts (solid blue line, dotted in inset) as $\rH$ increases, where the $\vP_*\to \vP_+$ connection is replaced by a $\vP_0\to \vP_+$ connection {(Fig.~\ref{fig:DYm7}(a), profiles at the blue diamond and the blue dot)} via a T-point, labeled TP. The latter state, being homoclinic, can be continued with PBC (dotted blue line) back towards lower values of $\rH$. One finds that the speed of these traveling states drops to zero (down-pointing red triangle) below which one finds single peak steady states on a $\vP_0$ background (dotted black line). If one instead increases $\rH$ and follows these slower fronts to the right, one encounters a different parity-breaking bifurcation (see left inset in Fig.~\ref{fig:DYm7}(a)); the right inset shows the complete steady state branch (dotted black line) on which these traveling states terminate {in terms of $A_{\rm max}$.} Thus the fast fronts are no longer connected with the slow fronts shown in Fig.~\ref{fig:speed}; the latter are not included in Fig.~\ref{fig:DYm7}(a).

Figure~\ref{fig:DYm7}(b) shows the termination of the single peak steady states at small values of $\rH$, and shows that these bifurcate from the left fold {$\rho_{\rm SN}^{+L}$} of the uniform states $\vP_+$ (Fig.~\ref{fig:bif_uni}, inset). This is a well-known possibility~\cite{yochelis2006reciprocal,burke2008classification,parra2016dark} and corresponds to the appearance of an infinitesimal bump on top of the uniform background state, whose amplitude then increases as one proceeds up along the dotted line.

\section{Discussion}\label{sec:discussion}

In this paper, we have identified an important phenomenon associated with front propagation that we have referred to as propagation failure. We have seen that propagation failure is associated with the presence of a fold on the front branch, and showed that this fold is only the first of many which are associated with the presence of a T-point in the equations for the front in the comoving frame. We showed that each turn around the spiral is associated with the addition of a spatial oscillation at the leading edge of the front. Following the front branch back out from the center of the spiral one finds a similar series of profiles albeit with an additional small amplitude precursor peak. Thus the net effect of the T-point is to generate the precursor peak. 

In fact we found two more T-points of this type, TP1 and TP2, both lying on disconnected branches of fronts that bifurcate from a branch of stationary states with no precursor and terminate back at almost the same parameter values on a nearby branch of stationary states with a precursor (Fig.~\ref{fig:DYm4_1P_rtrn_spiral2}). These states only differ in the structure of the dominant peak straddling the front, and may be part of a family of such T-points. {These T-points were uncovered using PBC instead of NBC but the mechanisms appear to be fundamentally the same. With PBC the solution travels when its left-right reflection symmetry is (spontaneously) broken, something that happens at parity-breaking bifurcations as in Fig.~\ref{fig:DYm4_1P_rtrn_spiral2}. In contrast, Fig.~\ref{fig:DYm4} is computed with NBC which can (and sometimes do) support different values of $A$ at either end of the domain. If this is the case, we may think of the moving solution as being generated via forced symmetry breaking due to unequal boundary values. In both cases it is the asymmetry that is responsible for the motion. As expected, our computations reveal that the speed in the latter (forced) case is larger ($v\sim 10^{-3}$) than in the former (spontaneous) case ($v\sim 10^{-5}$), despite the rather small difference between $\vP_*$ and $\vP_0$ at these parameter values.}

The addition of a precursor via a T-point arises in other parameter regimes. In Figs.~\ref{fig:DYm4_TP_1P} and \ref{fig:DYm4_2P_TP} we show two examples of similar transitions that arise when Eqs.~(\ref{eq:AI}) are solved with PBC. Here a passage through a pair of back-to-back T-points leads to the assimilation of a small precursor peak in the dominant peak, thereby changing a single dominant peak with a precursor into a double peak structure without a precursor, and a double peak with a precursor into a triple peak without one. As in the case of the spiral T-points these traveling states originate in parity-breaking bifurcations of stationary states. We expect that similar transitions involving yet more complex peaks can also be found.

The above results led us to investigate the properties of the steady states that give rise to the traveling states that are the primary focus of this work. In contrast to the foliated snaking present at small values of $\rH$ these states are all embedded in a $\vP_0$ background but also exhibit certain universal behavior. Tracking these states we found that the branch of symmetric stationary single peak states with a precursor undergoes a clockwise rotation in the $(\rH,A_{\rm max})$ plane during which the precursor peak amplitude grows to the same value as the dominant peak, thereby generating a 2P state consisting of two identical equispaced isolated peaks in the periodic domain. Double-peak states with a precursor undergo a similar rotation as $\rH$ varies, leading to a symmetric 3P state consisting of a symmetric double peak and an isolated single peak. Similar behavior for a triple peak states with a precursor also takes place. Note that the 2P state is essentially a 1P state computed on the half-domain $L/2$ and periodically replicated. It follows that we can construct an $n$P state consisting of $n$ identical equispaced isolated peaks in a domain of size $L$, all present at essentially the same value of $\rH$. Related behavior is found as part of the foliated snaking structure studied, for example, in Ref.~\cite{knobloch2021stationary} Throughout we compared our results for states with a precursor with the corresponding states without one. In certain respects the resulting bifurcation structure resembles the foliated snaking of equispaced peaks embedded in $\vP_*$~\cite{knobloch2021stationary} but here it involves a periodic multipeak state embedded in $\vP_0$ with a single peak located exactly half way between adjacent multipeak structures (Figs.~\ref{fig:DYm4_1P_stat} and \ref{fig:DYm4_2P_stat}).

Inevitably these multipeak states track one another but at very small values of $\rH$ the latter take on an intriguing form: they resemble a localized pattern state of the type exemplified by the Swift-Hohenberg equation~\cite{burke2006localized} (see Fig.~\ref{fig:DYm4_3P_stat}(c), {top} panel, computed at $\rH=2\times 10^{-5}$). This parameter value is close to those associated with the primary spiral T-point TP0 (Fig.~\ref{fig:DYm4}(a)) raising the possibility that the presence of these clumped peaks is associated with the oscillations at the leading edge of the traveling front displayed there.  Similar states involving additional peaks can also be found, all generated via the same clockwise rotation process in the $(\rH,A_{\rm max})$ plane. These states are unexpected for two reasons: no Turing or spatially periodic state is present at these parameter values and individual peaks with {monotonic profiles in general repell one another. Yet, despite this, coexisting states of this type are still present.}

Many of the states we have investigated are unstable in one spatial dimension. In particular, in 1D we found that for $\rH>\rf$ the only nontrivial stable state are the slow fronts that come in from large $\rH$. These fronts are of pushed type since they describe the invasion of a stable uniform state, here $\vP_0$, by another stable uniform state, here $\vP_*$. For $\rH<\rf$ these pushed fronts give way to a pulsating propagating front via a SNIPER bifurcation at $\rH=\rf$ and no uniformly traveling fronts, stable or unstable, are present. However, even these time-dependent states coexist with the uniform state $\vP_0$ which remains stable down to the Turing bifurcation at $\rT$ and below (see Fig.~\ref{fig:bif_uni}).

We conclude by reflecting on the significance of the results for the original motivation of improving our understanding of branching and more broadly for the modeling of biological systems. One of the key questions in attempts to describe key components of biological systems (e.g., reaction rates, interactions, and transport) in the governing equations is the minimal form that captures the essential details. Naturally, the desire is to simplify the model system as much as possible for the benefit of potential analysis. Yet, there is a question of robustness: how can one tell if insights provided by the analysis are model-specific or model-independent? In this paper, our motivation stemmed from a model suggested as a prototypical system for studying branching~\cite{meinhardt1976morphogenesis} and applied to pulmonary vasculature during lung development.\cite{yao2007matrix,guo2014branching} The key feature of the underlying process is front propagation delimiting the transition from undifferentiated cells to an irreversible commitment to differentiate.

In companion papers~\cite{knobloch2021stationary,knobloch2022instability} we have addressed the properties of localized solutions associated with the Turing bifurcation at $\rT$. Here we extended the scope of these studies to propagating fronts in this system and uncovered the mechanism behind the observed front propagation failure. Moreover, we showed that the mechanism is rather robust in the context of multiplicity of steady-state solutions (associated with chemical signaling) and diffusion constants (here of the differentiation field). The robustness arises from the organization of the front solutions, viewed as homoclinic or heteroclinic orbits in comoving coordinates, via different global T-point bifurcations, where each such point organizes the solutions within a different diffusivity regime. As a result, we believe that the unexpected presence of a SNIPER bifurcation of propagating fronts is a generic feature of this class of models that is likely also present in other multivariable reaction-diffusion models, thereby broadening the applicability of the present study to other biological systems exhibiting bistability, such as somitogenesis.\cite{baker2008mathematical}

\begin{acknowledgments}
We have benefited from discussions with Nicol\'as Verschueren. This work was supported in part by the National Science Foundation under grant DMS-1908891 (EK).
\end{acknowledgments}

\section*{DATA AVAILABILITY}
The data that support the findings of this study are available
from the corresponding author upon reasonable request.
\\

\begin{thebibliography}{36}
	\expandafter\ifx\csname natexlab\endcsname\relax\def\natexlab#1{#1}\fi
	\expandafter\ifx\csname bibnamefont\endcsname\relax
	\def\bibnamefont#1{#1}\fi
	\expandafter\ifx\csname bibfnamefont\endcsname\relax
	\def\bibfnamefont#1{#1}\fi
	\expandafter\ifx\csname citenamefont\endcsname\relax
	\def\citenamefont#1{#1}\fi
	\expandafter\ifx\csname url\endcsname\relax
	\def\url#1{\texttt{#1}}\fi
	\expandafter\ifx\csname urlprefix\endcsname\relax\def\urlprefix{URL }\fi
	\providecommand{\bibinfo}[2]{#2}
	\providecommand{\eprint}[2][]{\url{#2}}
	
	\bibitem[{\citenamefont{Whitesides and Grzybowski}(2002)}]{whitesides2002self}
	\bibinfo{author}{\bibfnamefont{G.~M.} \bibnamefont{Whitesides}}
	\bibnamefont{and}
	\bibinfo{author}{\bibfnamefont{B.}~\bibnamefont{Grzybowski}},
	\bibinfo{journal}{Science} \textbf{\bibinfo{volume}{295}},
	\bibinfo{pages}{2418} (\bibinfo{year}{2002}).
	
	\bibitem[{\citenamefont{Cross and Greenside}(2009)}]{cross2009pattern}
	\bibinfo{author}{\bibfnamefont{M.}~\bibnamefont{Cross}} \bibnamefont{and}
	\bibinfo{author}{\bibfnamefont{H.}~\bibnamefont{Greenside}},
	\emph{\bibinfo{title}{Pattern Formation and Dynamics in Nonequilibrium
			Systems}} (\bibinfo{publisher}{Cambridge University Press},
	\bibinfo{year}{2009}).
	
	\bibitem[{\citenamefont{Meron}(2015)}]{meron2015nonlinear}
	\bibinfo{author}{\bibfnamefont{E.}~\bibnamefont{Meron}},
	\emph{\bibinfo{title}{Nonlinear Physics of Ecosystems}}
	(\bibinfo{publisher}{CRC Press Boca Raton, FL}, \bibinfo{year}{2015}).
	
	\bibitem[{\citenamefont{Alonso et~al.}(2016)\citenamefont{Alonso, B{\"a}r, and
			Echebarria}}]{alonso2016nonlinear}
	\bibinfo{author}{\bibfnamefont{S.}~\bibnamefont{Alonso}},
	\bibinfo{author}{\bibfnamefont{M.}~\bibnamefont{B{\"a}r}}, \bibnamefont{and}
	\bibinfo{author}{\bibfnamefont{B.}~\bibnamefont{Echebarria}},
	\bibinfo{journal}{Reports on Progress in Physics}
	\textbf{\bibinfo{volume}{79}}, \bibinfo{pages}{096601}
	(\bibinfo{year}{2016}).
	
	\bibitem[{\citenamefont{Parkin et~al.}(2008)\citenamefont{Parkin, Hayashi, and
			Thomas}}]{parkin2008magnetic}
	\bibinfo{author}{\bibfnamefont{S.~S.} \bibnamefont{Parkin}},
	\bibinfo{author}{\bibfnamefont{M.}~\bibnamefont{Hayashi}}, \bibnamefont{and}
	\bibinfo{author}{\bibfnamefont{L.}~\bibnamefont{Thomas}},
	\bibinfo{journal}{Science} \textbf{\bibinfo{volume}{320}},
	\bibinfo{pages}{190} (\bibinfo{year}{2008}).
	
	\bibitem[{\citenamefont{Cross and Hohenberg}(1993)}]{ch93}
	\bibinfo{author}{\bibfnamefont{M.~C.} \bibnamefont{Cross}} \bibnamefont{and}
	\bibinfo{author}{\bibfnamefont{P.~C.} \bibnamefont{Hohenberg}},
	\bibinfo{journal}{Reviews of Modern Physics} \textbf{\bibinfo{volume}{65}},
	\bibinfo{pages}{851} (\bibinfo{year}{1993}).
	
	\bibitem[{\citenamefont{Burke and Knobloch}(2006)}]{burke2006localized}
	\bibinfo{author}{\bibfnamefont{J.}~\bibnamefont{Burke}} \bibnamefont{and}
	\bibinfo{author}{\bibfnamefont{E.}~\bibnamefont{Knobloch}},
	\bibinfo{journal}{Physical Review E} \textbf{\bibinfo{volume}{73}},
	\bibinfo{pages}{056211} (\bibinfo{year}{2006}).
	
	\bibitem[{\citenamefont{Knobloch}(2016)}]{knobloch2016localized}
	\bibinfo{author}{\bibfnamefont{E.}~\bibnamefont{Knobloch}},
	\bibinfo{journal}{IMA Journal of Applied Mathematics}
	\textbf{\bibinfo{volume}{81}}, \bibinfo{pages}{457} (\bibinfo{year}{2016}).
	
	\bibitem[{\citenamefont{Champneys et~al.}(2021)\citenamefont{Champneys,
			Al~Saadi, Bre{\~n}a-Medina, Grieneisen, Mar{\'e}e, Verschueren, and
			Wuyts}}]{champneys2021bistability}
	\bibinfo{author}{\bibfnamefont{A.~R.} \bibnamefont{Champneys}},
	\bibinfo{author}{\bibfnamefont{F.}~\bibnamefont{Al~Saadi}},
	\bibinfo{author}{\bibfnamefont{V.~F.} \bibnamefont{Bre{\~n}a-Medina}},
	\bibinfo{author}{\bibfnamefont{V.~A.} \bibnamefont{Grieneisen}},
	\bibinfo{author}{\bibfnamefont{A.~F.} \bibnamefont{Mar{\'e}e}},
	\bibinfo{author}{\bibfnamefont{N.}~\bibnamefont{Verschueren}},
	\bibnamefont{and} \bibinfo{author}{\bibfnamefont{B.}~\bibnamefont{Wuyts}},
	\bibinfo{journal}{Physica D} \textbf{\bibinfo{volume}{416}},
	\bibinfo{pages}{132735} (\bibinfo{year}{2021}).
	
	\bibitem[{\citenamefont{Bernitt et~al.}(2017)\citenamefont{Bernitt,
			D{\"o}bereiner, Gov, and Yochelis}}]{bernitt2017fronts}
	\bibinfo{author}{\bibfnamefont{E.}~\bibnamefont{Bernitt}},
	\bibinfo{author}{\bibfnamefont{H.-G.} \bibnamefont{D{\"o}bereiner}},
	\bibinfo{author}{\bibfnamefont{N.~S.} \bibnamefont{Gov}}, \bibnamefont{and}
	\bibinfo{author}{\bibfnamefont{A.}~\bibnamefont{Yochelis}},
	\bibinfo{journal}{Nature Communications} \textbf{\bibinfo{volume}{8}},
	\bibinfo{pages}{15863} (\bibinfo{year}{2017}).
	
	\bibitem[{\citenamefont{Hoon et~al.}(2012)\citenamefont{Hoon, Wong, and
			Koh}}]{hoon2012functions}
	\bibinfo{author}{\bibfnamefont{J.-L.} \bibnamefont{Hoon}},
	\bibinfo{author}{\bibfnamefont{W.-K.} \bibnamefont{Wong}}, \bibnamefont{and}
	\bibinfo{author}{\bibfnamefont{C.-G.} \bibnamefont{Koh}},
	\bibinfo{journal}{Molecular and Cellular Biology}
	\textbf{\bibinfo{volume}{32}}, \bibinfo{pages}{4246} (\bibinfo{year}{2012}).
	
	\bibitem[{\citenamefont{Itoh and Hasegawa}(2013)}]{itoh2013mechanistic}
	\bibinfo{author}{\bibfnamefont{T.}~\bibnamefont{Itoh}} \bibnamefont{and}
	\bibinfo{author}{\bibfnamefont{J.}~\bibnamefont{Hasegawa}},
	\bibinfo{journal}{The Journal of Biochemistry}
	\textbf{\bibinfo{volume}{153}}, \bibinfo{pages}{21} (\bibinfo{year}{2013}).
	
	\bibitem[{\citenamefont{Pismen}(2006)}]{pismen2006patterns}
	\bibinfo{author}{\bibfnamefont{L.~M.} \bibnamefont{Pismen}},
	\emph{\bibinfo{title}{Patterns and Interfaces in Dissipative Dynamics}}
	(\bibinfo{publisher}{Springer Science \& Business Media},
	\bibinfo{year}{2006}).
	
	\bibitem[{\citenamefont{Brauns et~al.}(2020)\citenamefont{Brauns, Halatek, and
			Frey}}]{brauns2020phase}
	\bibinfo{author}{\bibfnamefont{F.}~\bibnamefont{Brauns}},
	\bibinfo{author}{\bibfnamefont{J.}~\bibnamefont{Halatek}}, \bibnamefont{and}
	\bibinfo{author}{\bibfnamefont{E.}~\bibnamefont{Frey}},
	\bibinfo{journal}{Physical Review X} \textbf{\bibinfo{volume}{10}},
	\bibinfo{pages}{041036} (\bibinfo{year}{2020}).
	
	\bibitem[{\citenamefont{Ponedel and Knobloch}(2016)}]{ponedel2016forced}
	\bibinfo{author}{\bibfnamefont{B.~C.} \bibnamefont{Ponedel}} \bibnamefont{and}
	\bibinfo{author}{\bibfnamefont{E.}~\bibnamefont{Knobloch}},
	\bibinfo{journal}{The European Physical Journal Special Topics}
	\textbf{\bibinfo{volume}{225}}, \bibinfo{pages}{2549} (\bibinfo{year}{2016}).
	
	\bibitem[{\citenamefont{Yochelis}(2021)}]{yochelis2021nonlinear}
	\bibinfo{author}{\bibfnamefont{A.}~\bibnamefont{Yochelis}},
	\bibinfo{journal}{Chaos} \textbf{\bibinfo{volume}{31}},
	\bibinfo{pages}{051102} (\bibinfo{year}{2021}).
	
	\bibitem[{\citenamefont{Knobloch and Yochelis}(2021)}]{knobloch2021stationary}
	\bibinfo{author}{\bibfnamefont{E.}~\bibnamefont{Knobloch}} \bibnamefont{and}
	\bibinfo{author}{\bibfnamefont{A.}~\bibnamefont{Yochelis}},
	\bibinfo{journal}{IMA Journal of Applied Mathematics}
	\textbf{\bibinfo{volume}{86}}, \bibinfo{pages}{1066} (\bibinfo{year}{2021}).
	
	\bibitem[{\citenamefont{Knobloch and Yochelis}(2022)}]{knobloch2022instability}
	\bibinfo{author}{\bibfnamefont{E.}~\bibnamefont{Knobloch}} \bibnamefont{and}
	\bibinfo{author}{\bibfnamefont{A.}~\bibnamefont{Yochelis}},
	\bibinfo{journal}{Chaos} \textbf{\bibinfo{volume}{32}},
	\bibinfo{pages}{123129} (\bibinfo{year}{2022}).
	
	\bibitem[{\citenamefont{Meinhardt}(1976)}]{meinhardt1976morphogenesis}
	\bibinfo{author}{\bibfnamefont{H.}~\bibnamefont{Meinhardt}},
	\bibinfo{journal}{Differentiation} \textbf{\bibinfo{volume}{6}},
	\bibinfo{pages}{117} (\bibinfo{year}{1976}).
	
	\bibitem[{\citenamefont{Yao et~al.}(2007)\citenamefont{Yao, Nowak, Yochelis,
			Garfinkel, and Bostr{\"o}m}}]{yao2007matrix}
	\bibinfo{author}{\bibfnamefont{Y.}~\bibnamefont{Yao}},
	\bibinfo{author}{\bibfnamefont{S.}~\bibnamefont{Nowak}},
	\bibinfo{author}{\bibfnamefont{A.}~\bibnamefont{Yochelis}},
	\bibinfo{author}{\bibfnamefont{A.}~\bibnamefont{Garfinkel}},
	\bibnamefont{and} \bibinfo{author}{\bibfnamefont{K.~I.}
		\bibnamefont{Bostr{\"o}m}}, \bibinfo{journal}{Journal of Biological
		Chemistry} \textbf{\bibinfo{volume}{282}}, \bibinfo{pages}{30131}
	(\bibinfo{year}{2007}).
	
	\bibitem[{\citenamefont{Doedel et~al.}(2012)\citenamefont{Doedel, Champneys,
			Fairgrieve, Kuznetsov, Oldeman, Paffenroth, Sandstede, Wang, and
			Zhang}}]{doedel2012auto}
	\bibinfo{author}{\bibfnamefont{E.~J.} \bibnamefont{Doedel}},
	\bibinfo{author}{\bibfnamefont{A.~R.} \bibnamefont{Champneys}},
	\bibinfo{author}{\bibfnamefont{T.}~\bibnamefont{Fairgrieve}},
	\bibinfo{author}{\bibfnamefont{Y.}~\bibnamefont{Kuznetsov}},
	\bibinfo{author}{\bibfnamefont{B.}~\bibnamefont{Oldeman}},
	\bibinfo{author}{\bibfnamefont{R.}~\bibnamefont{Paffenroth}},
	\bibinfo{author}{\bibfnamefont{B.}~\bibnamefont{Sandstede}},
	\bibinfo{author}{\bibfnamefont{X.}~\bibnamefont{Wang}}, \bibnamefont{and}
	\bibinfo{author}{\bibfnamefont{C.}~\bibnamefont{Zhang}},
	\bibinfo{journal}{{AUTO}07p: {C}ontinuation and bifurcation software for
		ordinary differential equations, Concordia University,
		http://indy.cs.concordia.ca/auto}  (\bibinfo{year}{2012}).
	
	\bibitem[{\citenamefont{Glendinning and Sparrow}(1986)}]{glendinning1986t}
	\bibinfo{author}{\bibfnamefont{P.}~\bibnamefont{Glendinning}} \bibnamefont{and}
	\bibinfo{author}{\bibfnamefont{C.}~\bibnamefont{Sparrow}},
	\bibinfo{journal}{Journal of Statistical Physics}
	\textbf{\bibinfo{volume}{43}}, \bibinfo{pages}{479} (\bibinfo{year}{1986}).
	
	\bibitem[{\citenamefont{Zimmermann et~al.}(1997)\citenamefont{Zimmermann,
			Firle, Natiello, Hildebrand, Eiswirth, B{\"a}r, Bangia, and
			Kevrekidis}}]{zimmermann1997pulse}
	\bibinfo{author}{\bibfnamefont{M.~G.} \bibnamefont{Zimmermann}},
	\bibinfo{author}{\bibfnamefont{S.~O.} \bibnamefont{Firle}},
	\bibinfo{author}{\bibfnamefont{M.~A.} \bibnamefont{Natiello}},
	\bibinfo{author}{\bibfnamefont{M.}~\bibnamefont{Hildebrand}},
	\bibinfo{author}{\bibfnamefont{M.}~\bibnamefont{Eiswirth}},
	\bibinfo{author}{\bibfnamefont{M.}~\bibnamefont{B{\"a}r}},
	\bibinfo{author}{\bibfnamefont{A.~K.} \bibnamefont{Bangia}},
	\bibnamefont{and} \bibinfo{author}{\bibfnamefont{I.~G.}
		\bibnamefont{Kevrekidis}}, \bibinfo{journal}{Physica D}
	\textbf{\bibinfo{volume}{110}}, \bibinfo{pages}{92} (\bibinfo{year}{1997}).
	
	\bibitem[{\citenamefont{Sneyd et~al.}(2000)\citenamefont{Sneyd, LeBeau, and
			Yule}}]{sneyd2000traveling}
	\bibinfo{author}{\bibfnamefont{J.}~\bibnamefont{Sneyd}},
	\bibinfo{author}{\bibfnamefont{A.}~\bibnamefont{LeBeau}}, \bibnamefont{and}
	\bibinfo{author}{\bibfnamefont{D.}~\bibnamefont{Yule}},
	\bibinfo{journal}{Physica D} \textbf{\bibinfo{volume}{145}},
	\bibinfo{pages}{158} (\bibinfo{year}{2000}).
	
	\bibitem[{\citenamefont{Or-Guil et~al.}(2001)\citenamefont{Or-Guil, Krishnan,
			Kevrekidis, and B{\"a}r}}]{or2001pulse}
	\bibinfo{author}{\bibfnamefont{M.}~\bibnamefont{Or-Guil}},
	\bibinfo{author}{\bibfnamefont{J.}~\bibnamefont{Krishnan}},
	\bibinfo{author}{\bibfnamefont{I.}~\bibnamefont{Kevrekidis}},
	\bibnamefont{and} \bibinfo{author}{\bibfnamefont{M.}~\bibnamefont{B{\"a}r}},
	\bibinfo{journal}{Physical Review E} \textbf{\bibinfo{volume}{64}},
	\bibinfo{pages}{046212} (\bibinfo{year}{2001}).
	
	\bibitem[{\citenamefont{Romeo and Jones}(2003)}]{romeo2003stability}
	\bibinfo{author}{\bibfnamefont{M.~M.} \bibnamefont{Romeo}} \bibnamefont{and}
	\bibinfo{author}{\bibfnamefont{C.~K.} \bibnamefont{Jones}},
	\bibinfo{journal}{Physica D} \textbf{\bibinfo{volume}{177}},
	\bibinfo{pages}{242} (\bibinfo{year}{2003}).
	
	\bibitem[{\citenamefont{Champneys et~al.}(2007)\citenamefont{Champneys, Kirk,
			Knobloch, Oldeman, and Sneyd}}]{champneys2007shil}
	\bibinfo{author}{\bibfnamefont{A.~R.} \bibnamefont{Champneys}},
	\bibinfo{author}{\bibfnamefont{V.}~\bibnamefont{Kirk}},
	\bibinfo{author}{\bibfnamefont{E.}~\bibnamefont{Knobloch}},
	\bibinfo{author}{\bibfnamefont{B.~E.} \bibnamefont{Oldeman}},
	\bibnamefont{and} \bibinfo{author}{\bibfnamefont{J.}~\bibnamefont{Sneyd}},
	\bibinfo{journal}{SIAM Journal on Applied Dynamical Systems}
	\textbf{\bibinfo{volume}{6}}, \bibinfo{pages}{663} (\bibinfo{year}{2007}).
	
	\bibitem[{\citenamefont{Yochelis et~al.}(2022)\citenamefont{Yochelis, Flemming,
			and Beta}}]{yochelis2022versatile}
	\bibinfo{author}{\bibfnamefont{A.}~\bibnamefont{Yochelis}},
	\bibinfo{author}{\bibfnamefont{S.}~\bibnamefont{Flemming}}, \bibnamefont{and}
	\bibinfo{author}{\bibfnamefont{C.}~\bibnamefont{Beta}},
	\bibinfo{journal}{Physical Review Letters} \textbf{\bibinfo{volume}{129}},
	\bibinfo{pages}{088101} (\bibinfo{year}{2022}).
	
	\bibitem[{\citenamefont{Moreno-Spiegelberg
			et~al.}(2022)\citenamefont{Moreno-Spiegelberg, Arinyo-i Prats,
			Ruiz-Reyn{\'e}s, Matias, and Gomila}}]{moreno2022bifurcation}
	\bibinfo{author}{\bibfnamefont{P.}~\bibnamefont{Moreno-Spiegelberg}},
	\bibinfo{author}{\bibfnamefont{A.}~\bibnamefont{Arinyo-i Prats}},
	\bibinfo{author}{\bibfnamefont{D.}~\bibnamefont{Ruiz-Reyn{\'e}s}},
	\bibinfo{author}{\bibfnamefont{M.~A.} \bibnamefont{Matias}},
	\bibnamefont{and} \bibinfo{author}{\bibfnamefont{D.}~\bibnamefont{Gomila}},
	\bibinfo{journal}{Physical Review E} \textbf{\bibinfo{volume}{106}},
	\bibinfo{pages}{034206} (\bibinfo{year}{2022}).
	
	\bibitem[{\citenamefont{Raja et~al.}(2023)\citenamefont{Raja, {van Kan},
			Foster, and Knobloch}}]{raja2023}
	\bibinfo{author}{\bibfnamefont{M.}~\bibnamefont{Raja}},
	\bibinfo{author}{\bibfnamefont{A.}~\bibnamefont{{van Kan}}},
	\bibinfo{author}{\bibfnamefont{B.}~\bibnamefont{Foster}}, \bibnamefont{and}
	\bibinfo{author}{\bibfnamefont{E.}~\bibnamefont{Knobloch}},
	\bibinfo{journal}{arXiv:2303.00798}  (\bibinfo{year}{2023}).
	
	\bibitem[{\citenamefont{Strogatz}(2018)}]{strogatz2018nonlinear}
	\bibinfo{author}{\bibfnamefont{S.~H.} \bibnamefont{Strogatz}},
	\emph{\bibinfo{title}{Nonlinear Dynamics and Chaos: with Applications to
			Physics, Biology, Chemistry, and Engineering}} (\bibinfo{publisher}{CRC
		press}, \bibinfo{year}{2018}).
	
	\bibitem[{\citenamefont{Yochelis et~al.}(2006)\citenamefont{Yochelis, Burke,
			and Knobloch}}]{yochelis2006reciprocal}
	\bibinfo{author}{\bibfnamefont{A.}~\bibnamefont{Yochelis}},
	\bibinfo{author}{\bibfnamefont{J.}~\bibnamefont{Burke}}, \bibnamefont{and}
	\bibinfo{author}{\bibfnamefont{E.}~\bibnamefont{Knobloch}},
	\bibinfo{journal}{Physical Review Letters} \textbf{\bibinfo{volume}{97}},
	\bibinfo{pages}{254501} (\bibinfo{year}{2006}).
	
	\bibitem[{\citenamefont{Burke et~al.}(2008)\citenamefont{Burke, Yochelis, and
			Knobloch}}]{burke2008classification}
	\bibinfo{author}{\bibfnamefont{J.}~\bibnamefont{Burke}},
	\bibinfo{author}{\bibfnamefont{A.}~\bibnamefont{Yochelis}}, \bibnamefont{and}
	\bibinfo{author}{\bibfnamefont{E.}~\bibnamefont{Knobloch}},
	\bibinfo{journal}{SIAM Journal on Applied Dynamical Systems}
	\textbf{\bibinfo{volume}{7}}, \bibinfo{pages}{651} (\bibinfo{year}{2008}).
	
	\bibitem[{\citenamefont{Parra-Rivas et~al.}(2016)\citenamefont{Parra-Rivas,
			Knobloch, Gomila, and Gelens}}]{parra2016dark}
	\bibinfo{author}{\bibfnamefont{P.}~\bibnamefont{Parra-Rivas}},
	\bibinfo{author}{\bibfnamefont{E.}~\bibnamefont{Knobloch}},
	\bibinfo{author}{\bibfnamefont{D.}~\bibnamefont{Gomila}}, \bibnamefont{and}
	\bibinfo{author}{\bibfnamefont{L.}~\bibnamefont{Gelens}},
	\bibinfo{journal}{Physical Review A} \textbf{\bibinfo{volume}{93}},
	\bibinfo{pages}{063839} (\bibinfo{year}{2016}).
	
	\bibitem[{\citenamefont{Guo et~al.}(2014)\citenamefont{Guo, Chen, Zeng,
			Warburton, Bostr{\"o}m, Ho, Zhao, and Garfinkel}}]{guo2014branching}
	\bibinfo{author}{\bibfnamefont{Y.~A.} \bibnamefont{Guo}},
	\bibinfo{author}{\bibfnamefont{T.-H.} \bibnamefont{Chen}},
	\bibinfo{author}{\bibfnamefont{X.}~\bibnamefont{Zeng}},
	\bibinfo{author}{\bibfnamefont{D.}~\bibnamefont{Warburton}},
	\bibinfo{author}{\bibfnamefont{K.~I.} \bibnamefont{Bostr{\"o}m}},
	\bibinfo{author}{\bibfnamefont{C.-M.} \bibnamefont{Ho}},
	\bibinfo{author}{\bibfnamefont{X.}~\bibnamefont{Zhao}}, \bibnamefont{and}
	\bibinfo{author}{\bibfnamefont{A.}~\bibnamefont{Garfinkel}},
	\bibinfo{journal}{The Journal of Physiology} \textbf{\bibinfo{volume}{592}},
	\bibinfo{pages}{313} (\bibinfo{year}{2014}).
	
	\bibitem[{\citenamefont{Baker et~al.}(2008)\citenamefont{Baker, Schnell, and
			Maini}}]{baker2008mathematical}
	\bibinfo{author}{\bibfnamefont{R.~E.} \bibnamefont{Baker}},
	\bibinfo{author}{\bibfnamefont{S.}~\bibnamefont{Schnell}}, \bibnamefont{and}
	\bibinfo{author}{\bibfnamefont{P.~K.} \bibnamefont{Maini}},
	\bibinfo{journal}{Current Topics in Developmental Biology}
	\textbf{\bibinfo{volume}{81}}, \bibinfo{pages}{183} (\bibinfo{year}{2008}).
	
\end{thebibliography}

\end{document}